\documentclass[final,authoryear,3p,onecolumn]{elsarticle}
\usepackage[colorlinks=true,breaklinks=true]{hyperref}
\usepackage{amssymb}
\usepackage{amsthm}
\usepackage[fleqn]{amsmath}		
\usepackage{stmaryrd}			
\usepackage{boondox-cal}		
\usepackage{float,graphicx}		
\usepackage{subcaption}
\numberwithin{equation}{section}

\newcommand{\brr}{\frac{\partial a_1}{\partial \theta}}
\newcommand{\brs}{\frac{\partial a_2}{\partial \theta}}
\newcommand{\btr}{{\frac{\partial a_1}{\partial r}}}
\newcommand{\bts}{{\frac{\partial a_2}{\partial r}}}
\newcommand{\contr}{\!\boldsymbol{\cdot}\!}
\renewcommand{\div}{{\scriptstyle \bm{\nabla}} \contr}

\newcommand{\curl}{{\scriptstyle \bm{\nabla}} \times }

\newcommand{\grad}{{\scriptstyle \bm{\nabla}}}

\newcommand{\td}{ \overset{\bm .}}

\newcommand{\dtd}{ \overset{\bm{..}}}

\newcommand{\xdot}{\td{\bm{x}}}

\newcommand{\xddot}{\dtd{\bm{x}}}
\newcommand{\dcontr}{\boldsymbol{:}}
\newcommand{\psih}{\hat \psi}
\newcommand{\sigmae}{\overset{e}{\bm{\sigma}}}
\newcommand{\sigmai}{\overset{i}{\bm{\sigma}}}
\newcommand{\sigmasi}{\overset{i}{\sigma}}
\newcommand{\sigmase}{\overset{e}{\sigma}}
\newcommand{\sigmam}{\overset{m}{\bm{\sigma}}}
\newcommand{\fm}{\overset{m}{\bm{f}}}

\newcommand{\base}{\bm{e}}
\newcommand{\half}{\frac{1}{2}}
\newcommand{\tr}[1]{\text{tr}( {#1} )}
\renewcommand{\det}[1]{\text{det}( {#1} )}
\newcommand{\jump}[1]{\left\llbracket #1 \right\rrbracket}

\newcommand{\bm}[1]{\boldsymbol{#1}}
\newcommand{\cpx}[1]{{\bar {#1}}}

\let\oldtimes\times
\renewcommand{\times}{\!\oldtimes\!}

\usepackage{etoolbox}%
\makeatletter
\def\@mkboth#1#2{}
\newlength\appendixwidth
\preto\appendix{\addtocontents{toc}{\protect\patchl@section}}
\newcommand{\patchl@section}{%
    \settowidth{\appendixwidth}{\textbf{Appendix }}%
    \addtolength{\appendixwidth}{1.5em}%
    \patchcmd{\l@section}{1.5em}{\appendixwidth}{}{\ddt}
}
\makeatother

\date{Septembre 9, 2020}
\journal{JMPS}

\begin{document}

\hyphenation{Elec-tro-mag-net-ic--thermo-me-chan-i-cal}
\hyphenation{elec-tro}
\hyphenation{mag-neto}
\hyphenation{ther-mo}
\hyphenation{me-chan-i-cal}

\begin{frontmatter}



\title{A coupled electromagnetic-thermomechanical approach for the modeling of electric motors}

\author[label1]{N. Hanappier}
\author[label1,label3]{E. Charkaluk}
\author[label1,label3,label2]{N. Triantafyllidis\corref{cor}}

\address[label1]{Laboratoire de M\'ecanique des Solides (CNRS UMR 7649), Ecole Polytechnique, Institut Polytechnique de Paris}
\address[label3]{D\'{e}partement de M\'{e}canique, \'{E}cole Polytechnique, Route de Saclay, Palaiseau 91128, FRANCE}
\address[label2]{Aerospace Engineering Department \& Mechanical Engineering Department (emeritus), \\
The University of Michigan, Ann Arbor, MI 48109-2140, USA}
\cortext[cor]{Corresponding author: nicolas.triantafyllidis@polytechnique.edu}

\begin{abstract}

Future developments of lighter, more compact and powerful motors -- driven by environmental and sustainability considerations in the transportation industry -- involve higher stresses, currents and electromagnetic fields. Strong couplings between mechanical, thermal and electromagnetic effects will consequently arise and a consistent multiphysics modeling approach is required for the motors' design. Typical simulations -- the bulk of which are presented in the electrical engineering literature -- involve a stepwise process, where the resolution of Maxwell's equations provides the Lorentz and magnetic forces which are subsequently used as the external body forces for the resolution of Newton's equations of motion.  

The work presented here proposes a multiphysics setting for the boundary value problem of electric motors. 
Using the direct approach of continuum mechanics, a general framework that couples the electromagnetic, thermal and mechanical fields is 
derived using the basic principles of thermodynamics. 
Particular attention is paid to the derivation of the coupled constitutive equations for isotropic materials under small strain but arbitrary magnetization.
As a first application, the theory is employed for the analytical modeling of an idealized asynchronous motor for which we calculate the electric current, magnetic, stress and temperature fields as a function of the applied current and slip parameter. The different components of the
stress tensor and body force vector are compared to their purely mechanical counterparts due to inertia, quantifying 
the significant influence of electromagnetic phenomena.
\vspace{1cm}
\end{abstract}


\end{frontmatter}

\newpage
\tableofcontents

\section{Introduction}
\label{sec:introduction}

The increasing importance and market share of hybrid and purely electric vehicles, in the quest to reduce their carbon footprint, urges the electric motor industry to develop higher performance products with reduced manufacturing costs. New goals are set by various government agencies and industrial associations \citep{Lopez2019} in terms of efficiency,
reliability, power losses, power density, higher rotation velocity and reduced weight. Novel electric motor designs are needed to overcome these technological challenges in order to comply with the aforementioned technical objectives and appropriate modeling tools must be developed.

Modeling of electric motors has in the past been a topic studied predominantly by the electrical engineering community.
The focus has been on the calculation of the magnetic field and resulting torque and iron losses for different motor designs using both analytical, (e.g. see:
\cite{Boules1984, Zhu1993, LUBINIM2011}) and numerical (e.g. see:
\cite{Chari1971, Silvester1973, AbdelRazek1982, Arkkio1987, Huppunen2004}) methods. A particular class of analytical methods, termed \textit{``subdomain methods''} (e.g. see: \cite{Devillers2016}) constitute an approximate but efficient tool for evaluating the magnetic characteristics of motor concepts at the preliminary design stage.

In the late 90s, stress calculations in electric motors have appeared as a result of noise and vibrations concerns. As pointed out by \cite{Reyne1987}, the first difficulty encountered was the evaluation of the electromagnetic body forces, for which various authors gave different expressions, due to the absence of a consistent continuum electrodynamics framework. The multiplicity of the different formulations, direct as well as variational, for the thermomechanical-electromagnetic continuum, is a source of confusion. Different (albeit equivalent) expressions for the Maxwell stress and electromagnetic body forces can be obtained and are thus responsible for the difficulty in the correct modeling
of stresses in electric motors. For further discussion on this issue, the interested reader is referred to the article by \cite{KANKANALA2004} and book by \cite{Hutter2007}. The first FEM computations for stresses in electric motors used a stepwise, uncoupled, approach: electric currents and magnetic fields where calculated using a purely electromagnetic model; the electromagnetic body force vector 
was then introduced as the external body force in a purely mechanical model to calculate the resulting stress state (e.g. see: \cite{Reyne1988, Javadi1995}).

The above-described approximate methods are inadequate to deal with the true multiphysics nature of the electric motor problem. The magnetic fields and currents generate the body forces driving the motor. These become even more important in the ferrous materials with high magnetic susceptibility that are used to enhance and channel the magnetic flux for improved motor performance. Moreover, these materials have intrinsic strongly coupled magnetic and mechanical behavior, with the material magnetization influencing the stress state via the ``\textit{magnetostriction}'' phenomenon and the stress state of the material also impacting its magnetization 
via ``\textit{inverse magnetostriction} \citep{DanielHubert2020}. Moreover, strong currents influence temperature due to ohmic effects and so on. 
Recognizing these issues, recent work by \cite{Fonteyn2010PhD, Fonteyn2010, Fonteyn2010b} takes 
into account magnetoelastic coupling effects for the numerical stress calculation in electric motors. However several approximations are used (e.g. a small strain approximation involving non frame-indifferent invariants and the angular momentum balance principle is not imposed), thermal fields are not considered and resulting stresses are not compared to inertial terms, motivating the present study.

The goal of this work is a thermodynamically consistent formulation that couples the electromagnetic, thermal and mechanical effects for the boundary value problem of electric motors. On the theoretical side, general continuum mechanics theories coupling thermomechanical and electromagnetic effects in solids started back in the 1950s and 1960s. Although a literature review is beyond the scope of this study, a few comments are helpful to put in perspective the present work. As in \cite{Fonteyn2010, Fonteyn2010b}, the modeling approach followed here is the ``{\it direct}'' method\footnote{Also applied to the modeling of other electromagnetic problems such as Magneto-Rheological-Elastomers (e.g. see \cite{KANKANALA2004,Dorfmann2003}) or Electro-Magnetic Forming processes (e.g. see \cite{THOMAS2009}).} which uses conservation laws of continuum mechanics and the thermodynamics procedure introduced by \cite{COLEMAN1963} to obtain the problem's governing equations and constitutive laws; a very readable account is presented in the book by \cite{Kovetz2000}. For the electric motor applications of interest\footnote{Other applications use this approximation, such as ``{\it Electromagnetic Forming}''; e.g. see  \cite{THOMAS2009}.} the ``{\it eddy current}'' simplification of the problem is adopted (see \cite{Hiptmair2005} for a justification in linear materials) that neglects electric polarization and displacement currents for low frequency electric fields. This theory is subsequently used to obtain the analytical solution of an idealized asynchronous motor for which we calculate the electric current, magnetic, stress and temperature fields. The stress tensor and body force vector are compared to their purely mechanical counterparts due to inertia, quantifying  the significant influence of electromagnetic phenomena, a novelty in this area to the best of the author's knowledge. 

The presentation is organized as follows: following this introduction in Section~\ref{sec:introduction}, the general formulation for the boundary value problem for electric motors is given in Section~\ref{sec:general-formulation}, where particular attention is paid to the derivation of the coupled constitutive equations for isotropic materials under small strain but arbitrary magnetization. The analytical model of an idealized asynchronous motor is presented in Section~\ref{sec:motor},  where we calculate the magnetic
field for the rotor and airgap in addition to the temperature field, the magnetic, total and elastic stresses in the rotor and the torque as a function of the 
applied current and the slip parameter (equivalent to the mechanical torque). The results for three different rotor materials (electric steel, copper and aluminum)
using realistic geometric and operational regime values and material parameters are presented in Section~\ref{sec:results} and the work is
concluded with a critical review and suggestions for future work in Section~\ref{sec:conclusion}. The detailed derivations of the constitutive laws for isotropic
materials under small strain but arbitrary magnetization are given in \ref{appendix:smallstrainlinearization}, detailed expressions for some elastic stress solution
components are given in \ref{appendix:stresses} while the determination of the magnetostrictive coefficient is presented in \ref{appendix:magnetostrictionfit}.

\section{Boundary value problem for electric motors}
\label{sec:general-formulation}

The general formulation of the coupled electromagnetic-thermomechanical boundary value problem for electric motors is presented in this section.
Coordinate-free (dyadic) continuum mechanics notation is used with bold scripts referring to tensors, regular scripts to scalars. Eulerian fields are written using lowercase letters, while capital letters are used for their Lagrangian counterparts. A superposed dot $\td{f}$ denotes the total time derivative of field $f$. The method adopted is the current configuration, direct approach of continuum mechanics and tacitly assumes adequate smoothness of the fields involved. Unless stated otherwise, all field quantities are functions of the current position $\bm{x}$ and time $t$. Although the governing equations for electromagnetic continua are known (see \cite{Kovetz2000}), for self-sufficiency and clarity of the work a brief presentation is given in this section.

\subsection{General governing equations}
\label{sec:governing-equations}

The general equations of the problem can be distinguished in three groups, as presented in the subsections below: 
electromagnetics (Gauss and Amp\`ere), mechanics (conservation of mass, balance of
linear and angular momenta) and thermodynamics (conservation of energy and entropy inequality). In the applications considered 
the interfaces encountered are not moving with respect to matter, since they are either a free surface boundary or an interface between two different materials and hence in the sequel 
the interface velocity is the material velocity at the interface:  $\bm{v}_s = \xdot$. 

\subsubsection{Electromagnetics}
\label{sec:Maxwell-equations}

{\underline {\it Maxwell-Gauss law}} relates the \textit{electric displacement} $\bm d$ to the \textit{free electric charge density} $q$. The differential equation and the associated interface condition (in the absence of surface charges) are
\begin{equation}
\div \bm{d} = q\; ; \quad \bm{n} \contr \llbracket \bm{d} \rrbracket = 0 \; ,
\label{eq:gauss}
\end{equation}
where $\llbracket f \rrbracket$ denotes the jump of field $f$ across a boundary/interface surface with an outward normal $\bm{n}$. 

{\underline {\it Maxwell-Amp\`ere law}} links the \textit{h-field} (sometimes also called the \textit{magnetic field}) $\bm h$ to the time-rate of the \textit{electric displacement} $\bm d$ and the \textit{free total current density} $\bm{j}$. The differential equation and the associated interface condition are
\begin{equation}
\curl \bm{h} = \frac{\partial\bm{d}}{\partial t} + \bm{j}\; ; \quad \bm{n} \times \llbracket \bm{h} \rrbracket + (\bm{v}_s \contr \bm{n}) \jump{\bm d} = \bm{\kappa}  \; ,
\label{eq:ampere}
\end{equation}
where $\bm{v}_s$ is the \textit{velocity of the interface} and $\bm{\kappa}$ the corresponding \textit{surface current density}. The current density $\bm{j}$ consists of the \textit{conduction current density} $\mathbcal{j}$ augmented by the convection of free electric charges $q \xdot$, i.e. $\bm j = \mathbcal j + q\xdot$. 

{\underline{\it Maxwell-Faraday law}} relates the \textit{electric field} $\bm{e}$ to the time-rate of the \textit{magnetic field} 
(sometimes also called the \textit{magnetic flux}) $\bm{b}$. 
The differential equation and the associated interface condition are
\begin{equation}
\curl \bm{e} = - \frac{\partial\bm{b}}{\partial t}\; ; \quad \bm{n} \times \llbracket \bm{e} \rrbracket - (\bm{v}_s \contr \bm{n})\jump{\bm{b}} = 0 \; .  
\label{eq:faraday}
\end{equation} 

{\underline{\it No magnetic monopole law}} confirms the absence of signed magnetic charges (monopoles) -- hence the zero in its right-hand side, as compared to the Maxwell-Gauss law in \eqref{eq:gauss}. The corresponding differential equation and the associated interface condition are
\begin{equation}
\div \bm{b} = 0 \; ; \quad \bm{n}\contr\llbracket \bm{b} \rrbracket = 0 \; .  
\label{eq:no-monopole}
\end{equation}

It should be mentioned here that the first set of two equations -- Maxwell-Gauss and Maxwell-Amp\`ere -- result in the {\underline{\it charge conservation principle}} 
($\div \bm{j} + \partial q /\partial t = 0$) which thus need not be additionally enforced.

{\underline{\it Aether frame principle}} connects the fields $(\bm{d}, \bm{h})$ to $(\bm{e},\bm{b})$. For the electric motor applications of interest, the polarization of the material is assumed negligible, in contrast to its magnetization (electric motors  include magnets and high permeability materials). The corresponding relations are
\begin{equation}
	\bm{d}=\epsilon_0\bm{e}, \quad \bm{h}=\frac{1}{\mu_0}\bm{b} - \bm{m} \; ,
\label{eq:aether}
\end{equation}
where $\bm{m}$ is the \textit{magnetization} (per unit volume) of the material and $\epsilon_0$ and $\mu_0$ are respectively the \textit{electric permittivity}
and the \textit{magnetic permeability} of free space.

\subsubsection{Mechanics}
\label{sec:mechanics}

{\underline{\it Mass conservation}} is described by the following differential equation
\begin{equation}
\rho_0 = \rho J  \ \Longrightarrow  \  \td{\rho} + \rho (\div \xdot ) = 0 \; ,
\label{eq:massbalance}
\end{equation}
where $\rho_0$ and $\rho$ are respectively the reference and current \textit{mass densities} and $J \equiv \det{\partial \bm{x} / \partial \bm{X}}$ the volume change.
In the absence of a discontinuity propagating in the continuum the corresponding interface/boundary condition gives no additional information.

{\underline{\it Linear momentum balance}} requires the introduction of the \textit{generalized electromagnetic-mechanical momentum density} $\bm{g}$  
(e.g. \cite{Kovetz2000})
 -- instead of $\xdot$ for the purely mechanical problems -- to be determined subsequently and gives the following differential equation and boundary/interface
condition in the absence of mechanical surface tractions
\begin{equation}
 \rho \td{\bm{g}} = \div \bm{\sigma}  + \rho \bm{f}\; , \quad \bm{n}\contr\llbracket \bm{\sigma} \rrbracket = 0 \; . 
  \label{eq:linmombalance}
\end{equation}
The body force per unit mass $\bm{f}$ contains only \textit{external, purely mechanical body forces}, typically gravity. Electromagnetic forces are embedded in the \textit{total Cauchy stress} $\bm{\sigma}$ and in $\bm{g}$.

{\underline {\it Angular momentum balance}} The non-reciprocity of actions/reactions in an electromagnetic - thermomechanical continuum
implies an asymmetric stress tensor, thus requiring the introduction of the generalized momentum  $\bm g$, resulting in the following relation for the asymmetric total stress $\bm{\sigma}$\footnote{The wedge product of two vectors $\bm{a}$ and $\bm{b}$ is 
an antisymmetric rank two tensor, defined by $\bm{a}\wedge\bm{b}\equiv\bm{a}\bm{b}-\bm{b}\bm{a}$.}
\begin{equation}
 \rho \xdot \wedge \bm{g} =  \bm{\sigma} - \bm{\sigma}^{_T} \; .
\end{equation}
\label{eq:angmombalance}
As a check we note that for a purely mechanical theory where $\bm{g} = \xdot$, the Cauchy stress tensor is symmetric.

\subsubsection{Thermodynamics}
\label{sec:Thermodynamics}

Thus far the form of Maxwell laws and mechanics laws have the same expressions as in their corresponding purely electromagnetic and purely thermomechanical
counterparts; no electromagnetic body forces or body torques have been postulated. The coupling comes through the energy balance by adding an electromagnetic 
energy flux to the mechanical and thermal contributions, which allows
us to find the missing constitutive information involving the electromagnetic - thermomechanical coupling terms.
We denote by $\varepsilon$ the \textit{total specific energy of the continuum} (i.e. mechanical, electromagnetic and thermal) 
and by $\eta$ the \textit{specific entropy of the continuum}, each defined at a point $\bm{x}$ and time $t$. 

\underline{\it{Energy conservation}} for the generalized electromagnetic-thermomechanical continuum in local form and its associated boundary condition give
\begin{equation}
 \rho \td{\varepsilon} = \div \left( \bm{\sigma}\contr\xdot - \bm{q} - \mathbcal{e} \times \mathbcal{h} \right) + \rho(\bm{f}\contr\xdot + r)\; ; 
 \quad \bm{n}\contr\llbracket \bm{\sigma}\contr\xdot - \bm{q} - \mathbcal{e} \times \mathbcal{h} \rrbracket = 0 \; ,
\label{eq:engbalance}
\end{equation}
where $r$ is the \textit{internal heat source} per unit mass, $\bm{q}$ is the \textit{heat flux} and $\mathbcal{e} \times \mathbcal{h}$  -- also termed the 
\textit{Poynting vector} -- is the \textit{electromagnetic energy flux}, both fluxes leaving the continuum (hence their minus signs). The Poynting vector 
is the cross product of the \textit{electromotive force} $\mathbcal{e}$ by the \textit{magnetotomotive force} $\mathbcal{h}$
\begin{equation}
\mathbcal{e} \equiv \bm{e} + \xdot\times\bm{b}, \quad \mathbcal{h} \equiv \bm{h} - \xdot\times\bm{d} \; .
\label{eq:gal-eh}
\end{equation}

Let $\eta$ denote the \textit{specific entropy of the continuum}  at a point $\bm{x}$ and time $t$. 

\underline{\it{Entropy production inequality}}, written here in terms of the continuum's \textit{dissipation} $\mathcal{D}$ in local form and the associated boundary condition are
\begin{equation}
\mathcal{D} \equiv \rho T\td{\eta} - \rho r + T\div\left( \frac{ \bm{q}}{T} \right) \ge 0 \; ; \quad \bm{n}\contr\left \llbracket \frac{\bm{q}}{T}\right \rrbracket \ge 0 \; ,
\label{eq:entropyineq}
\end{equation}
where $T$ denotes the continuum's \textit{absolute temperature} field. Note that the adiabatic entropy source  and the adiabatic entropy flux have the same expressions 
as for the classical thermomechanics model: $\rho r/T$ and $-( \bm{n} \contr \bm{q})/T$ but $\eta$ and $\bm{q}$ may now also depend on the
electric and magnetic fields ($\bm{e}, \bm{b}$).

The stage is now set to exploit the requirement of a positive dissipation by applying the method of Coleman and Noll \citep{COLEMAN1963}
in order to obtain the problem's constitutive relations.

\subsection{Constitutive relations}
\label{sec:constitutive}

Instead of working with the total specific energy of the continuum $\varepsilon$, following \cite{Kovetz2000} we introduce the \textit{specific free energy of the solid} 
$\psi$,\footnote{Note that in the absence of electromagnetic fields, $\bm{g}$ reduces to $\xdot$ and $\psi$ to the Helmholtz specific free energy 
$\psi = \mathcal{u}-T \eta$, with $\mathcal{u} = \varepsilon - 1/2(\xdot \contr \xdot)$ the internal energy of the system, 
as expected from classical thermo-mechanics.} 
a function of the thermodynamic state variables: $\xdot$, $\bm{F}\equiv \partial \bm{x} / \partial \bm{X}$ the solid's \textit{deformation gradient}, $\bm{b}$, $\mathbcal{e}$, $T$, $\grad T$ and following \cite{THOMAS2009} $\bm{\xi}$, a set of \textit{internal variables} associated 
with the mechanical and magnetic dissipative processes in the solid
\begin{equation}
\psi(\xdot, \bm{F}, \bm{b}, \mathbcal{e}, T, \grad T, \bm{\xi}) \equiv \varepsilon - T\eta - \bm{g}\contr\xdot + \frac{1}{2}\xdot\contr\xdot - \frac{1}{\rho}\left[\frac{\epsilon_0}{2}\bm{e}\contr\bm{e} + \frac{1}{2\mu_0}\bm{b}\contr\bm{b} - \epsilon_0(\bm{e}\times\bm{b})\contr\xdot\right] \; .
\label{eq:freeengdef}
\end{equation}

Using the Coleman and Noll procedure, constitutive relations are deduced for $\eta,\ \bm{m},\ \bm{q},\ \bm{j},\ \bm{g},\ \bm{\sigma}$ and $\dot {\bm{\xi}}$, 
in terms of the thermodynamic state variables. These relations are distinguished in two categories: \textit{necessary constitutive relations (equalities)} obtained from 
\textit{reversible restrictions} -- involving terms multiplying the rates or gradients of the state variables that can assume arbitrary values -- and \textit{sufficient constitutive relations (inequalities)} deduced from \textit{non-reversible restrictions}, and more specifically from the dissipation inequality, once its reversible terms are removed.

{\underline{\it Constitutive equalities}} give the following results 
(in addition to $\partial\psi/\partial \mathbcal{e} = \partial\psi/\partial (\grad T)=\partial\psi/\partial \xdot = \bm{0}$) 
\begin{equation}
\begin{array}{rl}
\bm{\sigma} =&\!\!\! \displaystyle \rho \bm{F} \contr \bigg( \frac{\partial \psi}{ \partial \bm{F}} \bigg)^T + \epsilon_0\Big(\bm{e} \bm{e} - 
\frac{1}{2}(\bm{e}\contr\bm{e})\bm{I}\Big) + \frac{1}{\mu_0}\Big(\bm{b} \bm{b} - \frac{1}{2}(\bm{b}\contr\bm{b})\bm{I}\Big) - 
\Big( \bm{b} \bm{m} - (\bm{b}\contr\bm{m})\bm{I}\Big) + \xdot \epsilon_0(\bm{e}\times\bm{b}) \; , \vspace{0.2cm} \\
\bm{m} =&\!\!\! \displaystyle - \rho \frac{\partial\psi}{\partial \bm{b}} \; , \quad
\bm{g} = \xdot + \frac{1}{\rho}\epsilon_0 (\bm{e} \times \bm{b} )\; , \quad \eta = -\frac{\partial\psi}{\partial T} \; .
\end{array}
\label{eq:constitutive}
\end{equation}

Using the above results, in combination with \eqref{eq:engbalance} and \eqref{eq:freeengdef}, the dissipation inequality \eqref{eq:entropyineq} yields
\begin{equation}
 \mathcal{D} =  -\rho\frac{\partial\psi}{\partial \bm{\xi}} \contr \td{ \bm{\xi}} + \mathbcal{j}\contr\mathbcal{e} - \frac{\bm{q}}{T}\contr(\grad T) \ge 0 \; .
 \label{eq:dissipationineq}
\end{equation}

{\underline{\it Constitutive inequalities}} At this point no further details can be given
about a generalized \textit{Ohm's law} for the conduction current density $\mathbcal{j}$ and a generalized \textit{Fourier's law} for the heat flux $\bm{q}$, on
how they depend on the thermodynamic state variables, other than \eqref{eq:dissipationineq} has to be satisfied by 
\begin{equation}
\mathbcal{j} =  \mathbcal{\hat j} (\bm{F}, \bm{b}, T, \grad T, \bm{\xi}, \mathbcal{e}) \; , 
\quad \bm{q} =  \bm{\hat q}(\bm{F}, \bm{b}, T, \grad T, \bm{\xi}, \mathbcal{e})\; ,
\label{eq:constitutive_Ohm_Fourier}
\end{equation}
where it is assumed for simplicity that these vector fields are independent on $\xdot$.\footnote{No further assumption
is made here about the constitutive equation for the internal variables.} The well
known forms of these relations require further assumptions about linearity and decoupling between different physical mechanisms and will be discussed in
Subsection~\ref{sec:materials}.

Using the above-obtained constitutive results from \eqref{eq:constitutive}, we are now in a position to give a more concise 
than in \eqref{eq:freeengdef} expression for the solid's free energy
\begin{equation}
\rho \psi(\bm{F}, \bm{b}, T, \bm{\xi}) = \rho \varepsilon - \rho T \eta - \frac{\rho}{2}\xdot\contr\xdot - 
\left[\frac{\epsilon_0}{2}\bm{e}\contr\bm{e} + \frac{1}{2\mu_0}\bm{b}\contr\bm{b} \right] \; .
\label{eq:freeengdef_2}
\end{equation}
The above expression has a clear physical interpretation:  the solid's free energy density (per unit current volume) $\rho\psi$ is 
obtained from the corresponding total energy density 
$\rho\varepsilon$ of the continuum by subtracting the thermal contribution, the kinetic energy of the solid and the energy of the electromagnetic field.

One final restriction must be recalled, that of \textit{material frame indifference} which dictates the \textit{objectivity} of $\psi$, i.e. its invariance under
all translations and rigid body rotations of its arguments, dictating that
\begin{equation}
	\psi = \psih(\bm{C}, \bm{B}, T, \bm{\xi})\; ;\quad \bm{B} \equiv \bm{b}\contr\bm{F}\; ,\quad \bm{C} \equiv \bm{F}^{T}\contr\bm{F}\; ,
\label{eq:freeengrestrictions}
\end{equation}
where $\bm{C}$ the right Cauchy-Green deformation tensor. As it turns out, the use of $\psih$ is the most convenient for expressing the stress
tensor and its subsequent simplification for small strains.

\subsection{Potential formulation}
\label{sec:potentialformulation}

An alternative formulation of the two last Maxwell laws, \eqref{eq:faraday} and \eqref{eq:no-monopole}, involves the introduction of an electric scalar potential $\phi$ and a magnetic vector potential $\bm{a}$
\begin{equation}
\bm{e} = \displaystyle -\grad\phi - \frac{\partial\bm{a}}{\partial t}\; ; \quad \bm{b} = \displaystyle \curl\bm{a} \; .
\label{eq:empotentialseddycurrent}
\end{equation}
As defined, the two potentials $\phi$ and $\bm{a}$ are not unique and a \textit{gauge condition} needs to be additionally enforced, such as the 
\textit{Coulomb gauge}: $\div\bm{a}=0$.
For the problem at hand, the potential formulation leads to a lower number of unknowns, thus justifying its introduction in \eqref{eq:empotentialseddycurrent}.

\subsection{Eddy current approximation}
\label{sec:eddycurrentapprox}

A convenient approximation for certain applications of electromagnetism (electric motors, electromagnetic forming etc.) 
is the \textit{eddy current approximation}, which consists of ignoring the electric energy of the problem as compared to its magnetic counterpart
(e.g. \cite{THOMAS2009}). This assumption
neglects the free electric charges (and hence Gauss' equation \eqref{eq:gauss}) and results in ignoring the displacement current $\partial  {\bm d} /\partial t$ and the convection of electric charges $q\xdot$, and hence $\bm{j}=\mathbcal{j}$, in Maxwell-Amp\`ere's law \eqref{eq:ampere}. 
The simplified Maxwell-Amp\`ere's law and boundary condition, recalling also \eqref{eq:constitutive_Ohm_Fourier}$_1$, reduce to
\begin{equation}
\curl \bm{h} = \mathbcal{j} \; ; 
\quad \bm{n} \times \llbracket \bm{h} \rrbracket = \bm{\kappa} \; .
\label{eq:Maxwelllaws_eddycurrent}
\end{equation}
Note that the approximate charge conservation is $\div\mathbcal{j}=0$, which is automatically satisfied given \eqref{eq:Maxwelllaws_eddycurrent}$_1$.

For the mechanical governing equations, the eddy current approximation implies that electric field terms can be ignored compared
to their magnetic counterparts in the expression for the stress tensor in \eqref{eq:constitutive}$_1$ and in the linear momentum density in \eqref{eq:constitutive}$_3$, which now reduces to the classical mechanics condition $\bm{g}=\xdot$. As a consequence, the angular momentum balance \eqref{eq:angmombalance} now requires a symmetric total stress $\bm\sigma$, as found in \eqref{eq:constit_eddycurrent}. Note that other related works (see \cite{Fonteyn2010PhD,Fonteyn2010,Fonteyn2010b}) do not have a symmetric total stress. 

Taking into account \eqref{eq:Maxwelllaws_eddycurrent}, 
the simplified version of the linear momentum law \eqref{eq:linmombalance} is rewritten as
\begin{equation}
\rho \xddot = \div\left(2\rho\bm{F}\contr\frac{\partial {\psih}}{\partial\bm{C}}\contr\bm{F}^T \right) + \mathbcal{j} \times \bm{b} + \bm{m} \times (\curl \bm{b}) + (\div\bm{m})\bm{b} + \rho\bm{f}\; ; \quad \bm{n}\contr\llbracket \bm{\sigma} \rrbracket = 0 \; , 
\label{eq:LMB_eddycurrent}
\end{equation}
where $\mathbcal{j} \times \bm{b}$ are the Lorentz body forces, followed by the magnetic and the mechanical body forces.

The constitutive equalities under the eddy current approximation, written in terms of the specific free energy density $\psih$ in \eqref{eq:freeengrestrictions} take the form
\begin{equation}
\bm{\sigma} = 2\rho\bm{F}\contr\frac{\partial\psih}{\partial\bm{C}}\contr\bm{F}^T + \frac{1}{\mu_0}\Big(\bm{b} \bm{b} - \frac{1}{2}(\bm{b}\contr\bm{b})\bm{I}\Big) - \Big(\bm{m} \bm{b} + \bm{b} \bm{m} - (\bm{b}\contr\bm{m})\bm{I}\Big) \; , \quad \bm{m} = -\rho\bm{F}\contr\frac{\partial \psih}{\partial\bm{B}} \; .
\label{eq:constit_eddycurrent}
\end{equation}
The remaining constitutive relations, i.e. Ohm's and Fourier's laws \eqref{eq:constitutive_Ohm_Fourier}, the entropy
constitutive equality in \eqref{eq:constitutive}$_4$ and the dissipation inequality in \eqref{eq:dissipationineq} remain unaltered.

One more simplification is made possible by the eddy current approximation, consistent with ignoring the electric energy of the system 
(and hence Gauss's law), which allows the potential formulation for the electric field $\bm{e}$ to be expressed only in terms of the magnetic potential vector $\bm{a}$
\begin{equation}
\bm{e} = \displaystyle \bm{e}_{app}  - \frac{\partial\bm{a}}{\partial t} \; ,
\end{equation}
where $\bm{e}_{app}$ is an externally applied electric field (typically to the coil that drives the system, e.g. see \cite{THOMAS2009}). The magnetic field
is still given by $\bm{b} = \displaystyle \curl\bm{a}$ as in \eqref{eq:empotentialseddycurrent} and the electromotive force remains $\mathbcal{e} = \bm{e} + \xdot\times\bm{b}$, as defined in \eqref{eq:gal-eh}.

\subsection{Materials considered}
\label{sec:materials}

The eddy current boundary value problem formulated thus far is general, for it accounts for nonlinear magnetic and mechanical material response, 
both constitutive and kinematic (finite strains), as well as dissipative phenomena, i.e. plasticity, magnetic hysteresis etc., to be described by the
evolution laws for the internal variables $\bm{\xi}$. We assume that a typical electric motor in its steady-state regime experiences only small strains while it can also
sustain large magnetizations, often up to saturation level. The implications of these restrictions on the selected specific free energy and the resulting expressions
for the constitutive laws are given progressively below, as more assumptions are introduced from one step to the next.

\underline{\it Absence of mechanical and magnetic dissipation}
We consider material behavior that does not include plasticity or magnetic hysteresis, so internal variables $\bm{\xi}$ are not required
for material description. As a result, the specific free energy is a function of strain, magnetic field and temperature: $\psih(\bm{C}, \bm{B}, T)$.

\underline{\it Material isotropy}
Isotropy of the material response implies that its specific free energy is a function of six invariants (and temperature), 
i.e. $\psih(\bm{C}, \bm{B}, T)=\psih(I_1,I_2,I_3,J_1,J_2,J_3,T)$, where $I_i$ are the invariants of the right 
Cauchy-Green tensor $\bm{C}$ and $J_i$ are the coupled magneto-mechanical invariants of $\bm C$ and $\bm B$.

\underline{\it Decoupling of physical phenomena}
It is assumed that thermo-mechanical, thermo-magnetic couplings can be neglected, resulting in a separate thermal contribution $\psih_{th}$
constructed under the assumption of a constant specific heat coefficient $c_\epsilon$. It is further assumed that, in the absence of magnetic fields,
the free energy of the solid is $ \psih_e(I_1,I_2,I_3)$ and that the magneto-mechanical coupling is described by the magnetic interaction energy 
$ \psih_m(J_1,J_2,J_3)$.
\begin{equation}
\begin{array}{rl}
\psih(\bm{C}, \bm{B}, T) =&\!\!\! \psih_e(I_1,I_2,I_3) +  \psih_{m}(J_1,J_2,J_3) + \psih_{th} \; ;
\quad \psih_{th}= -c_{\epsilon}T [\ln \left(T/T_0\right) -1] \; , \vspace{0.2cm} \\
I_1 =&\!\!\! \tr{\bm{C}} \; , \quad I_2 = \half(\tr{\bm{C}}^2 - \tr{\bm{C}\contr\bm{C}})\; , \quad I_3 = \det{\bm{C}} \; , \vspace{0.2cm} \\
J_1 = &\!\!\! \bm{B}\contr\bm{C}^{-1}\contr\bm{B}\; , \quad J_2 = \bm{B}\contr\bm{B}\; , \quad J_3 = \bm{B}\contr\bm{C}\contr\bm{B} \; ,
\label{eq:decoupledenergy}
\end{array}
\end{equation}
where $T_0$ is a reference temperature.

The implication of isotropy and decoupling on the generalized Ohm and Fourier laws in \eqref{eq:constitutive_Ohm_Fourier} is discussed next. We assume that
the conduction current density $\mathbcal{j}$ depends solely on the electromotive force $\mathbcal{e}$ and that the heat flux $\bm{q}$ is only a function of the
temperature gradient $\grad T$
\begin{equation}\label{eq:isotorpiccurrentandheatflux}
	\mathbcal{j} = \gamma(\| \mathbcal{e}\|)\mathbcal{e}\; ; \quad  \bm{q} = - k(\| \grad T \|)\grad T\; ,
\end{equation}
where the scalar \textit{electrical conductivity} $\gamma(\| \mathbcal{e}\|) > 0$ and the scalar \textit{thermal conductivity} $k(\| \grad T \|) > 0$,
as dictated by the dissipation inequality \eqref{eq:dissipationineq}. The norm-dependence of these two scalar quantities is due to material isotropy.

\underline{\it Small strain approximation}
For the electric motor applications of interest here, we adopt the small strain approximation,
i.e. $\|\bm{\epsilon}\| \ll 1$, where $\bm{\epsilon} \equiv (1/2)(\grad\bm{u} + \bm{u}\grad)$.
Using Taylor series expansions in $\bm \epsilon$ about the reference configuration of the quantities involved up to first order in $\bm{\epsilon}$ 
and neglecting terms of order $\bm{\epsilon}\; \bm{b}$\footnote{The small strain constitutive expressions that include terms order $\bm{\epsilon}\; \bm{b}$ and the justification for the omission of these terms in \eqref{eq:smallearbitrb} are given in \ref{appendix:smallstrainlinearization}. In the completely analogous -- $\bm e \rightarrow \bm b, \ 
\bm p \rightarrow \bm m, \  \varepsilon_0 \rightarrow \mu_0^{-1}$ -- electroelastic problems neglecting the coupling terms is justified by assuming the small strain is of the same order as the square of the moderate electric fields, e.g. see \cite{Tian2012, Lefevre2017}.}, we obtain a total stress $\bm{\sigma}$ as the sum of a purely elastic part 
$\sigmae(\bm{\epsilon})$\footnote{The elastic part of the free energy $ \psih_e$ is independent of the magnetic field; 
upon linearization at $\bm{C}=\bm{I}$
one obtains the classical Lame constants $\lambda$ and $G$ appearing in \eqref{eq:smallearbitrb}.} and a purely magnetic part $\sigmam(\bm{b})$
\begin{equation}
\hspace{-0.25cm}
\begin{array}{l}
\displaystyle \bm{\sigma} = \overset{e}{\bm{\sigma}} + \overset{m}{\bm{\sigma}}\; ;\  \overset{e}{\bm{\sigma}} \equiv \lambda\tr{\bm{\epsilon}}\bm{I} + 2G\bm{\epsilon}\; ,
\  \overset{m}{\bm{\sigma}} \equiv \frac{1}{\mu_0}\left[\bm{bb} - \frac{1}{2}(\bm{b}\contr\bm{b})\bm{I}\right] - \frac{\chi(\|\bm{b}\|)}{\mu(\|\bm{b}\|)} \left[\bm{bb} - 
(\bm{b}\contr\bm{b})\bm{I}\right] + {\Lambda(\|\bm{b}\|)\over\mu(\|\bm{b}\|)}\bm{bb} \; , \vspace{0.2cm} \\
\displaystyle \bm{m} = {\chi(\|\bm{b}\|)\over\mu(\|\bm{b}\|)}\bm{b}\; ;\   {\chi(\|\bm{b}\|)\over\mu(\|\bm{b}\|)} 
= -2\rho_0\left[\frac{\partial \psih_m}{\partial J_1} + \frac{\partial  \psih_m}{\partial J_2} + \frac{\partial  \psih_m}{\partial J_3}\right]_{\bm{C}=\bm{I}}\; ,
\  {\Lambda(\|\bm{b}\|)\over\mu(\|\bm{b}\|)} = 2\rho_0\left[\frac{\partial \psih_m}{\partial J_2} + 2\frac{\partial  \psih_m}{\partial J_3} \right]_{\bm{C}=\bm{I}}\; ,
\label{eq:smallearbitrb}
\end{array}
\end{equation}
where $\chi(\|\bm{b}\|)$ is the material's \textit{magnetic susceptibility}, $\mu(\|\bm{b}\|)=\mu_0 [ 1+ \chi(\|\bm{b}\|)]$ its \textit{magnetic permeability} and 
$\Lambda(\|\bm{b}\|)$ a \textit{magneto mechanical coupling} coefficient\footnote{This coefficient gives the curvature of the strain vs magnetic field in 
a stress-free uniaxial magnetostriction experiment.}. It is important to note that at this stage our isotropic material model is valid for
small strains but arbitrary magnetization -- the typical case of interest in magnetic motors -- and that the corresponding magnetic susceptibility,
magnetic permeability and magnetomechanical coupling coefficient are functions of the norm of the magnetic field $\bm{b}$ (due to isotropy). We should also
mention another consequence of small strain: the density equals its reference counterpart, i.e. $\rho=\rho_0$, thus justifying its appearance \eqref{eq:smallearbitrb}.
A remark is in order at this point about the expressions presented in \eqref{eq:smallearbitrb}; they differ from similar expressions presented
by other authors (e.g. \cite{Aydin2017, Fonteyn2010PhD}) in view of our use of the objective invariants $J_k$ in our linearization procedure instead of their simplified, non-objective
counterparts. The interested reader can find the details of these lengthy derivations in \ref{appendix:smallstrainlinearization}.

\section{Application to an idealized asynchronous electric motor}
\label{sec:motor}

This section pertains to the steady-state regime solution of an idealized, asynchronous electric motor, consisting of a cylindrical rotor and
stator, as an application of the theory developed in Subsection~\ref{sec:eddycurrentapprox}. The solid cylindrical rotor geometry adopted here
for the sake of the analytical treatment of the boundary value problem, although uncommon in typical induction motors that have slots for
conducting wires, is used for high  frequency applications (see \cite{Gieras2012}). The novelty here lies in the analytical computation of the 
different body forces, stresses and temperature fields, performed using classical methods of elasticity. The results obtained  show how the analytical magnetic 
field computations presented by the electrical engineering community (e.g. \cite{LUBINIM2011, Gieras2012}, can be complemented by mechanics. 
An added advantage of this simplified analytical model is its use as a benchmark for verification in numerical codes.

To allow for an analytical solution, the motor geometry and the material behavior are considerably simplified using a 2D, plane strain framework 
and a homogeneous, linearized material response. The magnetic susceptibility $\chi$ and permeability $\mu$, the magneto mechanical coupling coefficient $\Lambda$, 
the electrical conductivity $\gamma$, the thermal conductivity $k$, the Lam\'e constants $\lambda$, $G$ and the mass density $\rho_0$
are all given constants. 
Details for the setting of the corresponding boundary value problem are given below, where the unknown fields to be determined are the scalar magnetic
potential $a$ ($\bm{a} = a \base_z$) in the rotor and the airgap, the rotor's temperature field $T$ and the Airy
stress potential $\phi$\footnote{Not to be confused with the electric potential, which is no longer needed.} of the elastic stress field $\sigmae$. 

\subsection{Problem description}
\label{sec:motor-descr}

\begin{figure}[H]
\centering
\includegraphics[width=.8\linewidth]{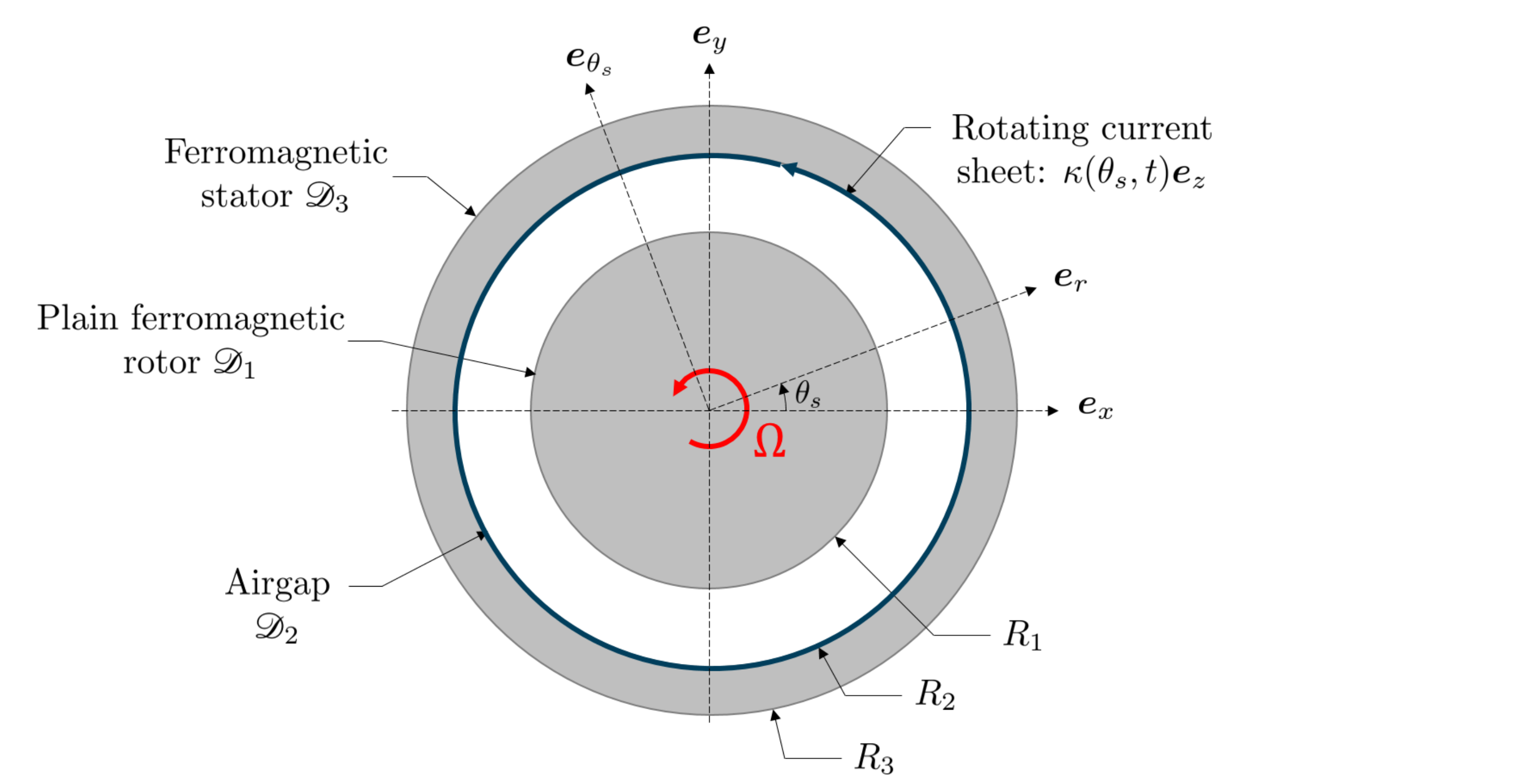}
\caption{Cross-section of the simplified electric motor, indicating rotor, airgap and stator domains and corresponding frames.}
\label{fig:motor1}
\end{figure}
The cross-section of the simplified induction motor is shown in Figure \ref{fig:motor1}; the motor is considered infinitely long in the normal
to the plane and under plane strain conditions. It is composed of a cylindrical ferromagnetic rotor (domain $\mathcal{D}_1: 0 \leq r \leq R_1$), 
surrounded by a cylindrical tubular stator (domain $ \mathcal{D}_3:  \ R_2 \leq r \leq R_3$), separated by an airgap 
(domain $ \mathcal{D}_2: \ R_1 \leq r \leq R_2$). Two different polar coordinate systems are used: the stator's fixed reference 
frame $\mathcal{S}(r,\theta_s,z)$  and the rotor's  moving frame $\mathcal{R}(r,\theta,z)$, where $\theta \equiv \theta_s - \Omega t$, 
with $\Omega$ the clockwise angular velocity of the rotor, as shown in Figure~\ref{fig:motor1}.

Following \cite{LUBINIM2011}, the motor is loaded by a current sheet of surface density $\bm{\kappa}$ perpendicular to the plane located on the internal radius of the stator. This current sheet models typical stator coils or windings supplied by a poly-phased (usually three-phased) alternating electric current of angular frequency $\omega$. The coils or windings are organized in $p$ pairs per phase and the applied surface current density is\footnote{For simplicity only the fundamental time harmonic of the current supply
is considered here.}
\begin{equation}
\bm{\kappa} = \kappa_0\cos(p\theta_s - \omega t)\base_z \; ,
\label{eq:currentsheet}
\end{equation}
with $\kappa_0$ the oscillation's amplitude in $A/m$. This current sheet rotates around the $z$-axis at the angular frequency $\omega/p$. It creates 
a rotating magnetic field of the same angular frequency, which triggers induced currents at the rotor. The interaction of the induced currents in
the rotor with the magnetic field create Lorentz forces that result in the rotor spinning at an angular frequency $\Omega$. 
Given that the phenomenon relies on induction, an angular frequency differential exists between the stator field and the rotor: $\Omega < {\omega}/{p}$. 
We thus define the relative angular frequency $\omega_r$ together with the slip parameter $s$
\begin{equation}
	\omega_r = \omega - p\Omega \; ; \quad s \equiv \frac{\omega_r}{\omega}\; ,	
	\label{eq:slip}
\end{equation}
where the angular velocities $\omega$ and $\Omega$ are constants, since the steady-state response of the motor is modeled.

Some additional assumptions are necessary to solve the problem.

{\underline{\it i) Infinite permeability, rigid stator}} It is assumed that the stator's strains are negligible -- thus guaranteeing a constant radius current sheet -- 
and that it has an infinite permeability, i.e. $\mu_3\longrightarrow \infty$, resulting in a negligible stator $h$-field
\begin{equation}
r > R_2: \ \bm{h}_3 = (\curl\bm{a}_3)/\mu_3 \approx {\bm 0}\; .
\end{equation}

{\underline{\it ii) Constant temperature airgap}} The air in the airgap is assumed to be maintained at a constant temperature $T_a$ by forced ventilation. 
Due to ohmic losses the rotor temperature rises, but a convective heat exchange discharges its excess
heat in the airgap. The corresponding radiation condition is
\begin{equation}
	r=R_1: \ \bm{q} \contr \base_r  = - k(\grad T)\contr \base_r = h_c(T(R_1)-T_a) \; ,
	\label{eq:heat-exchange}
\end{equation}
where $h_c$ is the convection coefficient and $T$ is the rotor temperature field.\footnote{The temperature field is only defined for the rotor, the $T_1$ notation is  
not used and the subscript $1$ is left out as superfluous.}

{\underline{\it iii) No external mechanical body forces}} No purely mechanical body forces, introduced in \eqref{eq:linmombalance} are considered,
i.e. $\bm{f}= \bm{0}$, since gravity effects are assumed negligible compared to inertia and magnetic contributions.

{\underline{\it iv) Constant velocity and acceleration}} Assuming a small slip $s$ ($\omega_r \ll \Omega$) and a small vibration amplitude, we can 
ignore the rates of the displacement 
$\td {\bm u}$ and $\dtd {\bm u}$ in the velocity and acceleration terms, by keeping only their $\Omega$-dependent contributions,
thus considerably simplifying the resulting algebra
\begin{equation}
\xdot \approx r \Omega \base_\theta \; , \quad \xddot \approx - r \Omega^2\base_r \; .
\label{eq:velocityapprox}
\end{equation}

One consequence is a constant inertia term $-\rho_0 r \Omega^2\base_r$ in the linear moment balance \eqref{eq:LMB_eddycurrent}. 
The other consequence of \eqref{eq:velocityapprox}$_1$ are the simpler expressions of the electromotive intensity $\mathbcal{e}$ defined in \eqref{eq:gal-eh}
and the material time derivative $\dot T$, when expressed in the moving rotor frame (recall $\theta \equiv \theta_s - \Omega t$)
\begin{equation}
\begin{array}{rlll}
\mathbcal{e}  &= \displaystyle \left[ -\frac{\partial \bm{a}}{\partial t} + \xdot \times {\bm b}\right]_{\mathcal{S}} &=  \displaystyle
 -\left[ \frac{\partial \bm{a}}{\partial t} + \Omega\frac{\partial \bm{a}}{\partial {\theta_s}}\right]_{\mathcal{S}} & = \displaystyle
 - \frac{\partial \bm{a}}{\partial t}\bigg|_\mathcal{R} \; , \vspace{0.2cm} \\ 
 \dot T & = \displaystyle \left[ \frac{\partial T}{\partial t} + \xdot \contr (\grad T)\right]_{\mathcal{S}} 
  & = \ \ \ \! \displaystyle \left[\frac{\partial T}{\partial t} + \Omega\frac{\partial T}{\partial {\theta_s}}\right]_{\mathcal{S}}
  & = \ \ \displaystyle \frac{\partial T}{\partial t}\bigg|_\mathcal{R}  \; .
 \end{array}
 \label{eq:small-strain-emf}
\end{equation}

Henceforth all equations are written in the rotor frame $\mathcal{R}$ and all field quantities are functions of $(r, \theta, t)$. These governing
equations and boundary conditions for the idealized, 2D motor are summarized below.
\begin{equation}
\begin{array}{rl}
	(r, \theta) \in {\mathcal D}_1 : &\!\!\!\! \curl\bm{b} = \mu\gamma\mathbcal{e}\; ; \quad (R_1,\theta) :   \base_r \times\jump{\bm{h}} = \bm{0}\; , 
	\  \base_r \contr\jump{\bm{b}} = \bm{0}\; , \vspace{0.2cm}\\
	(r, \theta) \in {\mathcal D}_2 : &\!\!\!\!   \curl\bm{b} = {\bm 0}\; ; \quad (R_2,\theta) :   \base_r \times \bm{h}  = \bm{\kappa}\; , \ 
	{\bm \kappa} = \kappa_0\cos(p\theta - \omega_r t)\base_z \; , \vspace{0.2cm} \\
	(r, \theta) \in {\mathcal D}_1 : &\!\!\!\!   \displaystyle \rho_0 c_\epsilon {{\partial T} \over {\partial t}} - k \grad^2 T = \gamma \mathbcal{e}\contr\mathbcal{e}\; ; 
	\quad  (R_1,\theta) :   \base_r \contr [k (\grad T)] = -h_c(T-T_a)\; ,  \vspace{0.2cm}   \\
	(r, \theta) \in {\mathcal D}_1 : &\!\!\!\!  \displaystyle \div(\sigmae + \sigmam ) = -\rho_0 r \Omega^2 \base_r \;  ;
	\quad   (R_1,\theta) :   \base_r \contr ({\bm{\sigmae}} +  \jump{\bm{\sigmam}}) = {\bm0} \; , \vspace{0.2cm} \\	
	(r, \theta) \in {\mathcal D}_1 \cup {\mathcal D}_2 : &\!\!\!\!  \bm{a} = a(r,\theta,t)\base_z\; ,\ \mathbcal{e} = \displaystyle - \frac{\partial \bm{a}}{\partial t}\; ,\
	\bm{b} = \curl\bm{a} = \frac{1}{r}\frac{\partial a}{\partial \theta}\base_r - \frac{\partial a}{\partial r}\base_\theta \; .
\end{array}
\label{eq:motorequations}
\end{equation}
In the boundary condition for the equilibrium equation \eqref{eq:motorequations}$_4$ there is no elastic stress field in the airgap
$\sigmae=\bm{0}$\footnote{The elastic stress field is only defined in the rotor, the $\sigmae_1$ notation is  
not used as unnecessary.}, in contrast to the magnetic stress field $\sigmam$ that exists in both the rotor and the airgap.

{\underline{\it v) External torque applied at rotor's center}} To balance the moment produced by the shear stresses, it is assumed that an 
external mechanical torque is applied at the center line of the rotor ($r=0$) along the $z$-axis. The resulting torque per unit rotor length $\mathcal{T}\base_z$ is
 
\begin{equation}
	\mathcal{T} = r^2 \int_0^{2\pi}\sigma_{r\theta}(r,\theta)\mathrm{d}\theta \; ,
	\label{eq:torque}
\end{equation}
and will be shown to be a constant, function of the relative angular frequency $\mathcal{T}(\omega_r)$ with $\mathcal{T}(0)=0$. 

\subsection{Dimensionless boundary value problem}
\label{sec:motor-formulation}

To guide the physical interpretation of the results, the following dimensionless variables and parameters of the problem are introduced
\begin{equation}
	\frac{r}{R_1} \rightarrow r, \quad \omega_r t \rightarrow t, \quad \frac{a}{\mu_0\kappa_0R_1} \rightarrow a, 
	\quad \frac{k(T-T_a)}{\gamma\omega_r^2(\mu_0\kappa_0)^2R_1^4} \rightarrow T,  
	\quad \frac{\bm{\sigma}}{\rho_0 R_1^2\Omega^2} \rightarrow \bm{\sigma}, \quad \zeta \equiv (R_2 - R_1)/R_1 \; .
	\label{eq:dimensionless}
\end{equation}
Henceforth, for simplicity the dimensionless variables and field quantities of the problem, $r,\; t,\; a, \; T,\; \bm{\sigma}$ are denoted by the same symbol
as their dimensioned counterparts.

The governing equations and the associated interface and boundary conditions (in the rotor frame) are given below,\footnote{Only the radius
of each domain of validity is recorded, since in all domains the angle $\theta\in[0,2\pi)$ and the time $t\in \mathbb {R}^+$.}
starting with the magnetic potential $a$
\begin{equation}
\begin{array}{ll}  \vspace{0.2cm}
\grad^2 a_1 = \alpha^2 \displaystyle \frac{\partial a_1}{\partial t} \; , \quad \alpha^2 \equiv \mu\gamma\omega_r R_1^2 \; ; & 0 \leq r \leq 1\; , \\  \vspace{0.2cm}
\displaystyle\frac{\partial a_1}{\partial r} = (1+\chi)\frac{\partial a_2}{\partial r}\; , \quad  
\frac{\partial a_1}{\partial \theta} = \frac{\partial a_2}{\partial \theta} \; ; & r=1 \; , \\ \vspace{0.2cm}
\grad^2 a_2 = 0 \; ; & 1 \leq r \leq 1 + \zeta\; , \\  \vspace{0.2cm}
\displaystyle\frac{\partial a_2}{\partial r} = \cos(p\theta - t) \; ; & r=1 + \zeta \; .
\end{array}
\label{eq:normalized-mag}
\end{equation}

The governing equation and boundary condition for the rotor's temperature field $T$ are
\begin{equation}
\begin{array}{ll}  \vspace{0.2cm}
\displaystyle {\mathcal F}^{-1}\frac{\partial T}{\partial t} - \grad^2 T = \left(\frac{\partial a_1}{\partial t}\right)^2 \; , 
\quad {{\mathcal F}} \equiv \frac{k}{\rho_0 c_\epsilon\omega_r R_1^2} \; ; & 0 \leq r \leq 1\; , \\
\displaystyle {\mathcal B}\frac{\partial T}{\partial r} + T = 0 \; , \quad {\mathcal B} \equiv \frac{k}{R_1h_c} \; ; & r=1 \; ,
\end{array}
\label{eq:normalized-temp}
\end{equation}
with ${\mathcal F}$ and ${\mathcal B}$ the ``{\it Fourier}'' and ``{\it Biot}'' dimensionless coefficients respectively.

Finally, the governing equations and boundary conditions for the rotor's elastic stress field $\sigmae$ are\footnote{Henceforth 
the rotor's body force is denoted by $\bm f$, taking the symbol used in \eqref{eq:linmombalance} for the purely mechanical body force.} 
\begin{equation}
\begin{array}{ll}
\displaystyle \div {\sigmae} =  {\bm{f}} \; ,\quad {\bm{f}} \equiv s_j \alpha^2\frac{\partial \bm{a}_1}{\partial t}\times(\curl \bm{a}_1) 
- s_m\grad\left(\|\curl\bm{a}_1\|^2\right) -  r \base_r \; ; & 0 \leq r \leq 1\; ,  \vspace{0.2cm} \\ 
\displaystyle s_j \equiv s_0 \frac{1 + \Lambda}{1 + \chi} \; ,  
\quad s_m \equiv {s_0\over 2}{{\chi + \Lambda}\over{1 + \chi}} \;  ,
\quad s_0 \equiv \frac{\mu_0\kappa_0^2}{\rho_0 R_1^2\Omega^2} \; . & \vspace{0.2cm} \\  
\displaystyle \sigmase_{rr} = \frac{s_0}{2} \left[ \left(\brs\right)^2 -\left(\bts\right)^2 \right] - (\frac{s_j}{2} + s_m)  \left(\brr \right)^2 
+ (\frac{s_j}{2} - s_m)   \left(\btr \right)^2   \; ; & r=1 \; , \vspace{0.2cm} \\ 
\displaystyle \sigmase_{r\theta} = - s_0 \brs\bts + s_j \brr\btr \; ; & r=1 \; .
\end{array}
\label{eq:normalized-stress}
\end{equation}
$s_0$ is an equivalent of the ``{\it Stuart}'' number for magnetic fluids and gives the ratio of Maxwell over inertia stress magnitudes. The dimensionless coefficients $s_j$ and $s_m$ appearing in the expressions for the total stress in the rotor ${\bm{\sigma}}_1$ (sum of the elastic $\sigmae$ and the magnetic $\sigmam$ components respectively) depend on its magnetic properties while the total stress tensor in the airgap ${\bm{\sigma}}_2$ (Maxwell stress in vacuum) depends only on $s_0$. The corresponding expressions for the magnetic field and the total stress in each domain are given by
\begin{equation}
\begin{array}{ll}
\displaystyle {\bm{\sigma}}_1 = \sigmae +\sigmam  \; , \ \sigmam  = \displaystyle  s_j \bm{b}_1\bm{b}_1 + (s_m-{s_j\over2})(\bm{b}_1\contr\bm{b}_1)\bm{I} \; ,\quad  
{\bm b}_1 = \curl {\bm{a}}_1 \; ; & 0 \leq r \leq 1\; , \vspace{0.2cm} \\
{\bm{\sigma}}_2 = \sigmam = \displaystyle  s_0  [ \bm{b}_2\bm{b}_2 - \half(\bm{b}_2\contr\bm{b}_2)\bm{I}] \; ,\quad  {\bm b}_2 = \curl {\bm{a}}_2 \; ;  & 1 \le r \le 1 + \zeta \; . 
\end{array}
\label{eq:total-stress}
\end{equation}

We first solve \eqref{eq:normalized-mag} to find the magnetic potential $a$, thus obtaining the ohmic dissipation for the heat equation 
\eqref{eq:normalized-temp}, which is then used to determine the rotor's
temperature field $T$. The magnetic potential gives the body forces for the linear momentum balance in
\eqref{eq:normalized-stress}, thus providing the rotor's elastic field $\sigmae$.

\subsection{Magnetic Potential}
\label{sec:mag-pot}

Solving the linear problem in \eqref{eq:normalized-mag} subject to the harmonic loading in \eqref{eq:currentsheet}, is more efficiently done
in the complex domain, where the magnetic potential $a_k(r,\theta,t)$ takes the form
\begin{equation}
a_k(r,\Theta) =  \Re\left\{ \cpx{a}_k(r)\exp(-i\Theta) \right\} = A_k(r) \cos\Theta + B_k(r) \sin\Theta \; ,\ k=1,2 \; ; \quad \Theta \equiv p\theta-t \; ,
\label{eq:MP-form}
\end{equation}
where $\cpx{a}_k(r)=A_k(r) + iB_k(r)$ is the complex\footnote{Complex quantities are henceforth denoted by an overbar $(\cpx{\ })$.} magnetic potential amplitude that depends only on $r$. 

In the rotor domain, \eqref{eq:normalized-mag} results in a Bessel differential equation for the complex amplitude $\cpx{a}_1(r)$  
\begin{equation}
r^2\frac{\mathrm{d}^2\cpx{a}_1}{\mathrm{d}r^2} + r\frac{\mathrm{d}\cpx{a}_1}{\mathrm{d}r} + ({\cpx\alpha}^2r^2 - p^2)\cpx{a}_1 = 0 \ 
\Longrightarrow \ \cpx{a}_1(r) = \cpx{A}J_{p}(\cpx{\alpha}r) \; ;\quad {\cpx\alpha}^2 \equiv - i\alpha^2 \; ,
\label{eq:MP-rotor}
\end{equation}
where the constant $\alpha^2$ is defined in \eqref{eq:normalized-mag} and $J_p$ denotes a Bessel function of the first kind. The above expression 
for $\cpx{a}_1$ accounts for the fact that there is no singularity in $r=0$, and hence explains the absence of a Bessel function of the second kind in the general solution. 

In the airgap domain, \eqref{eq:normalized-mag} gives a Laplace equation for the complex amplitude $\cpx{a}_2(r)$
\begin{equation}
r^2\frac{\mathrm{d}^2\cpx{a}_2}{\mathrm{d}r^2} + r\frac{\mathrm{d}\cpx{a}_2}{\mathrm{d}r} - p^2\cpx{a}_2 = 0 \ 
\Longrightarrow \ \cpx{a}_2(r) = \cpx{B}r^p + \cpx{C}r^{-p} \; .
\label{eq:MP-air}
\end{equation}
The complex-valued constants $\cpx{A}$, $\cpx{B}$ and $\cpx{C}$ appearing in \eqref{eq:MP-rotor} and \eqref{eq:MP-air} are determined
using the interface and boundary conditions in \eqref{eq:normalized-mag}, and are found to be
\begin{equation} 
\begin{array}{l}
\vspace{0.2cm}
\displaystyle \cpx{A} = {{2 \mathcal h} \over {J_p(\cpx{\alpha})+ \cpx{\mathcal g} }} \; , \\ \vspace{0.2cm}
\displaystyle \cpx{B} = \mathcal h \left[ 1 - {{2 \cpx{\mathcal g}}\over [J_p(\cpx{\alpha}) + \cpx{\mathcal g}][1 + (1+\zeta)^{-2p}]} \right] \; , \\ \vspace{0.2cm} 
\displaystyle \cpx{C} = \mathcal h \left[ 1 - {{2 \cpx{\mathcal{g}}}\over [J_p(\cpx{\alpha})+ \cpx g][1 + (1+\zeta)^{2p}]} \right] \; ,\\
\displaystyle \cpx{\mathcal g} \equiv  \left[ {J_p(\cpx{\alpha}) - \displaystyle{{\cpx{\alpha}}\over{p}} J_{p+1}(\cpx{\alpha})} \right]
\left[{{(1+\zeta)^{p} + (1+\zeta)^{-p}} \over {(1+\zeta)^{p}-(1+\zeta)^{-p}}}\right] {1 \over {1 + \chi}} \; , \quad \mathcal h \equiv {{ (1+\zeta)} \over {p[(1+\zeta)^{p} - (1+\zeta)^{-p}]}} \; .
\end{array}
\label{eq:MP-coefs}
\end{equation}

Using \eqref{eq:MP-coefs}, the sought real amplitudes $A_k(r)$ and $B_k(r)$ in \eqref{eq:MP-form} 
are given in terms of their complex counterparts found in \eqref{eq:MP-rotor} and \eqref{eq:MP-air},
i.e. $A_k(r) = \Re\left\{ \cpx{a}_k(r)\right\}, \  B_k(r) = \Im\left\{\cpx{a}_k(r)\right\} ; \ k=1,2$.

\subsection{Rotor temperature}
\label{sec:temp}

From the linearity of the governing equations for the temperature field in \eqref{eq:normalized-temp} and the magnetic potential 
solution in the rotor in \eqref{eq:MP-rotor}, the forcing term in the conduction equation is found to be: 
$(\partial a_1/\partial t)^2 = 0.5[(B_1(r))^2+(A_1(r))^2] + 0.5[(B_1(r))^2-(A_1(r))^2]\cos(2\Theta) - A_1(r)B_1(r)\sin(2\Theta)$. The use of
superposition and complex formulation lead to the following rotor temperature field $T(r,\theta,t)$
\begin{equation} 
T(r,\Theta) = T_0(r) + \Re\left\{{\cpx T}(r) \exp(-i2\Theta) \right\}\; ; \quad \Theta \equiv p\theta-t \; ,
\label{eq:Temp-rotor}
\end{equation}
where the function $T_0(r)$ is real and ${\cpx T}(r)$ is complex. The real function $T_0(r)$ is found from  \eqref{eq:normalized-temp} to be
\begin{equation}
\frac{\mathrm{d}^2 T_0}{\mathrm{d}r^2} + \frac{1}{r} \frac{\mathrm{d}T_0}{\mathrm{d}r}  = - {{B_1}^2(r)+{A_1}^2(r) \over {2}} 
\ \Longrightarrow \  T_0(r) =  c_0 - \frac{1}{2}\int_0^r\left(\frac{1}{r}\int_0^r [B_1^2+A_1^2] r \mathrm{d}r\right)\mathrm{d}r \; ,
\label{eq:Temp-T0}
\end{equation}
with the unknown constant $c_0$ to be determined from the boundary condition.

Solving for the complex function $\cpx{T}(r)$ is reduced to solving a Bessel differential equation with a forcing term through
the superposition of a homogeneous and a particular solution $\cpx{T}_p(r)$, as follows
\begin{equation}
\begin{array}{l} \vspace{0.2cm}
 \displaystyle  r^2 \frac{\mathrm{d}^2 \cpx T}{\mathrm{d}r^2} + r \frac{\mathrm{d}\cpx T}{\mathrm{d}r} + \left({\cpx \beta}^2 r^2 - (2p)^2 \right) \cpx T = 
r^2\frac{{\cpx a_1}^2}{2} \quad \Longrightarrow \quad \cpx T(r) = \cpx c J_{2p}(\cpx \beta r) + \cpx T_p(r) \; ;
\quad  {\cpx\beta}^2 \equiv - \frac{2i}{\mathcal F} \; , \\
\cpx T_p(r) = \displaystyle {\pi \over 4} \left[ - J_{2p}(\cpx \beta r) \int_0^r Y_{2p}(\cpx \beta r) {\cpx a_1}^2 r \mathrm{d}r +
Y_{2p}(\cpx \beta r) \int_0^r J_{2p}(\cpx \beta r) {\cpx a_1}^2 r \mathrm{d}r \right] \; ,
\end{array}
\label{eq:Temp-Tbar}
\end{equation}
where the unknown constant $\cpx c$ in the homogeneous part of the  solution will be specified from the boundary condition. In solving \eqref{eq:Temp-Tbar}
we made use of the fact that the solution is bounded at $r=0$, and hence there is no contribution from the Bessel function of
the second kind $Y_{2p}$ to the homogeneous part of the solution. However, $Y_{2p}$ does enter under the integrals in the expressions for the particular
solution $\cpx T_p (r)$ as seen above.

Finally, the boundary condition at $r=1$ in \eqref{eq:normalized-temp} splits into two boundary conditions: one for $T_0(r)$ that gives
$c_0$ and the other for $\cpx T(r)$ that provides $\cpx c$
\begin{equation}
\begin{array}{l}
 \vspace{0.2cm}
c_0 = \displaystyle \frac{1}{2}\left[ \int_0^{1}\left(\frac{1}{r}\int_0^r [(B_1(r))^2+(A_1(r))^2] r \mathrm{d}r\right)\mathrm{d}r + {\mathcal B} \int_0^{1} [(B_1(r))^2+(A_1(r))^2]r \mathrm{d}r \right] \; ,  \\
\cpx c  = \displaystyle {\pi \over 4} \left[ \int_0^1 Y_{2p}(\cpx \beta r) {\cpx a_1}^2(r) r \mathrm{d}r -
{{Y_{2p}(\cpx \beta)+ {\mathcal B} \cpx\beta Y'_{2p}(\cpx \beta) }\over{J_{2p}(\cpx \beta)+ {\mathcal B} \cpx\beta J'_{2p}(\cpx \beta) }}  \int_0^1 J_{2p}(\cpx \beta r) {\cpx a_1}^2(r) r \mathrm{d}r  \right] \; ,
\label{eq:Temp-const}
\end{array} 
\end{equation}
where $J'_{2p}$ and $Y'_{2p}$ denote the derivatives of the first and second kind Bessel functions of order $2p$ with respect to their argument. 

Having determined $T_0(r)$ and $\cpx T(r)$, one can find from \eqref{eq:Temp-rotor} the rotor
temperature field $T(r,\Theta)$. 

\subsection{Rotor stresses}
\label{sec:stresses}

The principle of superposition is used again for determining the rotor's elastic stress field $\sigmae$. Recalling the
definitions for $\bm{f}$ in \eqref{eq:normalized-stress}
and the solution for the magnetic potential $a_1$ in \eqref{eq:MP-form} and \eqref{eq:MP-rotor}, the body forces can be 
expressed as  $\bm{f}(r,\Theta) = \bm{N}(r) +  \grad V(r,\Theta)$\footnote{Given the electromagnetic part of the forcing $\overset{m}{\bm f} = -\div\overset{m}{\bm\sigma}$ in \eqref{eq:normalized-stress}, it is tempting to choose $\overset{e}{\bm\sigma} = -\overset{m}{\bm\sigma}$ as a particular solution to the electromagnetic forcing $\overset{m}{\bm f}$. However, this particular solution is ineligible as it does not satisfy the compatibility condition (see \cite{BARBER2009}), thus leading to the proposed approach.}, where $\bm{N}(r)$ is not derivable from a potential 
(non-conservative part of the force field), while the remaining terms are derivable from a potential $V(r,\Theta)$. 
\begin{equation}
\hspace{-1cm}
\begin{array}{rl}
\vspace{0.2cm}
\bm{f} = & \!\!\!\! \displaystyle \bm{N} + \grad V\; ; \\ \vspace{0.2cm}
\bm{N}=  & \!\!\!\! \displaystyle - \frac{s_j\alpha^2}{2}\frac{p}{r}(A_1^2+B_1^2)\base_\theta\; ; \
V(r,\Theta) = V_0(r) + V_{cs}(r,\Theta) \; , \ V_{cs} = V_s(r) \sin(2\Theta) + V_c(r) \cos(2\Theta) \; ,
\\  \vspace{0.2cm}
V_0(r) =& \!\!\!\! \displaystyle - {r^2 \over 2}
+ \frac{s_j\alpha^2}{2}\int_0^r (A_1B_1' - A_1'B_1)\mathrm{d}r
- \frac{s_m}{2}\left( \frac{p^2}{r^2}(A_1^2 + B_1^2) + (A_1'^2+B_1'^2) \right) \; ,
\\  \vspace{0.2cm}
V_c(r) = & \!\!\!\! \displaystyle -  \frac{s_j\alpha^2}{2}A_1B_1 
- \frac{s_m}{2}\left( \frac{p^2}{r^2}(B_1^2 - A_1^2) + (A_1'^2-B_1'^2) \right) \; ,
\\  \vspace{0.2cm}
V_s(r) = & \!\!\!\! \displaystyle  \frac{s_j\alpha^2}{2}\frac{(A_1^2-B_1^2)}{2} - s_m \left(\frac{p^2}{r^2}A_1B_1 + A_1'B_1'\right) \; . 
\end{array} 
\label{eq:potentials}
\end{equation}
Consequently, the rotor's elastic stress field $\sigmae$ is decomposed as follows
\begin{equation}
\sigmae(r,\Theta) = \sigmae^{_N}(r) +  \sigmae^{_V}(r,\Theta) + \sigmae^{_h}(r,\Theta) \; ,  \quad 
\left\{\begin{split}
 & \div \sigmae^{_N} = \bm{N} \; , \\
 & \div \sigmae^{_V} = \grad V \;  ,\\
 & \div \sigmae^{_h} = {\bm 0} \;  ,
\end{split}\right.
\label{eq:stress-equation-with-potentials}
\end{equation}
where each one of the constituent fields $\sigmae^{_N},\;  \sigmae^{_V},\; \sigmae^{_h}$ corresponds, in view of \eqref{eq:smallearbitrb}, 
to a compatible elastic strain field, i.e. derivable from a displacement field. By abuse of terminology we call these elastic stress fields 
\textit{elastically compatible}. 

Using the expression for $\bm{N}(r)$ from \eqref{eq:potentials}, an elastically compatible particular solution for $\sigmae^{_N}(r)$
is found\footnote{Because we look for a particular solution only, integration constants are discarded.\label{foot16}} by solving the tangential equilibrium ODE,
\begin{equation}
\begin{array}{l} \vspace{0.2cm}
\displaystyle \frac{\mathrm{d} \sigmase^{_N}_{r \theta}}{\mathrm{d}r}  + \frac{2}{r} \sigmase^{_N}_{r\theta} = -\frac{s_j\alpha^2}{2}\frac{p}{r}(A_1^2+B_1^2)
\ \Longrightarrow \ \sigmase^{_N}_{r\theta} = -\frac{s_j\alpha^2}{2}\frac{p}{r^2}\int_0^r r(A_1^2+B_1^2)\mathrm{d}r \; .
\end{array}
\label{eq:sigmae0}
\end{equation}

An elastically compatible particular  solution for the elastic stress field $\sigmae^{_V}$ is found using the Airy stress function method 
in polar coordinates (see \cite{BARBER2009}). The components of $\sigmae^{_V}$ can be expressed in
terms of a stress potential $\phi_{_V}$ as follows
\begin{equation}
\displaystyle \sigmase^{_V}_{rr} = \frac{1}{r}\frac{\partial\phi_{_V}}{\partial r} + \frac{1}{r^2}\frac{\partial^2\phi_{_V}}{\partial \Theta^2} + V \; ,\  \sigmase^{_V}_{\theta\theta} = \frac{\partial^2\phi_{_V}}{\partial r^2} + V \; , \  \sigmase^{_V}_{r\theta} = -\frac{\partial}{\partial r}\left(\frac{1}{r}\frac{\partial\phi_{_V}}{\partial \Theta}\right)\; ; \quad
\displaystyle \nabla^2\phi_{_V} = - {{1-2\nu}\over{1-\nu}} V \; . 
\label{eq:linksigma-phiV}
\end{equation}
The stress potential $\phi_{_V}$ is found (see footnote \ref{foot16}), by solving the Laplacian in \eqref{eq:linksigma-phiV} with the help of \eqref{eq:potentials}
\begin{equation}
\displaystyle \phi_{_V}(r,\Theta) = - {{1-2\nu}\over{1-\nu}}\left(\int_0^r\frac{1}{r}\int_0^r V_0 r \mathrm{d}r\ \mathrm{d}r + 
\frac{r^{2p}}{4p}\int_0^r V_{cs} r^{-2p+1}\mathrm{d}r - \frac{r^{-2p}}{4p}\int_0^r V_{cs} r^{2p+1} \mathrm{d}r \right) \; .  
\label{eq:particular-phi}
\end{equation}

The components of the elastically compatible  homogeneous solution stress field $\sigmae^{_h}$ are expressed in terms of the potential $\phi_h$ 
\begin{equation}
\displaystyle \sigmase^{_h}_{rr} = \frac{1}{r}\frac{\partial\phi_{_h}}{\partial r} + \frac{1}{r^2}\frac{\partial^2\phi_{_h}}{\partial \Theta^2} \; ,\  \sigmase^{_h}_{\theta\theta} = \frac{\partial^2\phi_{_h}}{\partial r^2} \; , \  \sigmase^{_V}_{r\theta} = -\frac{\partial}{\partial r}\left(\frac{1}{r}\frac{\partial\phi_{_V}}{\partial \Theta}\right)\; ; \quad
\displaystyle \nabla^4\phi_{_h} = 0 \; . 
\label{eq:linksigma-phih}
\end{equation}
Solving the biharmonic equation for $\phi_{_h}$ in \eqref{eq:linksigma-phih} we obtain\footnote{The solution is extracted from the general \cite{Michell1899} solution. Given the form of the body forces and boundary conditions, only those terms consistent with a solution of the form $\sigmae = \sigmase_0(r) + \sigmase_c(r)\cos(2\Theta) + \sigmase_s(r)\sin(2\Theta)$ are kept. Also, all terms leading to stress singularities in $r=0$ are excluded except for the $\Phi_{02}\theta$ term required for the torque at $r=0$. \label{Fourier-components}}
\begin{equation}
\phi_h(r,\Theta) = \Phi_{01}\frac{r^2}{2} + \Phi_{02}\theta + \left( \Phi_{c1} r^{2p} + \Phi_{c2} r^{2p+2} \right)\cos(2\Theta) + \left( \Phi_{s1} r^{2p} + \Phi_{s2} r^{2p+2} \right)\sin(2\Theta) \; .
\label{eq:homogeneous-phi}
\end{equation}
The final expressions for $\sigmae^{_h}, \sigmae^{_V}$ are obtained from \eqref{eq:linksigma-phih} and \eqref{eq:linksigma-phih}.
The six constants $\Phi_{01}, \Phi_{02}, \Phi_{c1}, \Phi_{c2}, \Phi_{s1}, \Phi_{s2}$ are determined by the $r=1$ boundary conditions in \eqref{eq:normalized-stress}
\begin{equation}
\begin{array}{ll}
\displaystyle \sigmase^{_V}_{rr}(1) + \sigmase^{_h}_{rr}(1) = \frac{s_0}{2} \left[ \left(\brs\right)^2 -\left(\bts\right)^2 \right] - (\frac{s_j}{2} + s_m)  \left(\brr \right)^2 
+ (\frac{s_j}{2} - s_m)   \left(\btr \right)^2  \; , \vspace{0.2cm} \\ 
\displaystyle \sigmase^{_N}_{r\theta}(1) + \sigmase^{_V}_{r\theta}(1) + \sigmase^{_h}_{r\theta}(1) = - s_0 \brs\bts + s_j \brr\btr  \; .
\end{array}
\label{eq:stress-bc}
\end{equation}
From the decomposition in radial, cosine and sine terms (see footnote \ref{Fourier-components}), result three equations for the normal and three equations for the tangential boundary conditions, thus uniquely determining the sought constants.
The full expressions for the stress at the rotor (elastic and magnetic components) can be then determined from \eqref{eq:total-stress}
and \eqref{eq:stress-equation-with-potentials} but are too cumbersome to be recorded here; the components of $\sigmae^{_V}$
and $\sigmae^{_h}$ are given in \ref{appendix:stresses}.
\subsection{Rotor torque}
\label{sec:torque}

We are now in a position to give the expression for the torque/unit length $\mathcal{T}$. Recalling \eqref{eq:torque} and using the results for the
stress field obtained above, one has
\begin{equation}
\mathcal{T} =  4\pi\rho_0 \Omega^2 R_1^4 s_0 p \left( \frac{\mathcal h}{\| J_p(\cpx{\alpha}) + 
\bar {\mathcal g} \|} \right)^2\Im\left\{ J_p(\cpx{\alpha})\cpx{\alpha}^*J_{p+1}(\cpx{\alpha})^* \right\} \; ,
\label{eq:torque-detail1}
\end{equation}
where $(\ )^*$ denotes complex conjugation. This result gives the torque in terms of geometry, applied current (poles, amplitude and frequency), 
magnetic and electric properties and density of the rotor. Remarkably, $\mathcal{T}$ is independent of the mechanical properties of the rotor, i.e. 
its shear modulus $G$ and Poisson ratio $\nu$.

As the torque is a function of slip velocity $\omega_r$, it is instructive to find from \eqref{eq:torque-detail1} the initial slope of the $\mathcal{T}(\omega_r)$ curve.
Using asymptotics of the Bessel functions with respect to $\cpx{\alpha}$ for 
$\| \cpx{\alpha} \|^2 = \alpha^2 = \omega_r \mu \gamma R_1^2<\!\!< 1$, one obtains
\begin{equation}
\mathcal{T} \approx \omega_r {{2\pi\gamma}\over{p(1+p)}} \left[{{\mu_0 \kappa_0 (1 + \chi)(1 + \zeta) R_1 } \over
{ (1+\zeta)^p+ (1+\zeta)^{-p} +(1+\chi) [(1+\zeta)^p - (1+\zeta)^{-p}]}}\right]^2 + O(\omega_r)^2\; .
\label{eq:torque-gen}
\end{equation}
One should keep in mind that the above expression gives only the initial slope of the $\mathcal{T}(\omega_r)$ curve, but depending on the problem,
the range of validity of this linear approximation can be very small.

\section{Results and discussion}
\label{sec:results}

Although we solve an idealized motor, the results presented here correspond to materials, geometries and operating parameters found in the 
electrical engineering literature. The dimensionless quantities introduced in \eqref{eq:dimensionless} allow a direct
comparison of the results to related physically meaningful quantities.
\subsection{Material, geometry and operating parameters}
\label{sec:values}

The motor geometry and operating parameters used in the calculations are shown in Table~\ref{tab:values}. The study covers three materials typically found in electric motors: electrical steel, copper and aluminum. Despite the different  motor architecture, the same values as in \cite{LUBINIM2011} are used whenever possible. The peak value of the current sheet is presently reduced to $1.3\times10^4 A/m$ -- from $8\times10^4 A/m$ in \cite{LUBINIM2011} -- in order to keep the maximum value of the magnetic field in the steel rotor below saturation,\footnote{The chosen current sheet amplitude results in a maximum magnetic field of about $1.3 T$ for the base case motor (steel rotor), roughly corresponding to the onset of magnetic field saturation for typical electrical steels (e.g. M400-50A), see \cite{Rekik2014}.} phenomenon not accounted for here. 

Unfortunately, not all needed parameters can be found for a particular electric steel, thus requiring the use of experimental
data from the open literature for comparable materials.
The value for the magneto-mechanical coupling coefficient $\Lambda$ is fitted from \cite{Aydin2017}, for the no-prestressed case, as detailed  in \ref{appendix:magnetostrictionfit}. A typical value for the magnetic susceptibility $\chi = 4000$ for electric steel is adopted, while the elastic constants $\nu$ and $E$ are taken from \cite{Belahcen2006}. The rest of the material parameters -- not given in \cite{Belahcen2006} and \cite{Aydin2017} -- are taken from the open literature, as it is also done for the case of copper and aluminum, where we assume negligible magnetic effects ($\chi = \Lambda = 0$).

The base case motor, which serves as a benchmark, is made of electric steel, has an airgap parameter $\zeta = 0.05$ and a slip parameter $s= 0.02$.
The rest of the geometric and operating parameters are kept fixed, independently of the rotor material, as shown below in Table~\ref{tab:values}. 
\begin{table}[H]
\begin{center}
\begin{tabular}{l l l l}
\hline 
\textbf{Geometry} \\ 
\hline
Rotor radius $R_1$ & $6$ cm & &	 \\
Airgap parameter $\zeta=(R_2-R_1)/R_1$ & $0.05$ (base case)  & & \\
Number of pole pairs $p$ & 2 & &	 \\
\hline \textbf{Operating parameters} \\ \hline
Peak value of current sheet $\kappa_0$ & $1.3\times10^4$ A/m  & &	 \\
Angular velocity of current supply $\omega$ 	& $100\pi$ rad/s  & & \\
Slip $s = \omega_r/\omega$ & $2$\% (base case)  & &  \\
External temperature $T_a$ & $20^\circ$C & &  \\
Convection coefficient $h_c$ &  $40$ W/m$^2$/K & &  \\
\hline
\textbf{Material properties} 	& \textbf{Electrical steel} 	& \textbf{Copper}		& \textbf{Aluminum} \\ 
\hline
Electric conductivity $\gamma$ 	& $2.67 \times 10^6$ S/m & $5.96 \times 10^7$ S/m	& $3.5 \times 10^7$ S/m \\
Magnetic susceptibility $\chi$ 		& $4,000$  			& $\approx 0$ 			& $\approx 0$ \\
Magneto-mechanical coupling $\Lambda$ &  $-1,800$              & $\approx 0$                    & $\approx 0$  \\
Mass density $\rho_0$ 			& $7,650\ \text{kg/m}^3$ 	&$8,940\ \text{kg/m}^3$    & $2,700\ \text{kg/m}^3$ \\
Young's modulus $E$ 			& $183 \times 10^9$ Pa	& $117 \times 10^9$ Pa 	& $69 \times 10^9$ Pa  \\
Poisson ratio $\nu$ 				& $0.34$				& $0.33$ 				& $0.32$ \\
Specific heat capacity $c_\epsilon$   & $480$\ J/kg/K 		& $385$ \ J/kg/K                &  $921$\ J/kg/K \\
Thermal conductivity $k$ 		        & $45$\ W/m/K                  & $397$ \ W/m/K               &  $225$ \ W/m/K\\
\hline
\end{tabular}
\caption{Motor geometry, operating parameters and rotor material properties \label{tab:values}}
\end{center}
\end{table}

As discussed in Subsection~\ref{sec:motor-descr}, the equations are solved in the rotor frame $\mathcal{R}$ 
and all field quantities are functions of $(r, \Theta)$, where $\Theta = p\theta -t$ and $p$
the motor pole number (here taken $p=2$). The results here are a snapshot
of these rotating fields at $t=0$ and are presented by plotting the corresponding field quantity at $(r, \theta)$.

\subsection{Magnetic field in rotor and airgap}
\label{sec:results-mag}

Magnetic field calculations for realistic geometries are routine for the electrical engineering community. The results for the
current simple motor geometry are presented here solely for the purpose of explaining the resulting force and strain fields. 

\begin{figure}[H]
    \centering
    \begin{subfigure}[t]{0.32\linewidth}
        \centering
        \includegraphics[width=\linewidth]{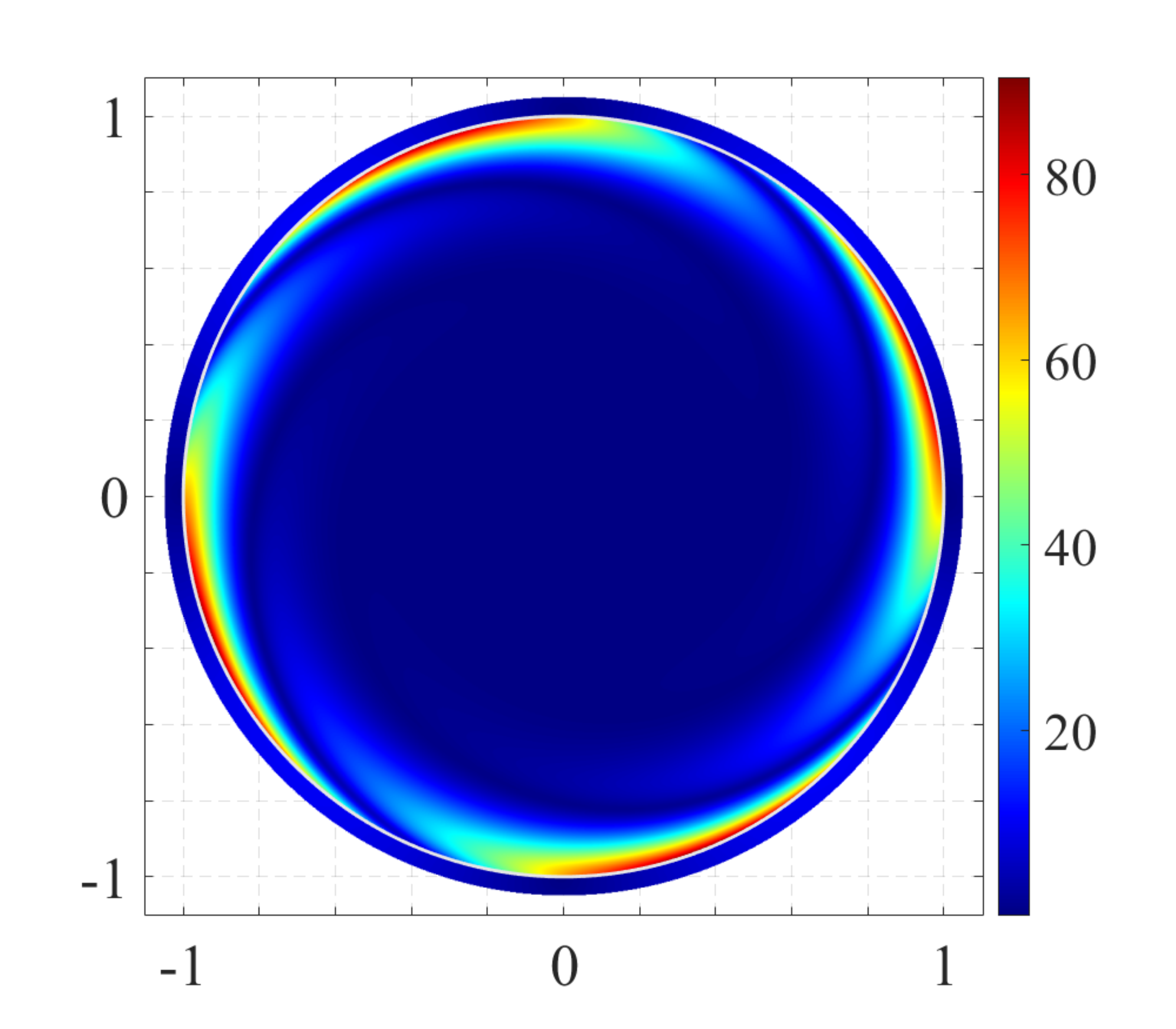}
        \caption{$\| \bm b \|$ for $s = 0.02$}
    \end{subfigure}%
    ~ 
    \begin{subfigure}[t]{0.32\linewidth}
        \centering
        \includegraphics[width=\linewidth]{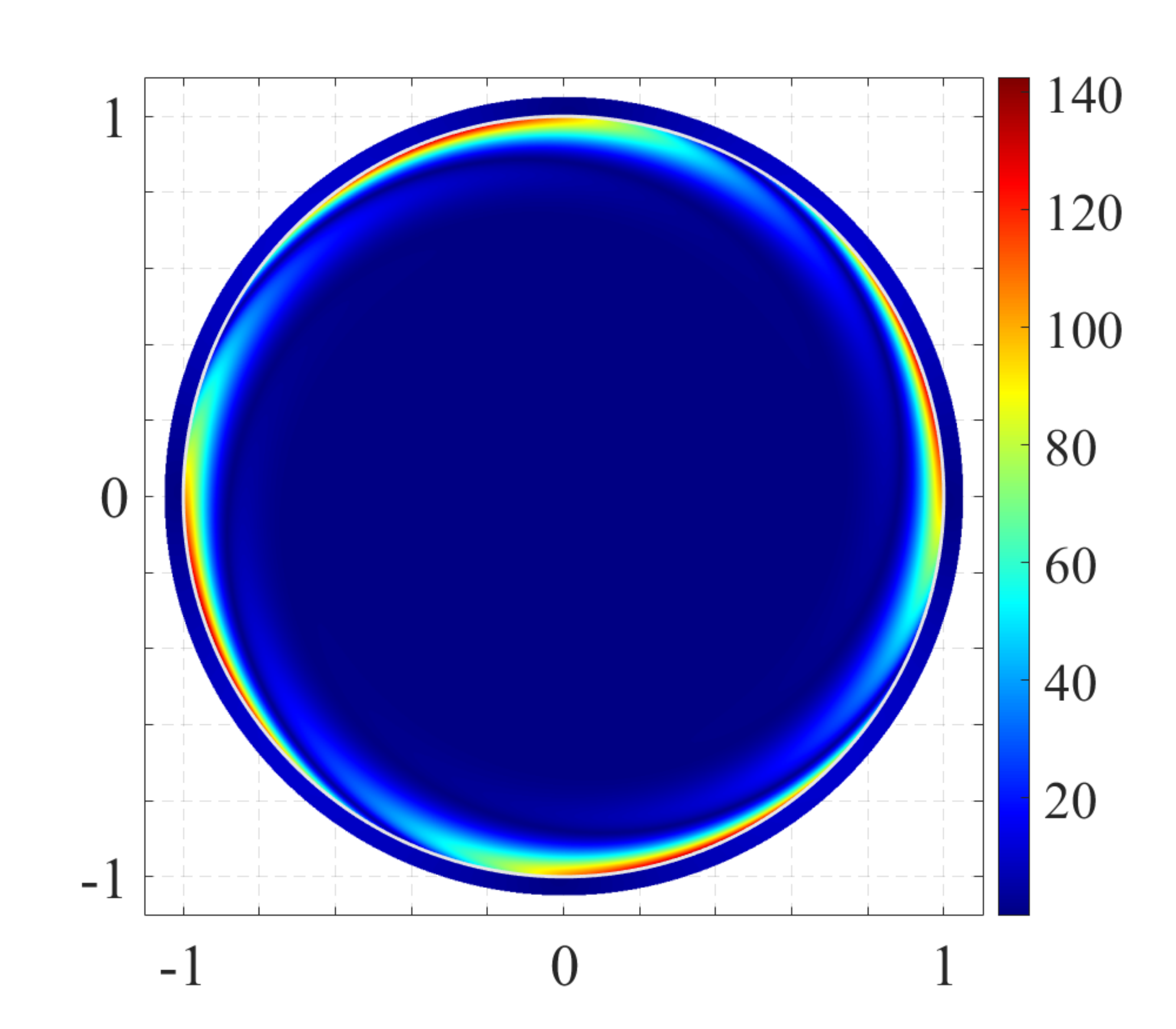}
        \caption{$\| \bm b \|$ for $s = 0.05$}
    \end{subfigure}
    ~
        \centering
    \begin{subfigure}[t]{0.32\linewidth}
        \centering
        \includegraphics[width=\linewidth]{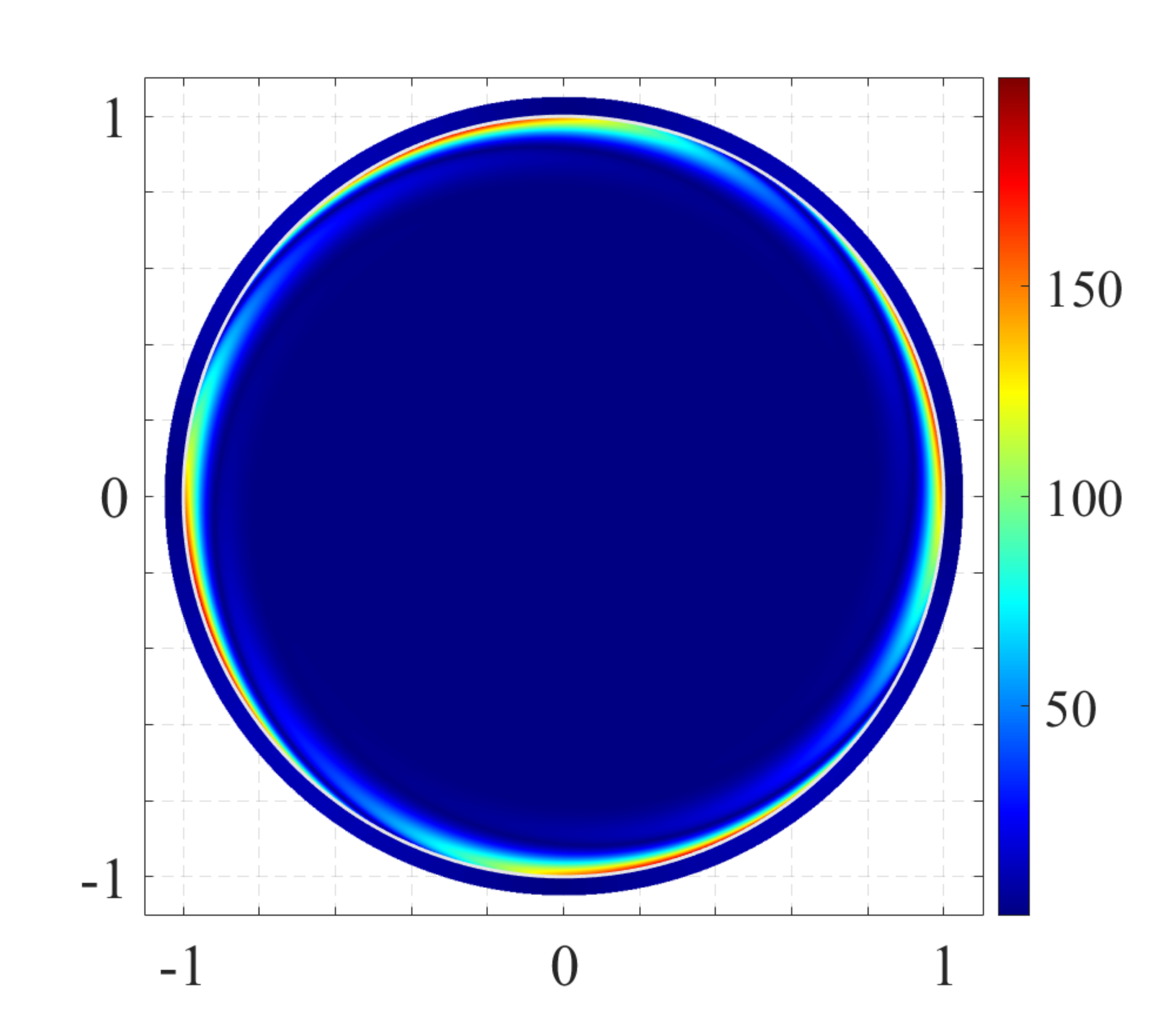}
        \caption{$\| \bm b \|$ for $s = 0.10$}
    \end{subfigure}%
    \caption{Magnetic field norm $\| \bm b \|$ for a steel rotor (normalized by $\mu_0 \kappa_0$), for different values of the slip parameter $s$.}
    \label{fig:magfield_slip}
\end{figure}

\begin{figure}[H]
	\centering
    \includegraphics[width=0.32\linewidth]{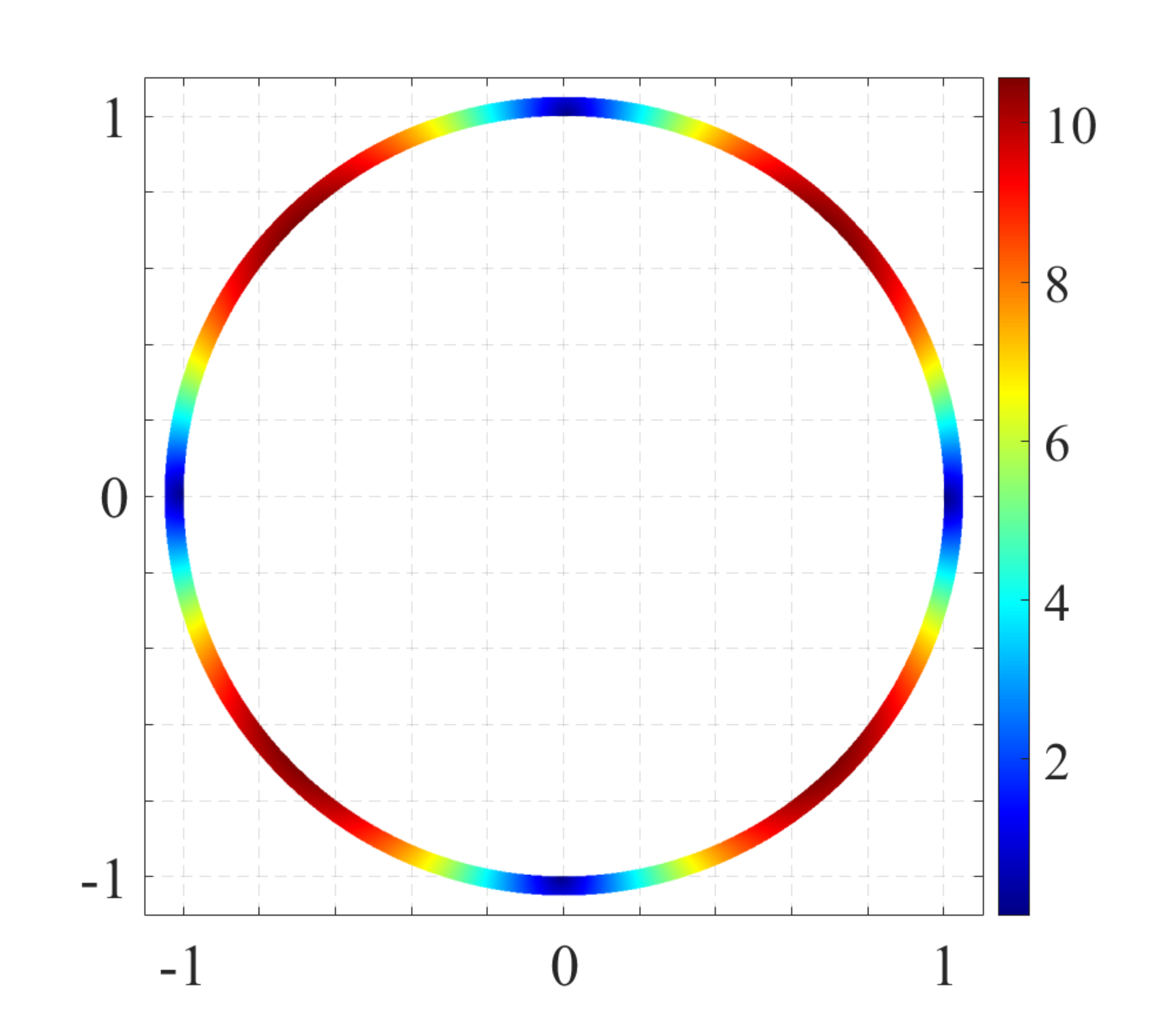}
    \caption{Magnetic field norm $\| \bm b \|$ in the airgap region (normalized by $\mu_0 \kappa_0$) for the base case motor in Figure~\ref{fig:magfield_slip}(a).}
    \label{fig:airgap_mag_field}
\end{figure}

The magnetic field plots in Figures~\ref{fig:magfield_slip} and \ref{fig:airgap_mag_field} show the contours of 
the dimensionless (normalized by $\mu_0\kappa_0$) magnetic field 
$\|\bm{b}\| = (b_r^2 +b_\theta^2)^{1/2}$ for three different values of the slip parameter $s = 0.02,\; 0.05,\; 0.10$ in the case of a steel rotor
with an airgap parameter $\zeta = 0.05$. Notice that the magnetic field increases away from the center and peaks in a localized
zone near the rotor periphery.  As the slip $s$ (equivalently the relative velocity $\omega_r$) increases, the localized high magnetization zone narrows,
(e.g. see \cite{Jackson1999} that the skin depth $\delta=(2/{\gamma\omega_r\mu})^{1/2}$). The four localized magnetic field zones are a result of the number of poles ($p=2$).

The high permeability of the rotor material ($\chi =4000$ for electric steel) drastically increases its magnetic field, 
thus masking the variations of the considerably smaller -- by one order of magnitude -- strength of the
 magnetic field in the airgap in Figures~\ref{fig:magfield_slip}. To remedy this, Figure \ref{fig:airgap_mag_field} shows only the airgap 
 magnetic field (hiding the rotor magnetic field) for the $s=0.02$ slip motor of Figure~\ref{fig:magfield_slip}(a).

\begin{figure}[H]
    \centering
    \begin{subfigure}[t]{0.32\linewidth}
        \centering
        \includegraphics[width=\linewidth]{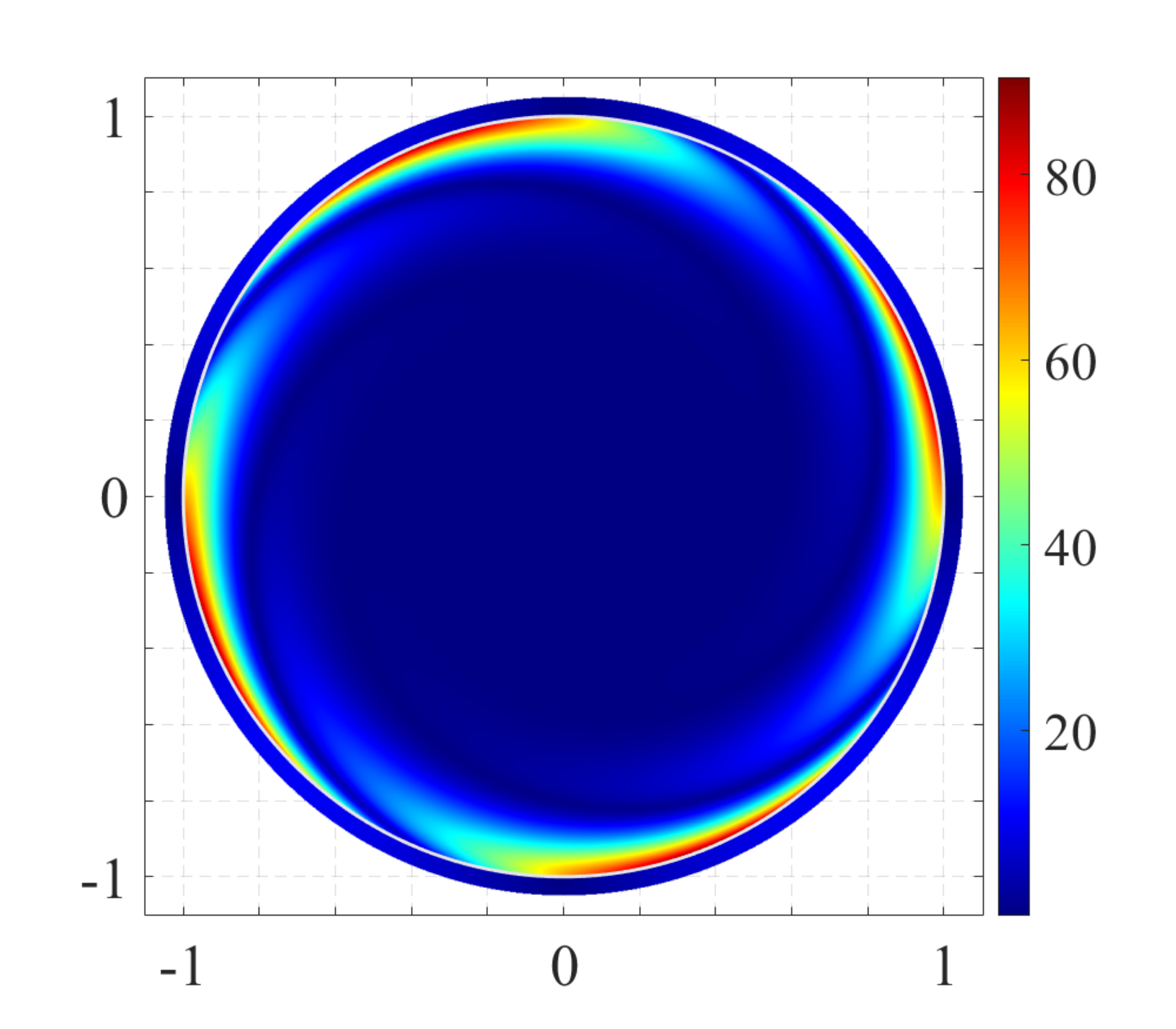}
        \caption{$\| \bm b \|$ for $\zeta = 0.02$}
    \end{subfigure}%
    ~ 
    \begin{subfigure}[t]{0.32\linewidth}
        \centering
        \includegraphics[width=\linewidth]{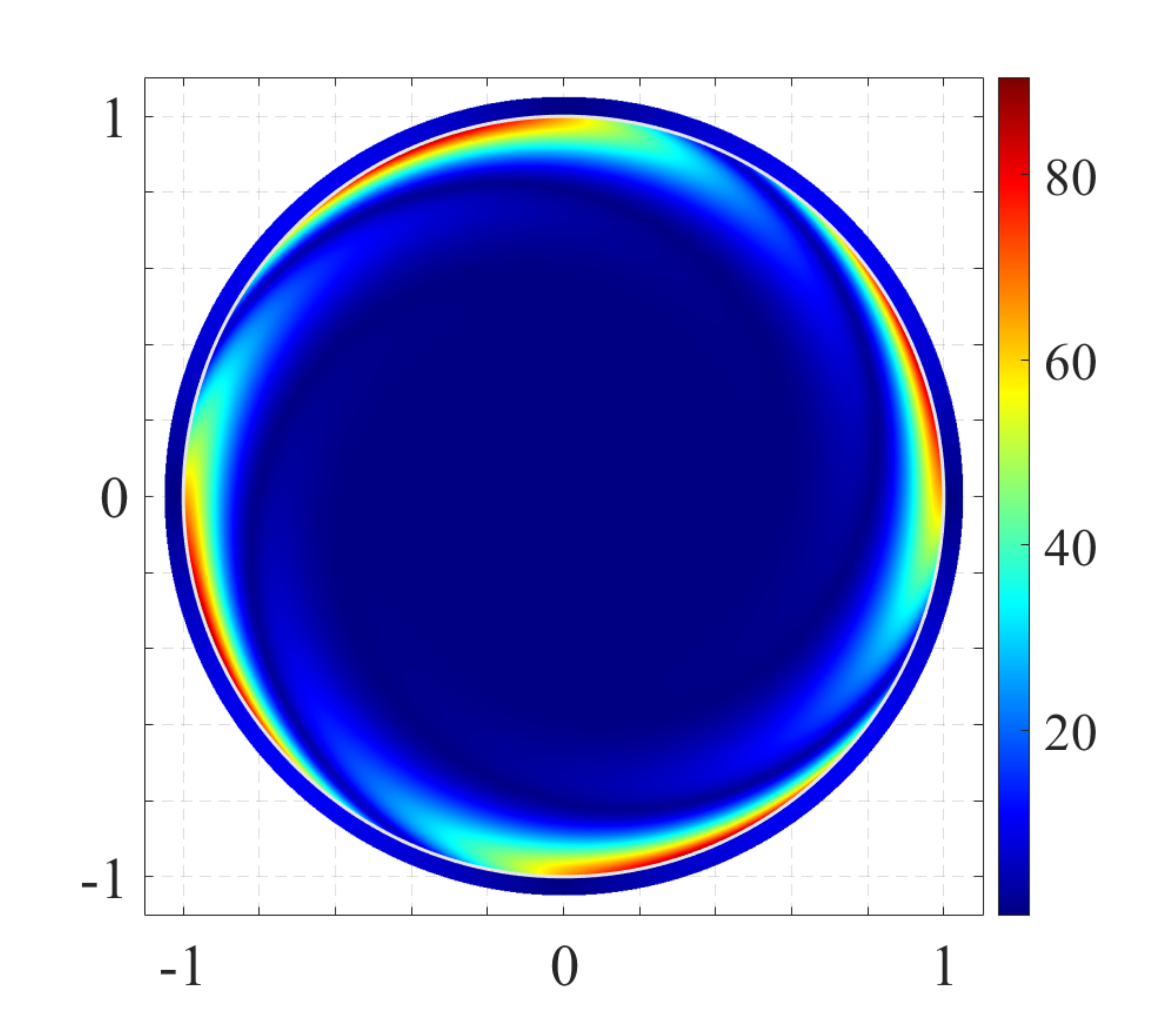}
        \caption{$\| \bm b \|$ for $\zeta = 0.05$}
    \end{subfigure}
    ~
        \centering
    \begin{subfigure}[t]{0.32\linewidth}
        \centering
        \includegraphics[width=\linewidth]{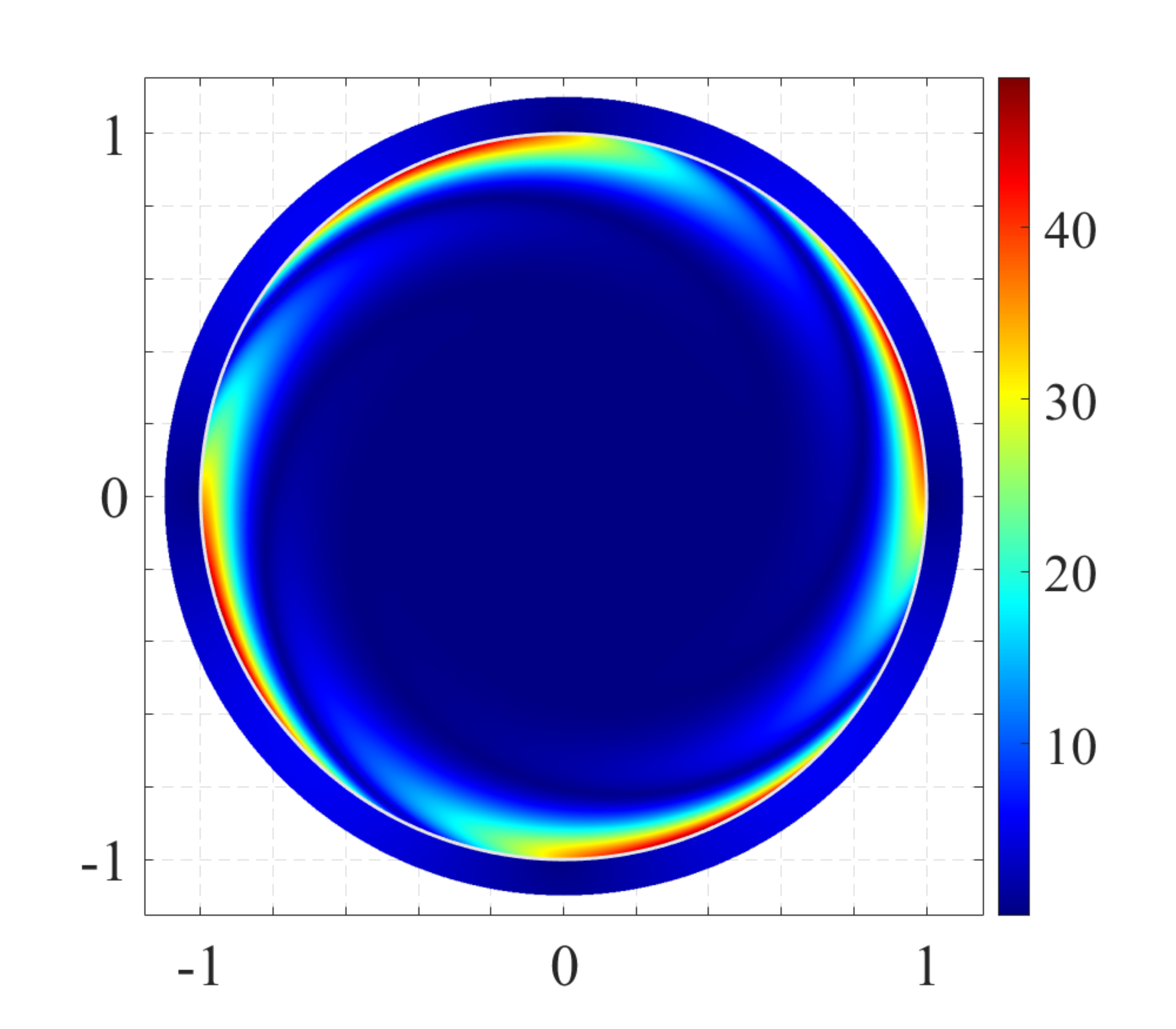}
        \caption{$\| \bm b \|$ for $\zeta=0.10$}
    \end{subfigure}%
    \caption{Magnetic field norm $\| \bm b \|$ for a steel rotor (normalized by $\mu_0 \kappa_0$), for different values of the airgap parameter $\zeta$.}
    \label{fig:magfield_airgap}
\end{figure}

The influence of changing motor geometry is presented in Figure~\ref{fig:magfield_airgap} for three different airgap parameters 
$\zeta=0,02,\; 0.05,\; 0.10$ in a steel rotor and a slip value $s=0.02$. As expected,
Reducing the airgap size does not affect the distribution of the magnetic field, but increases drastically the maximum strength of the field.

\begin{figure}[H]
    \centering
    \begin{subfigure}[t]{0.32\linewidth}
        \centering
        \includegraphics[width=\linewidth]{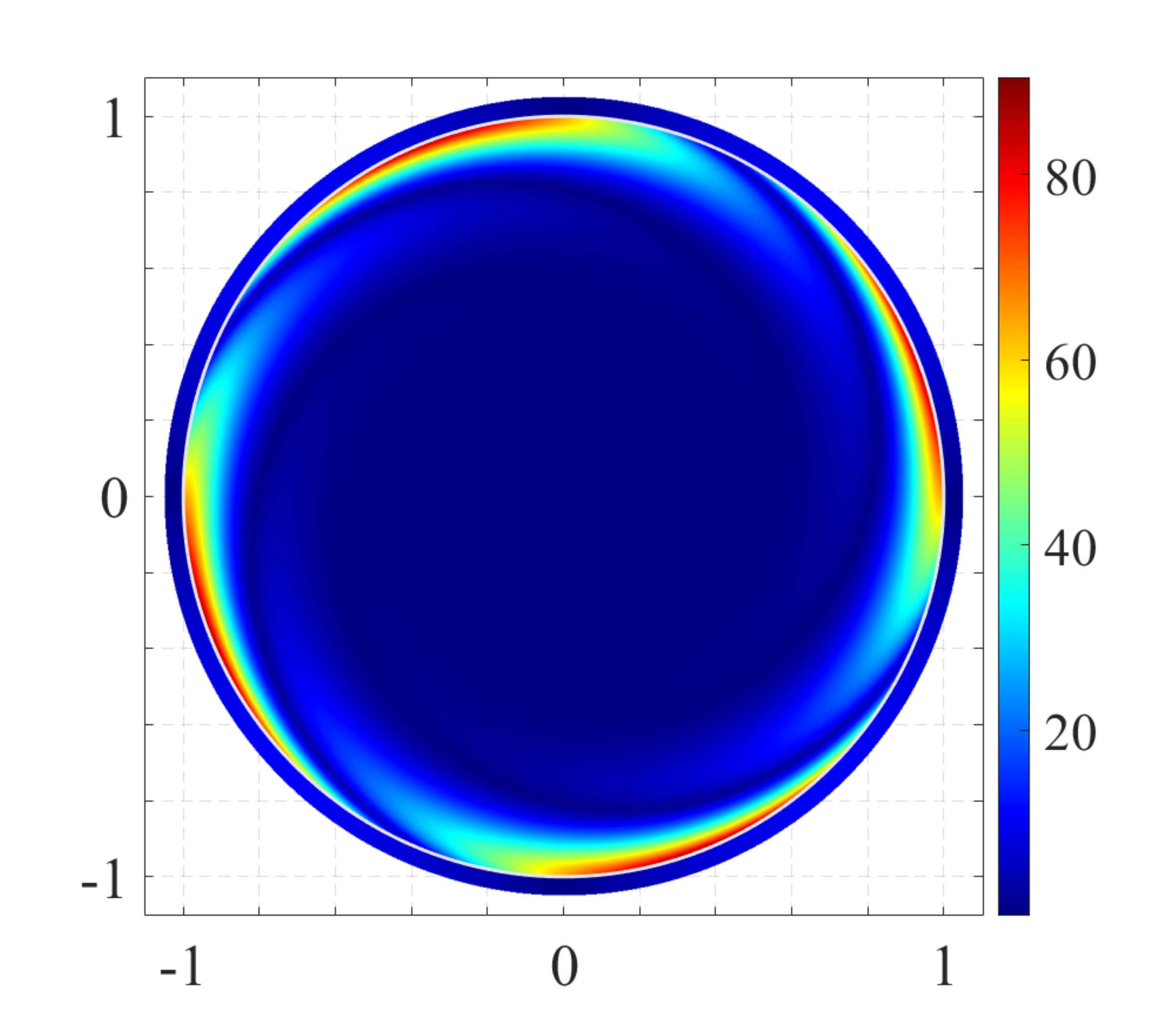}
        \caption{$\| \bm b \|$ -- steel}
    \end{subfigure}%
    ~ 
    \begin{subfigure}[t]{0.32\linewidth}
        \centering
        \includegraphics[width=\linewidth]{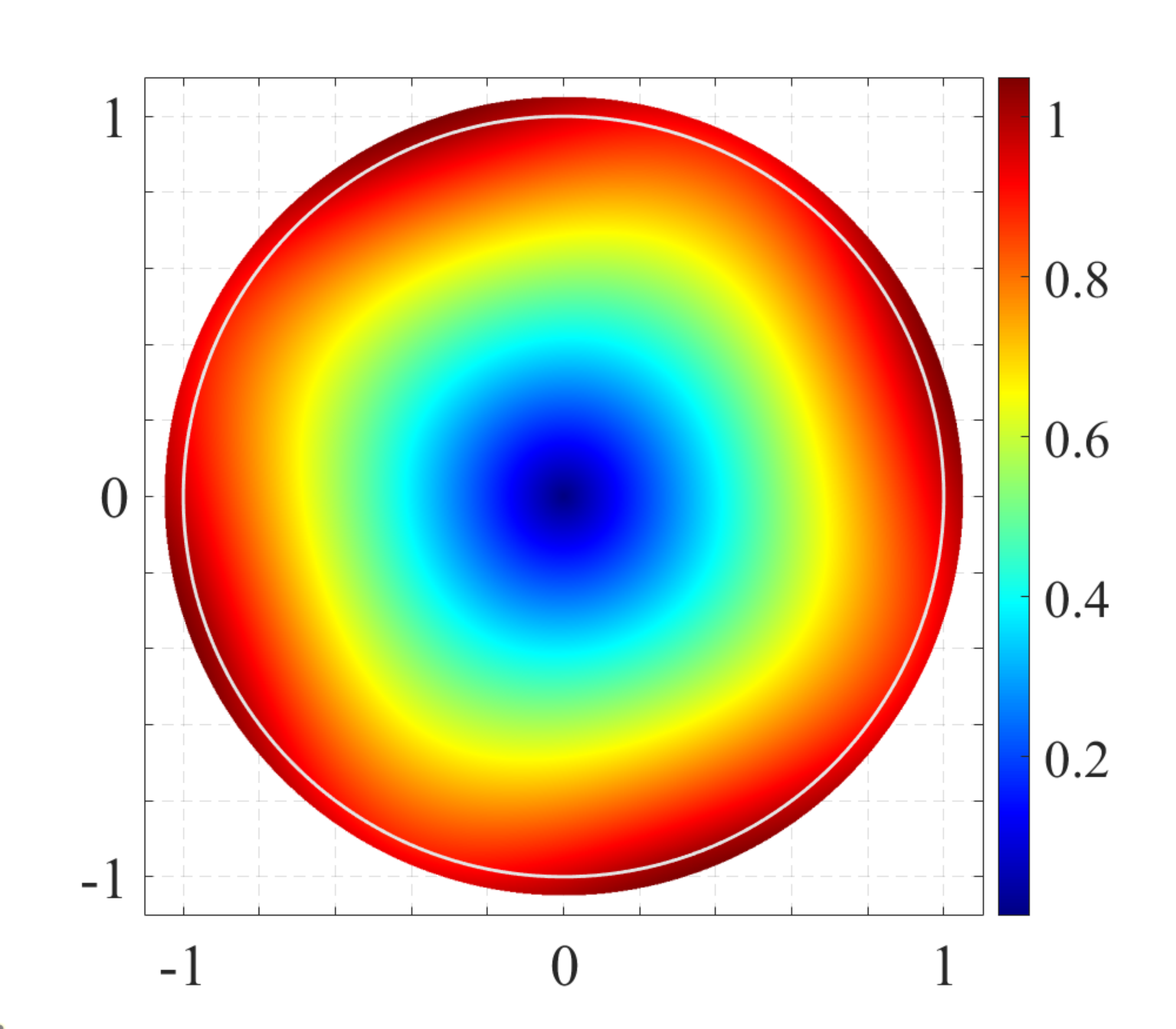}
        \caption{$\| \bm b \|$ -- copper}
    \end{subfigure}%
    ~ 
    \begin{subfigure}[t]{0.32\linewidth}
        \centering
        \includegraphics[width=\linewidth]{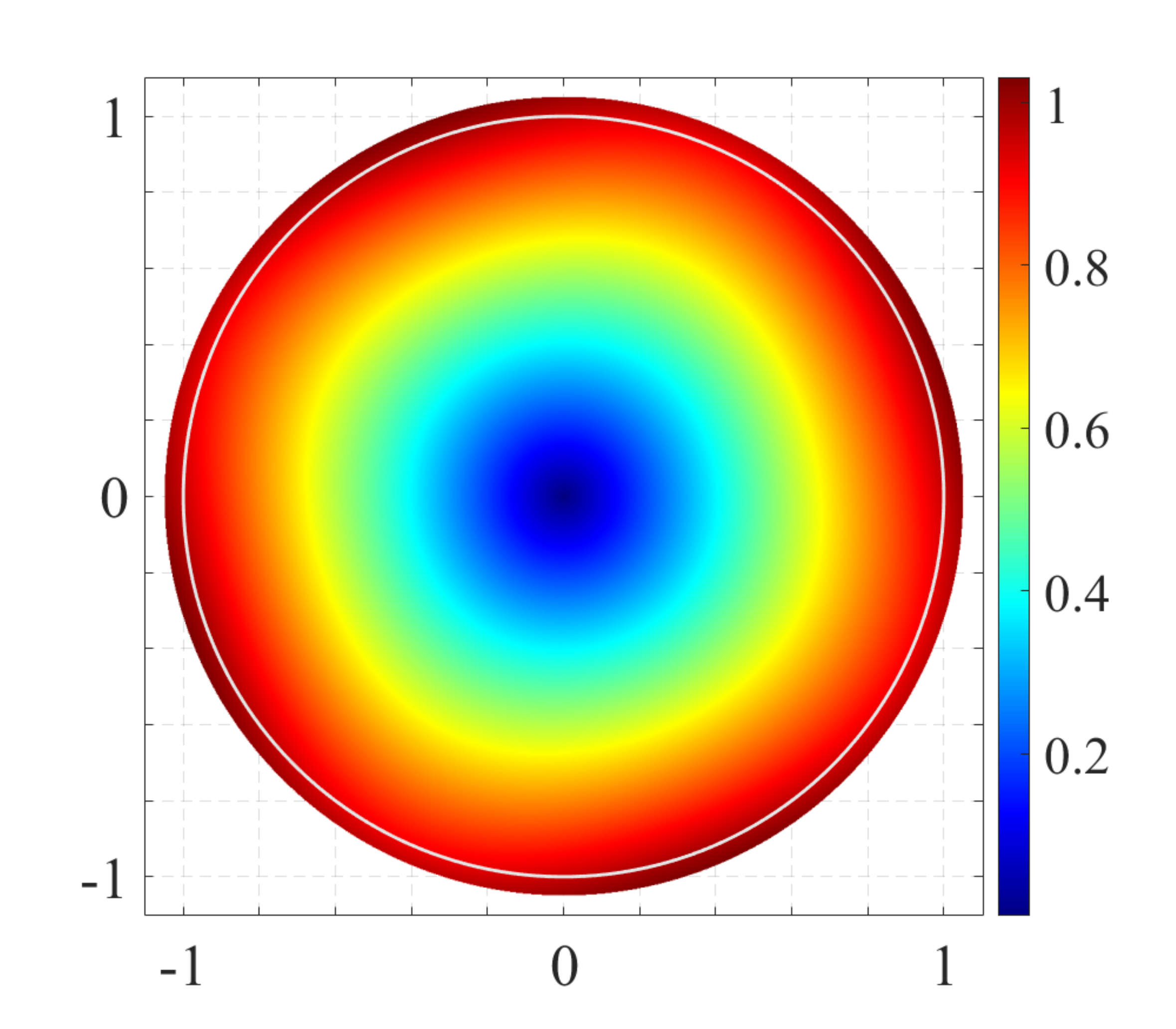}
        \caption{$\| \bm b \|$ -- aluminum}
    \end{subfigure}
    \caption{Magnetic field norm $\| \bm b \|$ (normalized by $\mu_0 \kappa_0$), for different rotor materials in motors with $s=0.02,\; \zeta=0.05$.}
    \label{fig:magfield_material}
\end{figure}

Comparison of the magnetic fields for different rotor materials is presented in Figure~\ref{fig:magfield_material}, where the results for the
high magnetic susceptibility steel are contrasted to the non-magnetic copper and aluminum rotors. The slip and airgap parameters are kept at their default value $s=0.02$, $\zeta=0.05$. 
Notice that for both the copper and aluminum rotors the maximum value of the magnetic field is two orders of magnitude less than in steel.
One can also observe that the normalized magnetic field for aluminum and copper reaches its maximum 
value at the rotor boundary, given the absence of magnetization in these materials. The slightly larger extent for the maximum magnetic field 
zone for the copper rotor, is attributed to its higher electrical conductivity which results in higher induced currents than in aluminum.

\subsection{Rotor temperature field}

\begin{figure}[H]
    \centering
    \begin{subfigure}[t]{0.325\linewidth}
        \centering
        \includegraphics[width=\linewidth]{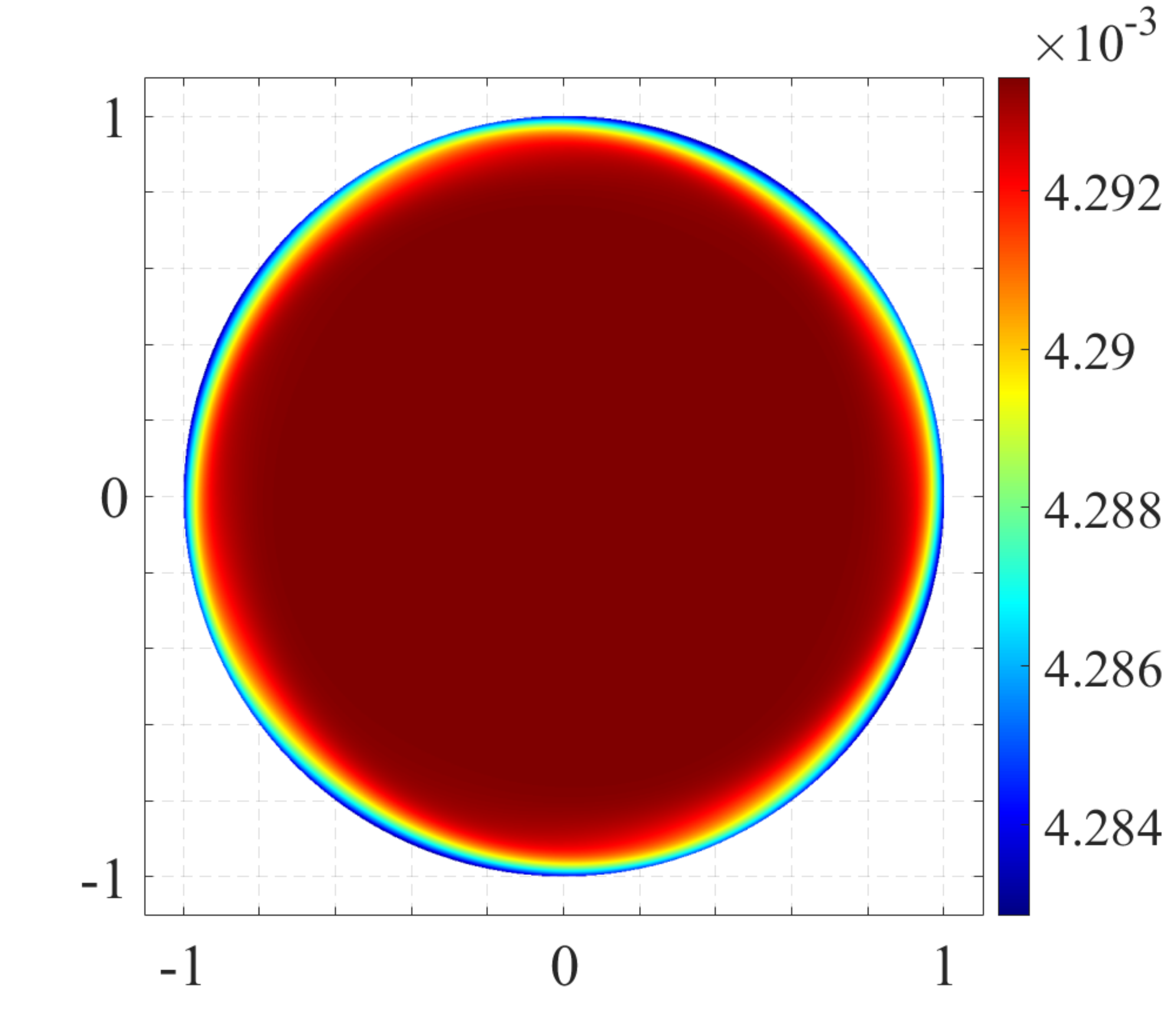}
        \caption{Full temperature field $T(r, \Theta)$}
    \end{subfigure}%
    ~ 
        \centering
    \begin{subfigure}[t]{0.325\linewidth}
        \centering
        \includegraphics[width=\linewidth]{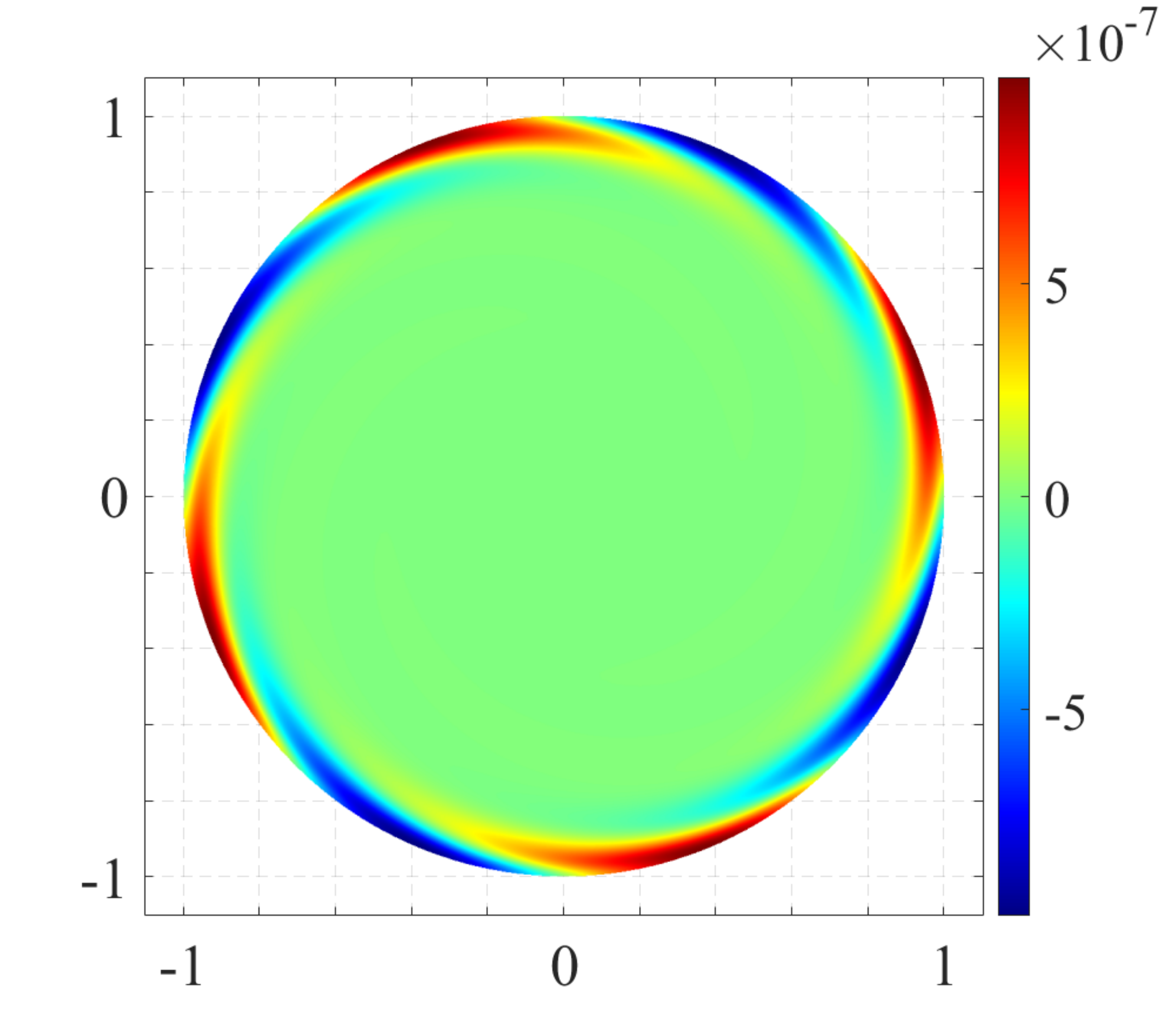}
     \caption{$\Theta$-variation $\Re\left\{\cpx {T}(r) \exp(-i 2\Theta)\right\}$}
    \end{subfigure}%
    \caption{Normalized temperature increase for the steel rotor (base case); (a) full field and (b) angular variation.}
    \label{fig:NoramlizedTemp_steel}
\end{figure}

The full-field dimensionless temperature\footnote{Here T denotes absolute temperature in $^\circ$K and not its normalized counterpart defined 
in \eqref{eq:dimensionless}.} $(T - T_a)/T_a \rightarrow T(r, \Theta)$ for the base case steel rotor is  
presented in Figure~\ref{fig:NoramlizedTemp_steel}; the normalization with respect to the reference temperature $T^a$ adopted here as a more physically
meaningful choice. Since the mean field dominates, the $\Theta$-dependent variations are completely masked by the scale used to plot Figure~\ref{fig:NoramlizedTemp_steel}(a). The $\Theta$-dependent variation $\Re\left\{\cpx {T}(r) \exp(-i 2\Theta)\right\}$, whose amplitude is four orders of magnitude lower than the mean, is plotted by itself in Figure~\ref{fig:NoramlizedTemp_steel}(b). According to the values given in Table~\ref{tab:values} for the thermal characteristics of the idealized motor, the almost uniform temperature increase of the rotor is a mere $0.086 ^{\circ}C$ from an ambient airgap temperature of $20 ^{\circ}C$, with the maximum temperature occurring at the center.

\begin{figure}[H]
    \centering
    \begin{subfigure}[t]{0.32\linewidth}
        \centering
        \includegraphics[width=\linewidth]{./Images/Steel/temperature/RelativeTemperatureDifference_T.pdf}
        \caption{$T(r, \Theta)$ -- steel}
    \end{subfigure}%
    ~ 
    \begin{subfigure}[t]{0.32\linewidth}
        \centering
        \includegraphics[width=\linewidth]{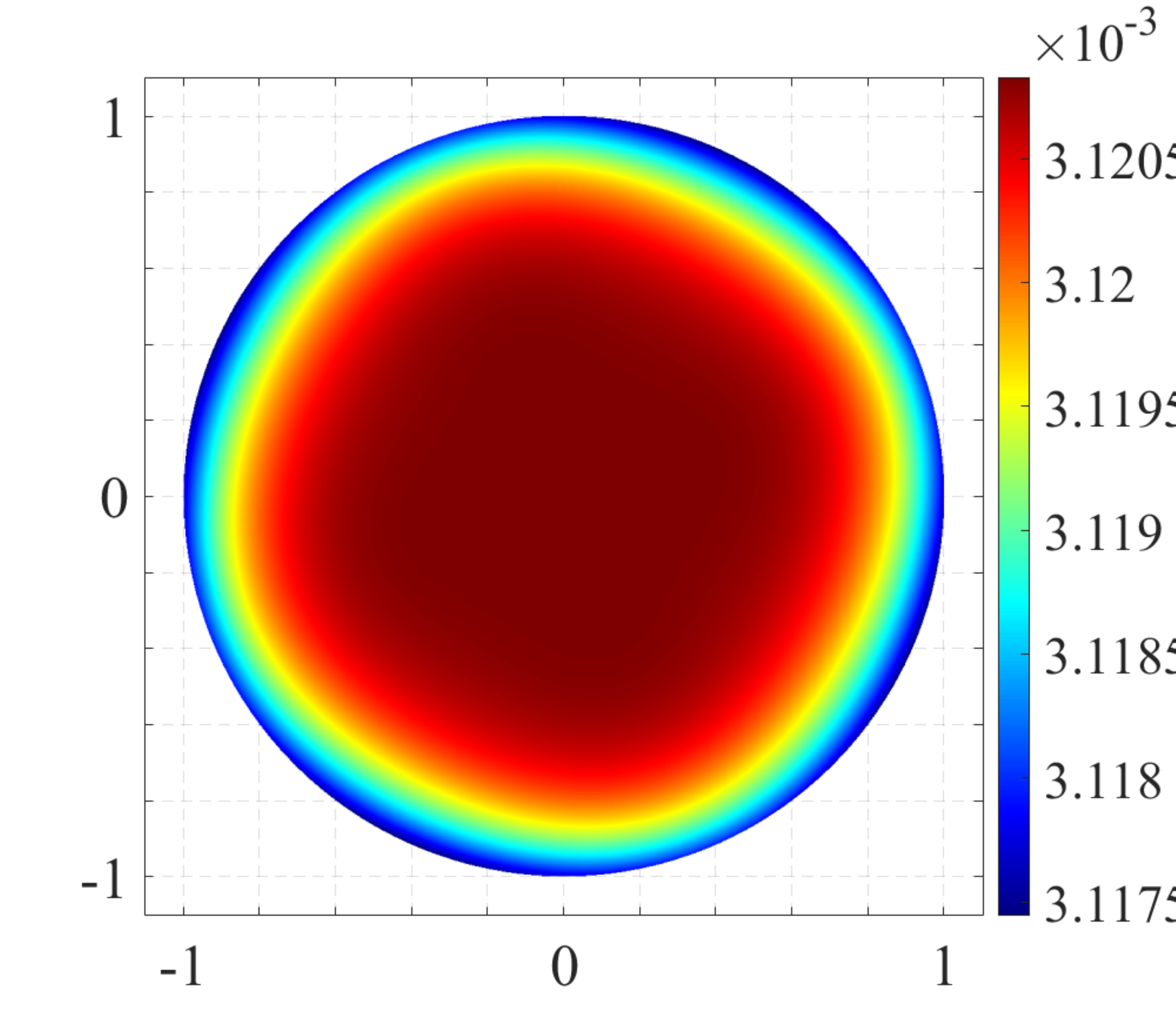}
        \caption{$T(r, \Theta)$ -- copper}
    \end{subfigure}%
    ~ 
        \centering
    \begin{subfigure}[t]{0.32\linewidth}
        \centering
        \includegraphics[width=\linewidth]{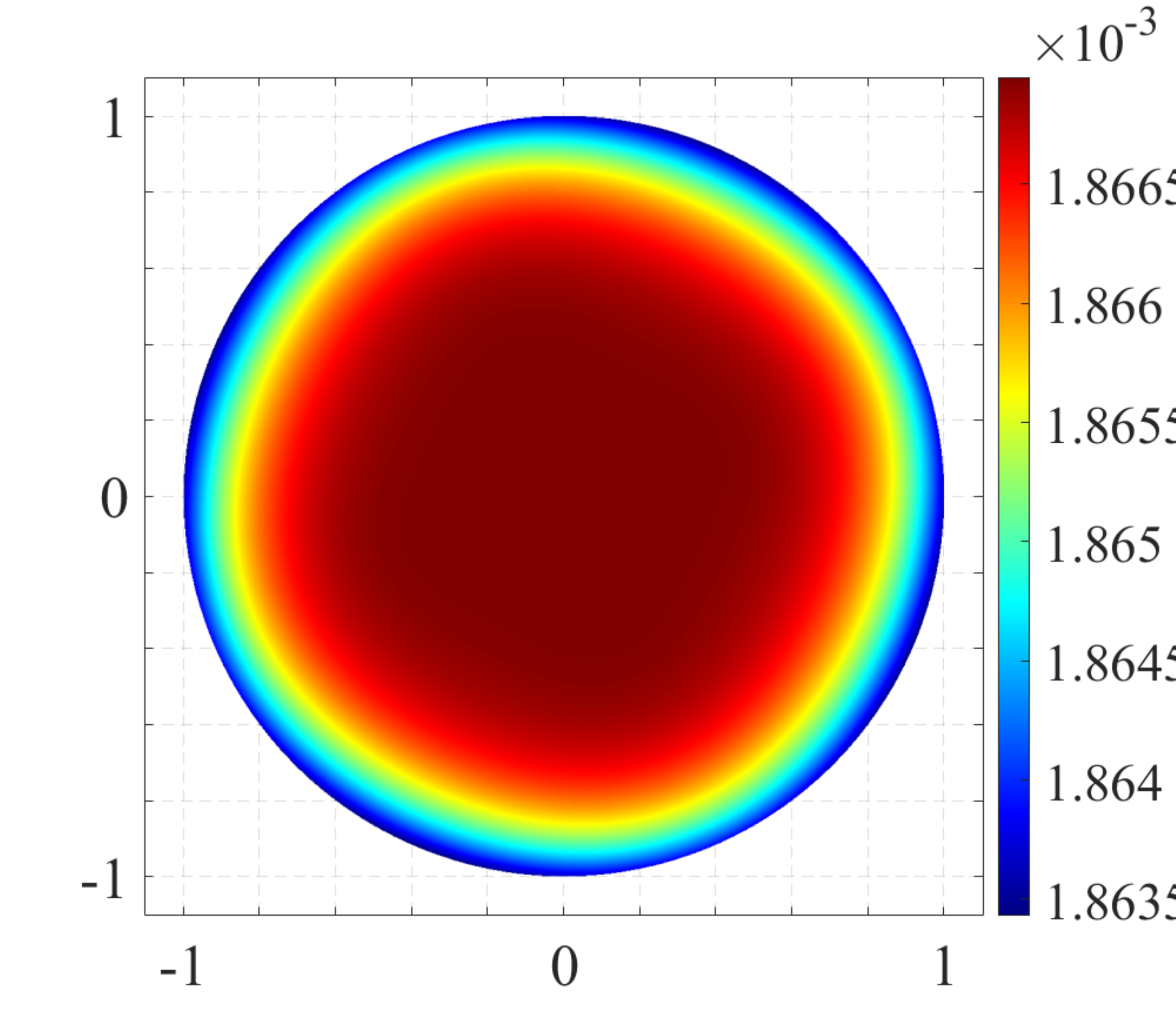}
        \caption{$T(r, \Theta)$ -- aluminum}
    \end{subfigure}%
    \caption{Normalized temperature increase $T$ for the base case motor: (a) steel, (b) copper and (c) aluminum rotors.}
    \label{fig:NormalizedTemp_fecopalu}
\end{figure}

The influence of the rotor material on the dimensionless temperature increase $T(r, \Theta)$ in the base case motor is presented next 
in Figure~\ref{fig:NormalizedTemp_fecopalu}. In comparing the results for steel in (a), copper in (b) and aluminum in (c), we notice that
the temperature increase is almost uniform over the rotor, with the highest increase $0.086 ^{\circ}C$ occurring in steel,  $0.062 ^{\circ}C$
for copper and $0.037 ^{\circ}C$ for aluminum. 

Ohmic dissipation is the sole dissipation mechanism considered, as discussed in the first remark of Subsection~\ref{sec:materials} and
depends on the relative frequency $\omega_r$. The relatively low frequency used (about $1Hz$, we consider $\omega_r$ at $2\%$ slip)  explains the very low temperature increase found here.

\subsection{Rotor current density, Lorentz and magnetic body forces}

\begin{figure}[H]
    \centering
    \begin{subfigure}[t]{0.32\linewidth}
        \centering
        \includegraphics[width=\linewidth]{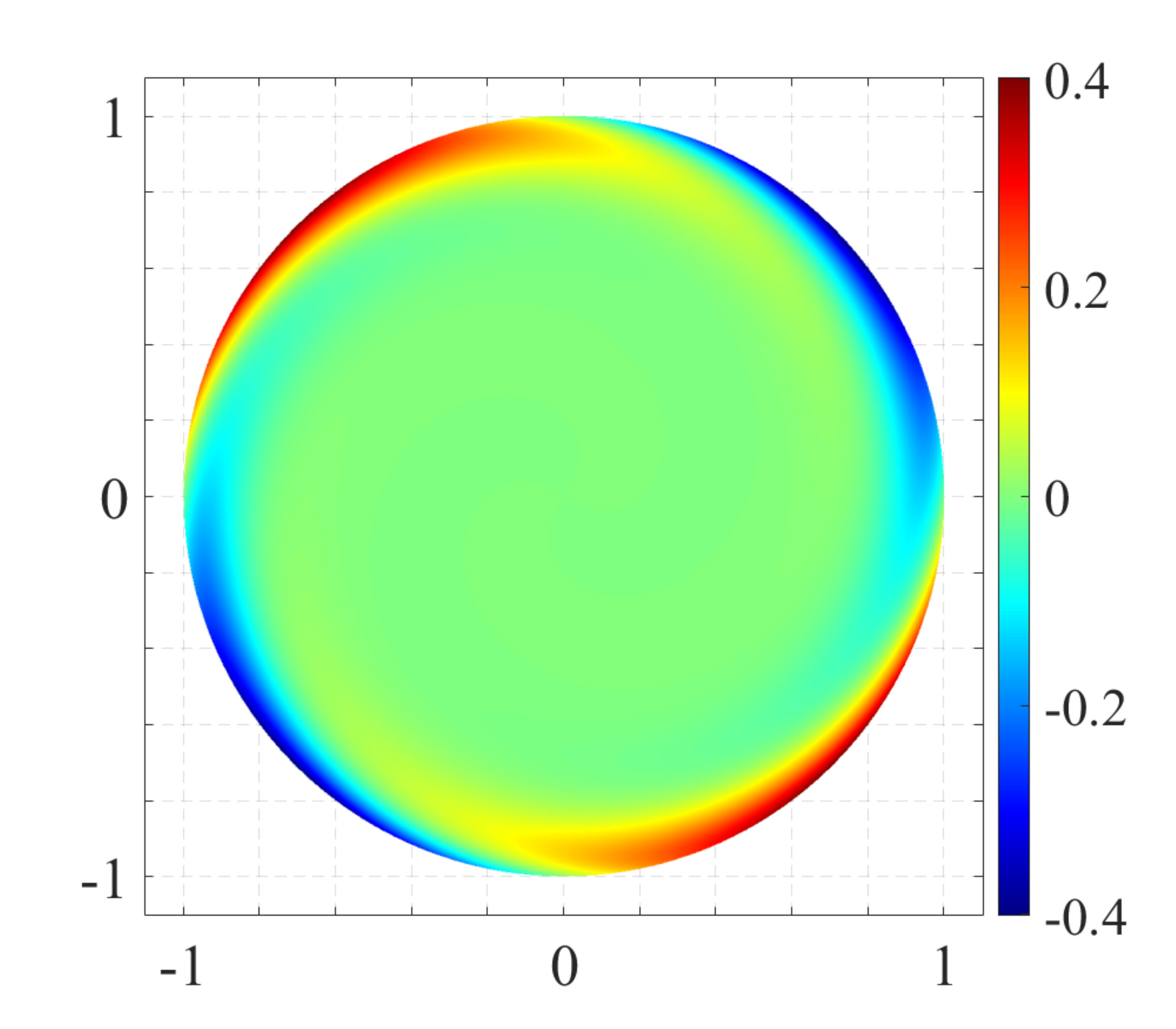}
        \caption{$j(r, \Theta)$ - steel}
    \end{subfigure}%
    ~ 
    \begin{subfigure}[t]{0.32\linewidth}
        \centering
        \includegraphics[width=\linewidth]{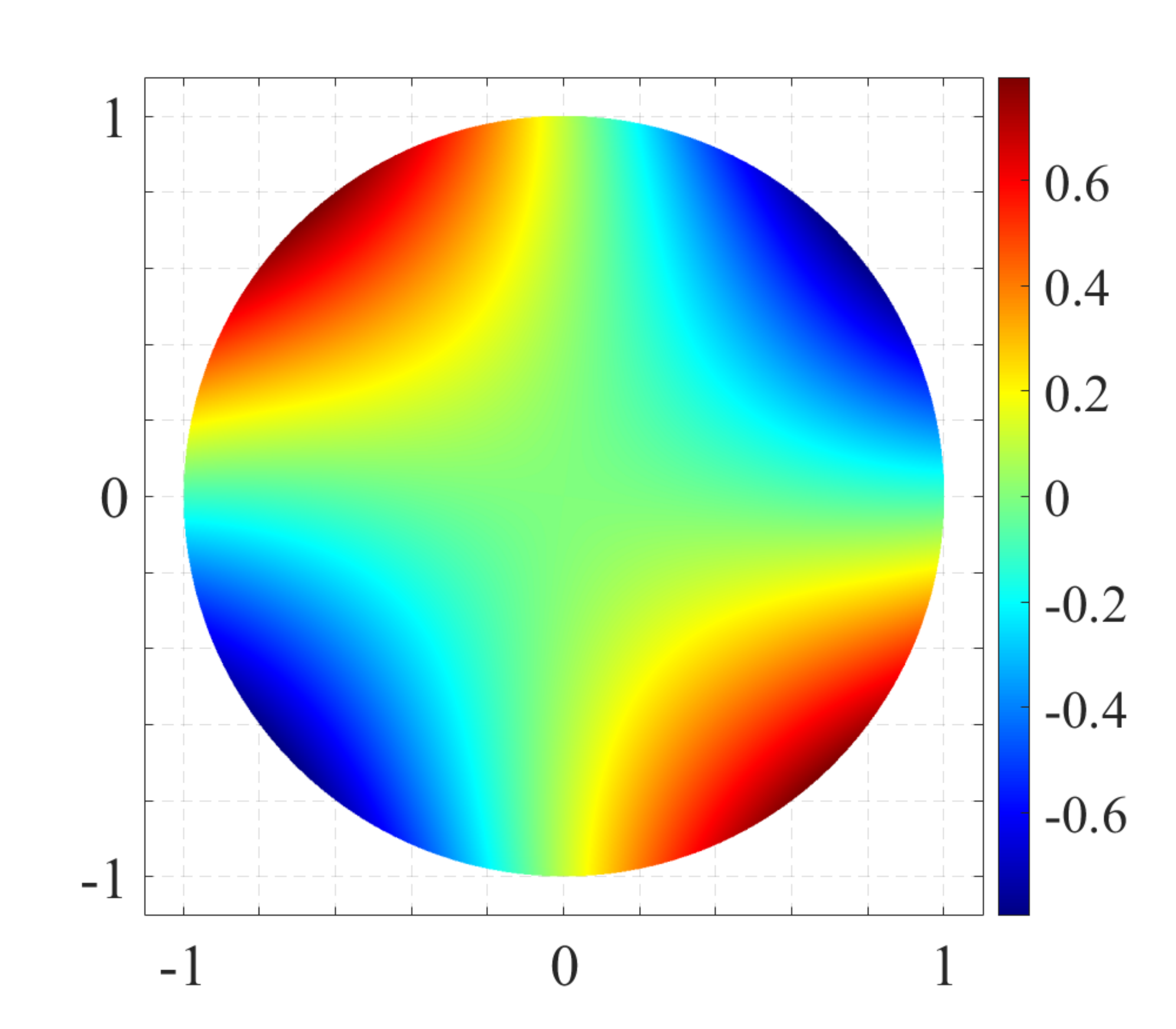}
        \caption{$j(r, \Theta)$ - copper}
    \end{subfigure}
    ~
        \centering
    \begin{subfigure}[t]{0.32\linewidth}
        \centering
        \includegraphics[width=\linewidth]{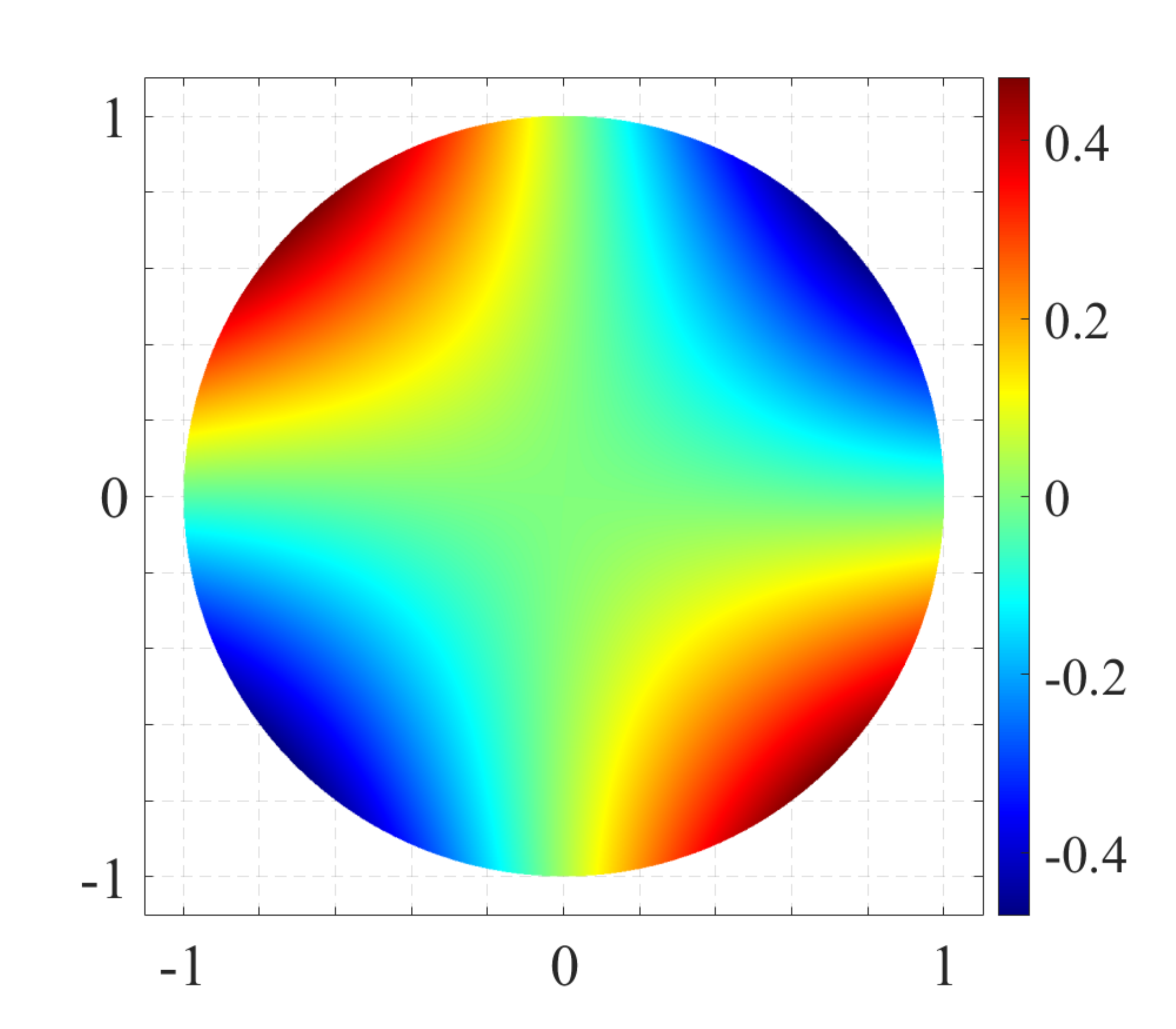}
        \caption{$j(r, \Theta)$ - aluminum}
    \end{subfigure}%
    \caption{Current density $j$ (normalized by $\kappa_0/R_1$), for the base case motor: (a) steel, (b) copper and (c) aluminum rotors.}
    \label{fig:currents}
\end{figure}

{\underline{\it Current density}} 
The dimensionless current density field  $j = j_z = -\gamma(\partial a/\partial t)$, (normalized by $\kappa_0/R_1$) for the base case motor
is presented in Figure~\ref{fig:currents} for steel (a), copper (b) and aluminum (c) rotors, respectively. The currents for steel are forming thin plumes near the rotor surface because of the high magnetic permeability that concentrates the magnetic field at the rotor-airgap interface -- see
Figure~\ref{fig:magfield_material}(a) -- limiting its penetration into the rotor. The current distribution for copper and aluminum rotors is very similar, given the absence of magnetization. Notice in Figure~\ref{fig:currents} that the maximum current values for steel are the lowest 
while the corresponding ones for copper are the highest, as expected by the different rotor material conductivities 
according to Table~\ref{tab:values}.

{\underline{\it Lorentz, magnetization and magnetostricive body forces}} The different components of the magnetic
body force $\fm$, defined as the divergence of the magnetic stress $\sigmam$ in \eqref{eq:smallearbitrb}, are
\begin{equation}
	\fm  \equiv \div \sigmam = \bm{j}\times\bm{b} + \bm{m}\contr (\bm{b}\grad) + \frac{\Lambda}{\mu}\bm{b}\contr(\grad\bm{b}) \; ; 
	\quad \bm{m}=\frac{\chi}{\mu}\bm{b} \; , 
	\label{eq:em-forces}
\end{equation}
where $\mu=\mu_0(1+\chi)$. The three different magnetic body force components in \eqref{eq:em-forces} are:  
the \textit{Lorentz body force:} $\bm{j}\times\bm{b}$, 
a \textit{magnetization body force:} $ \bm{m}\contr  (\bm{b}\grad)$ and a \textit{magnetostriction force:} 
$(\Lambda / \mu)\bm{b}\contr(\grad\bm{b})$. The last two components are absent in non-magnetic copper and aluminum ($\chi\approx\Lambda\approx 0$).

\begin{figure}[H]
    \centering
    \begin{subfigure}[t]{0.32\linewidth}
        \centering
        \includegraphics[width=\linewidth]{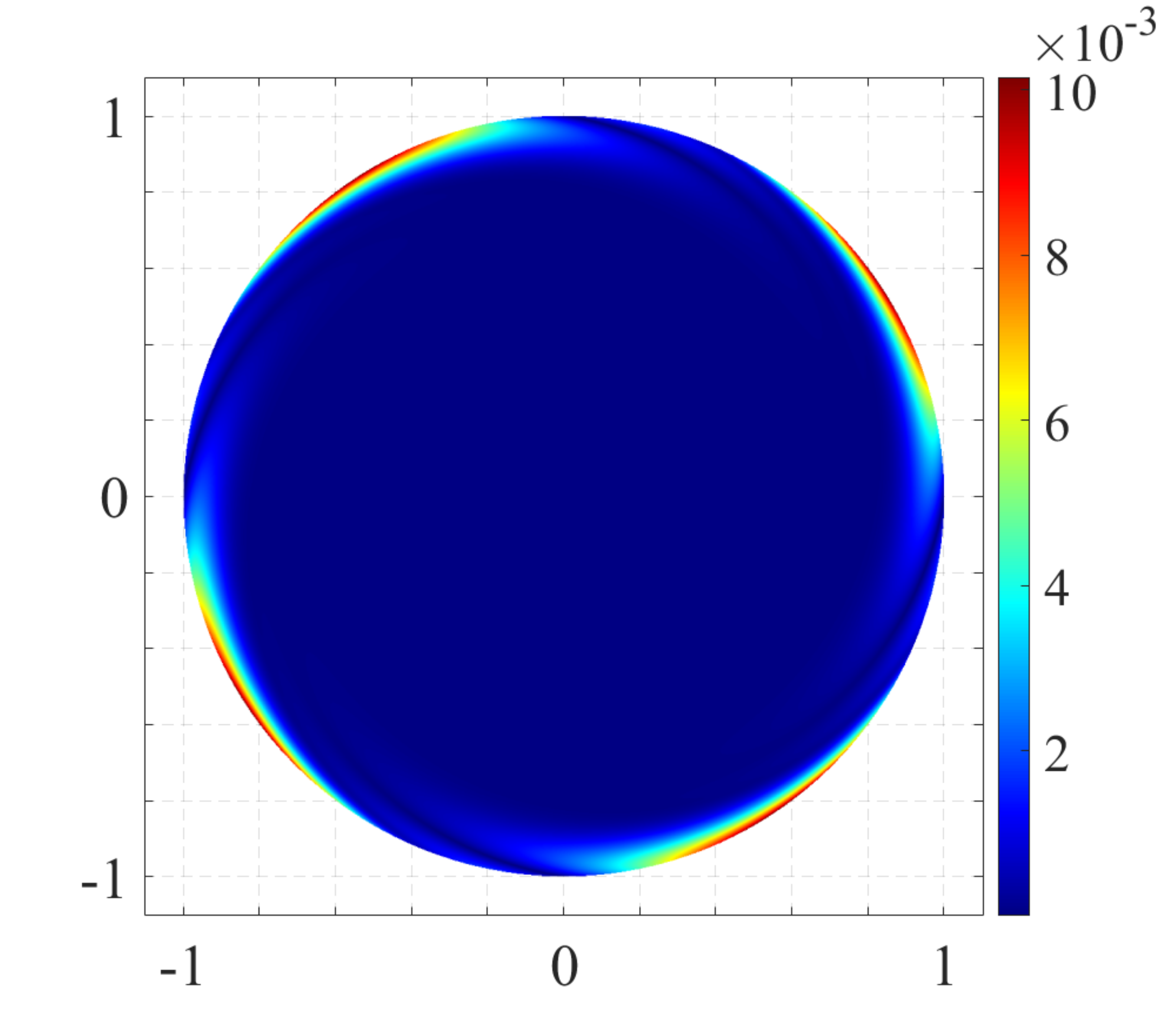}
        \caption{Lorentz $\| \bm{j}\times\bm{b} \|$}
    \end{subfigure}%
    ~ 
    \begin{subfigure}[t]{0.32\linewidth}
        \centering
        \includegraphics[width=\linewidth]{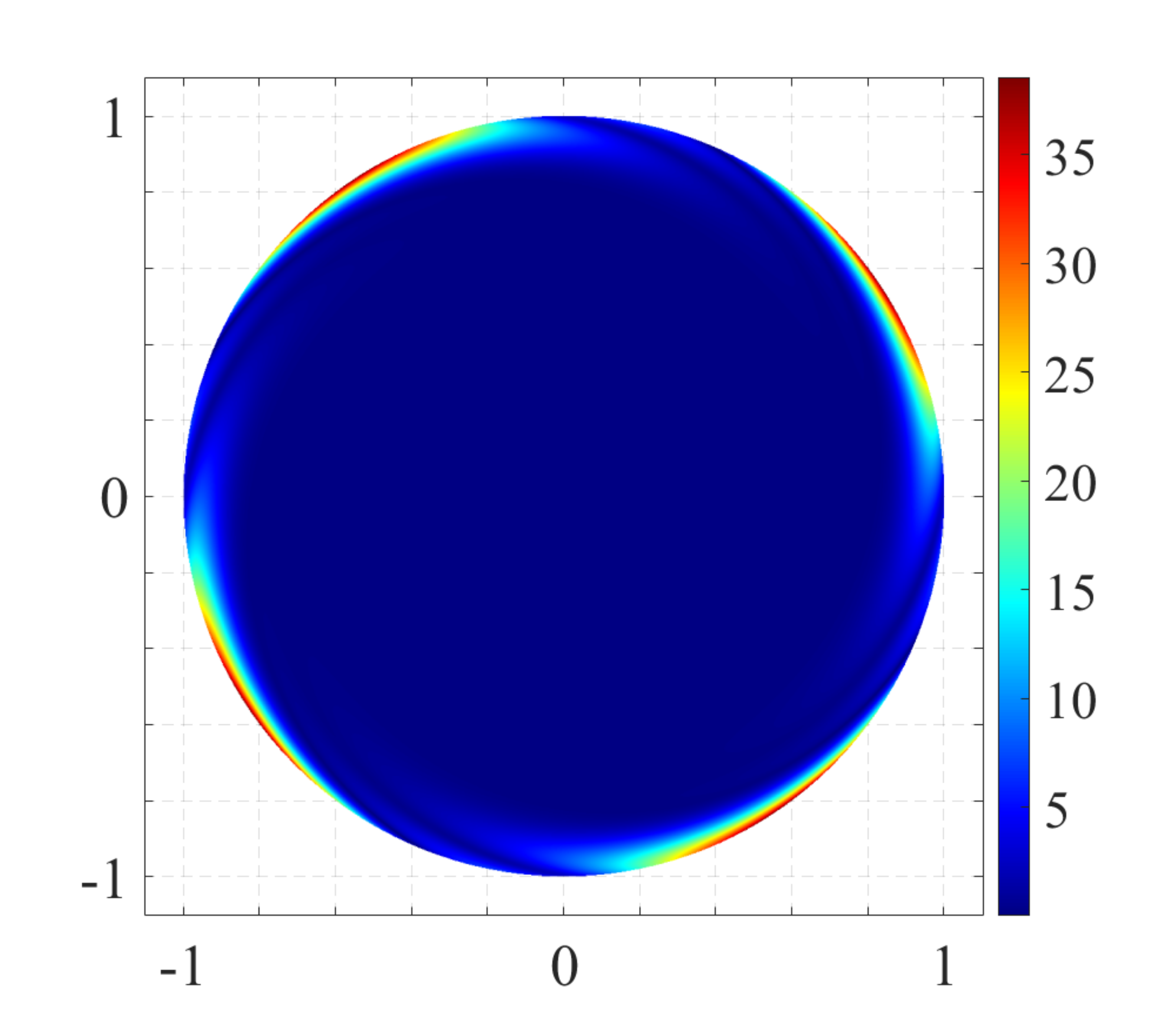}
        \caption{Magnetization $\|  \bm{m}\contr  (\bm{b}\grad) \|$}
    \end{subfigure}
    ~
        \centering
    \begin{subfigure}[t]{0.32\linewidth}
        \centering
        \includegraphics[width=\linewidth]{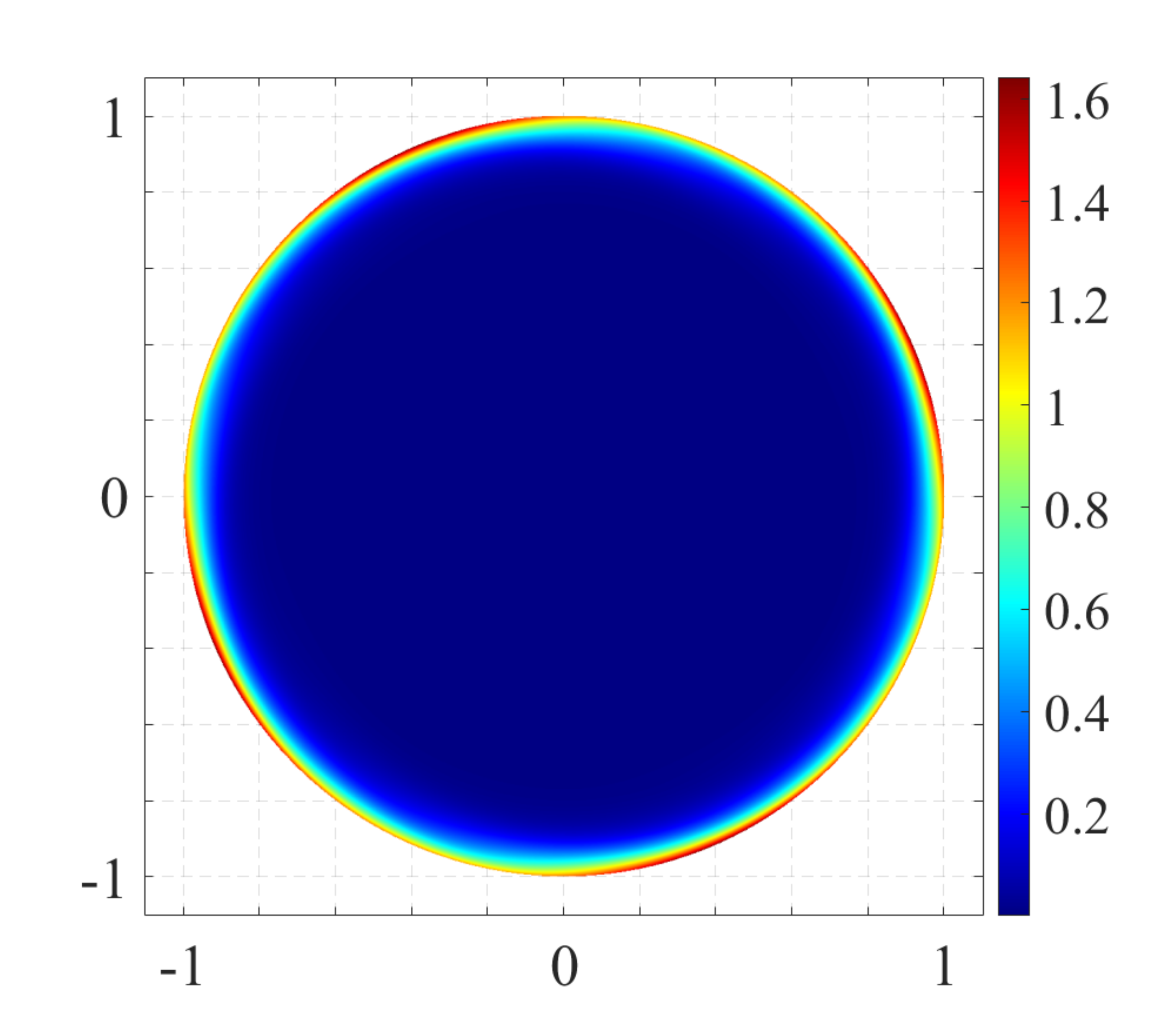}
        \caption{Magnetostriction  $\|(\Lambda / \mu)\bm{b}\contr(\grad\bm{b})\|$}
    \end{subfigure}%
    \caption{Comparison of the different magnetic body forces (normalized by $\rho_0 R_1 \Omega^2$) 
    for the base case motor with a steel rotor.}
    \label{fig:forces_steel}
\end{figure}

Figure \ref{fig:forces_steel} shows the amplitude of the three different components of the electromagnetic force,
(normalized by the amplitude of the centrifugal force density $\rho_0 R_1 \Omega^2$), for the base case
motor with a steel rotor case. The first important observation is that the Lorentz forces are negligible, with their maximum value of the order of
$1\%$ of the inertial forces. A straightforward dimensional analysis indicates $\| \bm{j} \| \approx \| \bm{b}\|  /(\mu R_1)$, giving 
 $\| \bm{j} \times \bm{b}\| \approx \| \bm{b} \| ^2 / (\mu R_1)$ for the Lorentz component of the body force,
compared to the magnetic $\chi \| \bm b \| ^2 / (\mu R_1)$ and magnetostrictive $\Lambda \| \bm b \| ^2 / (\mu R_1)$ components. 

Observe that the magnetization force is larger than its inertial counterpart -- up to approximately forty times at the rotor's edge
due to the highest magnetic field gradients there, according to Figure~\ref{fig:magfield_material}(b) -- pointing to the importance of accounting for magnetization body forces in electric motor models. The magnetostrictive forces are not negligible and peak at about $160\%$ of their inertial counterpart 
(or about $5\%$ of the maximum magnetization forces), a somewhat surprising result in view of the same order $\chi$ and $\Lambda$ coefficients from 
Table~\ref{tab:values} but explained by the different expressions for the corresponding forces in \eqref{eq:em-forces}.
\begin{figure}[H]
    \centering
    \begin{subfigure}[t]{0.32\linewidth}
        \centering
        \includegraphics[width=\linewidth]{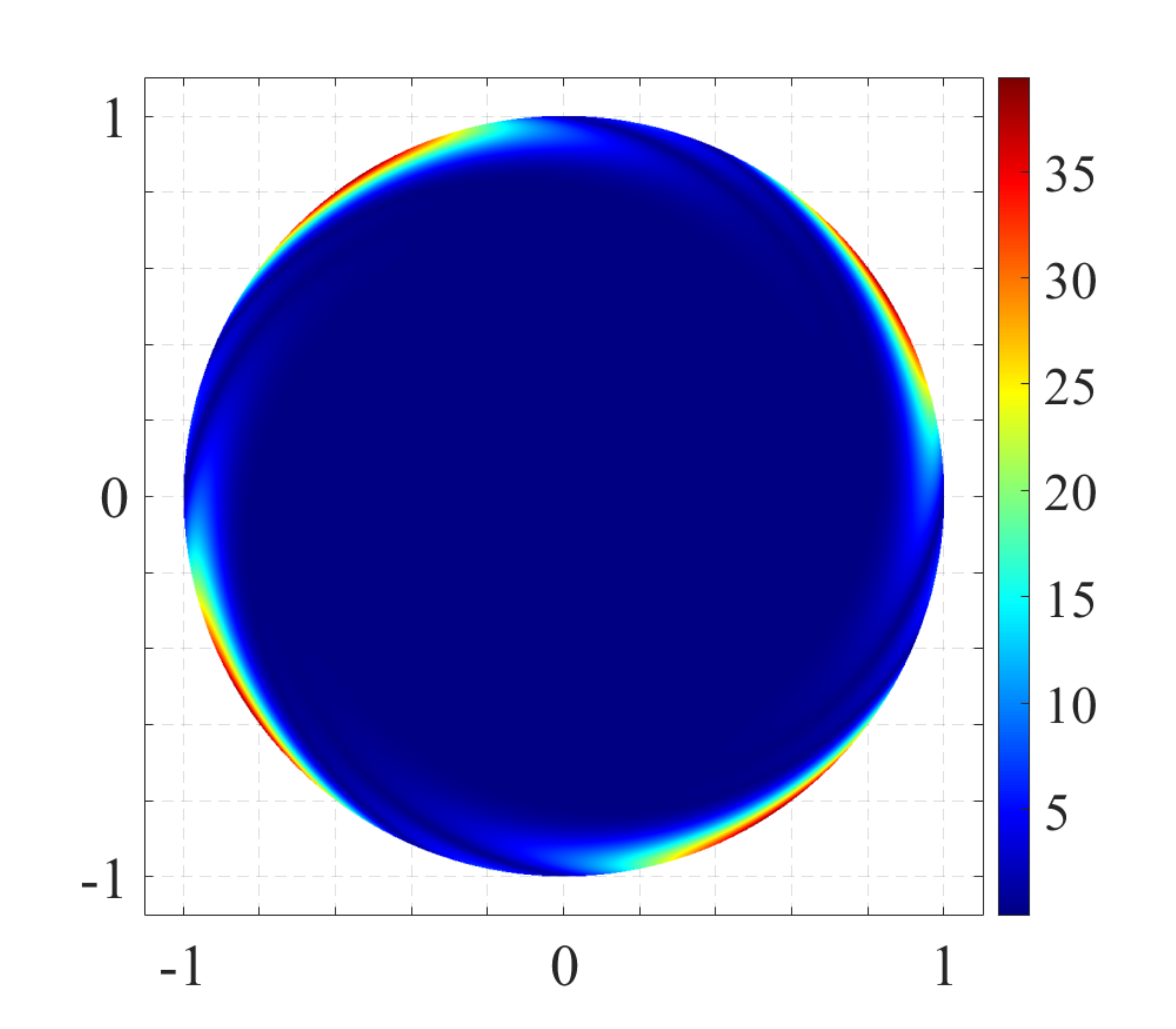}
        \caption{$\| \fm \|$ -- steel}
    \end{subfigure}%
    ~ 
    \begin{subfigure}[t]{0.32\linewidth}
        \centering
        \includegraphics[width=\linewidth]{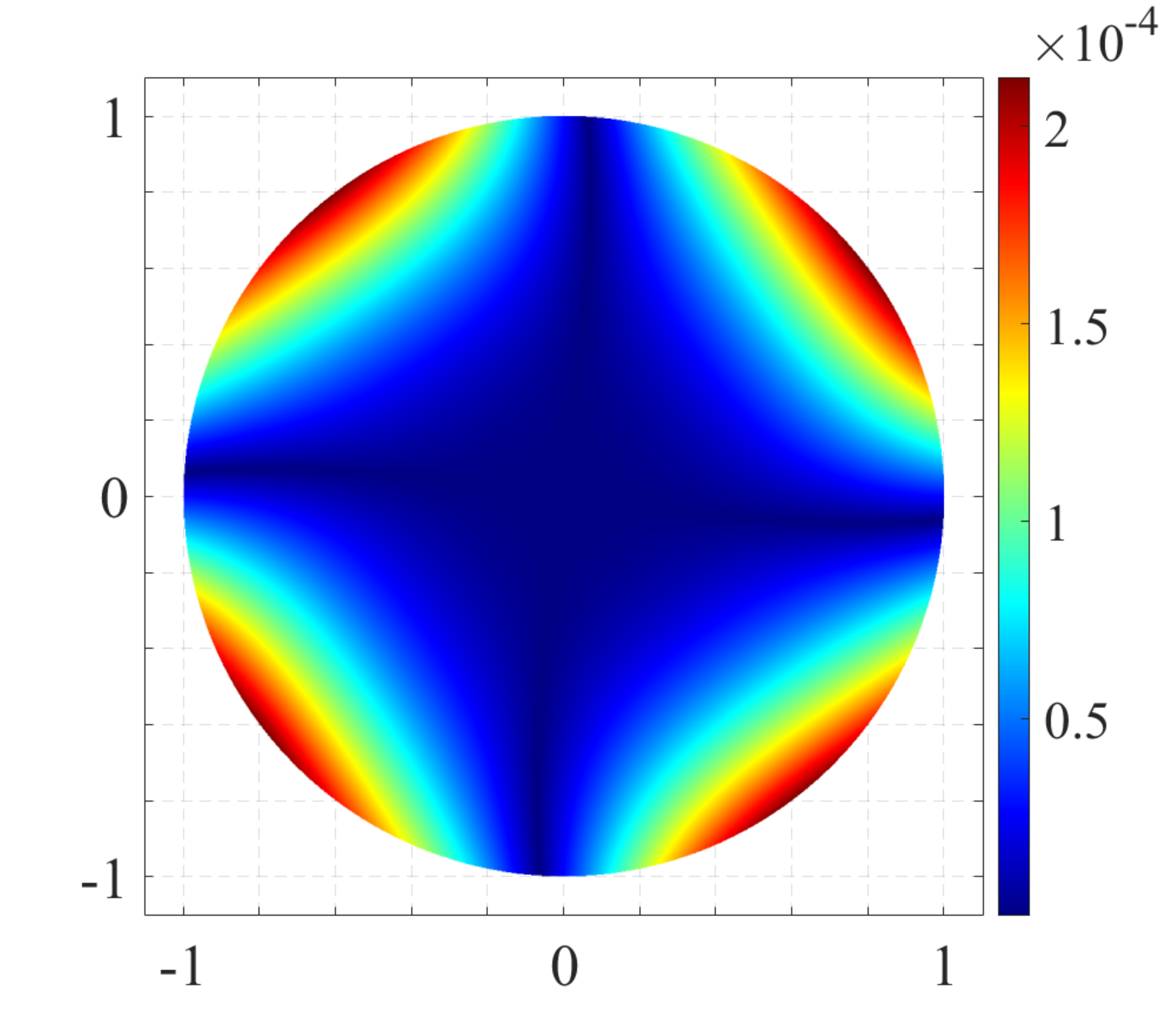}
        \caption{$\| \fm \|  = \| \bm{j} \times \bm{b} \|$ -- copper}
    \end{subfigure}
    ~
        \centering
    \begin{subfigure}[t]{0.32\linewidth}
        \centering
        \includegraphics[width=\linewidth]{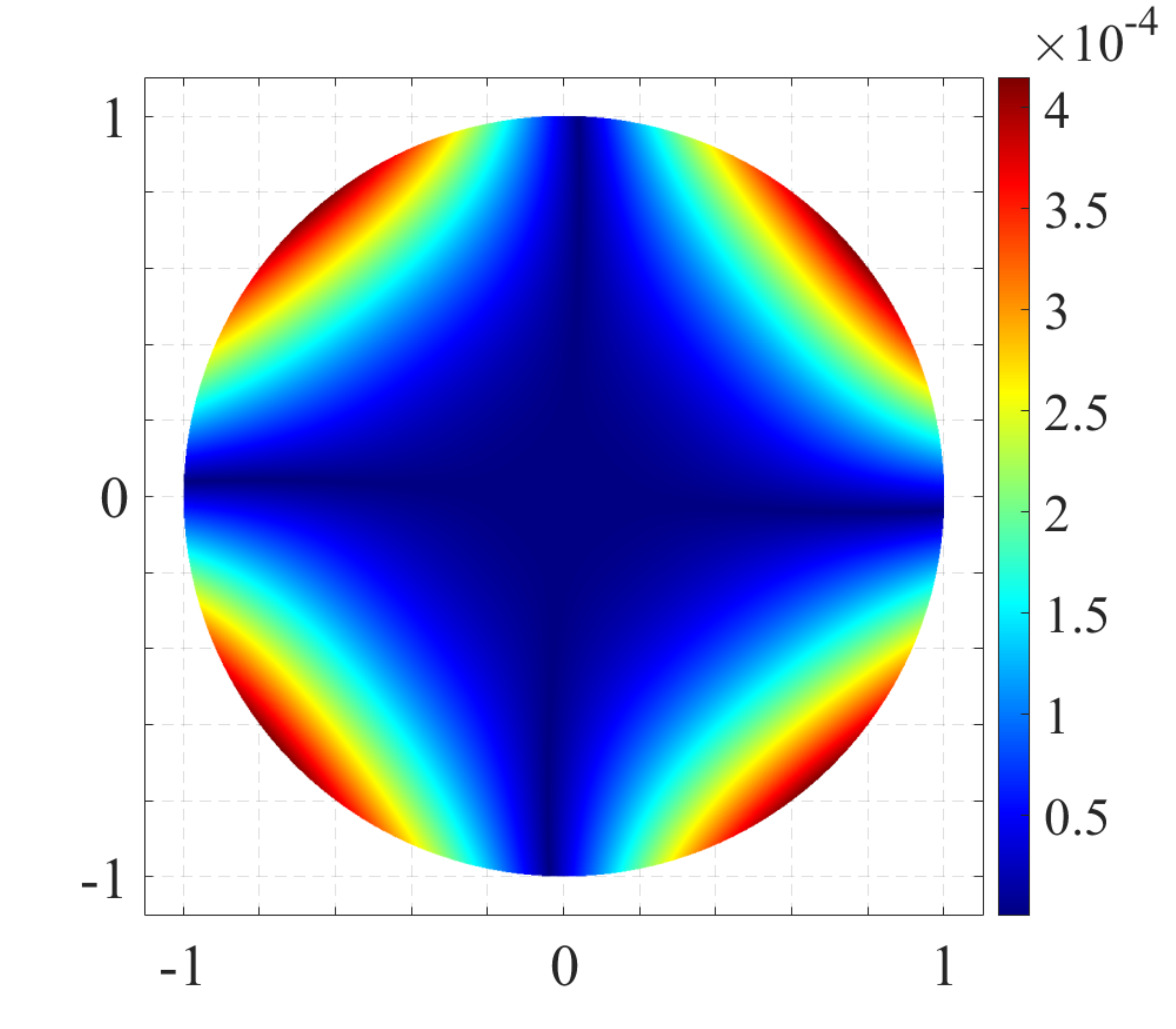}
        \caption{$\| \fm \|   = \| \bm{j} \times \bm{b} \|$ -- aluminum}
    \end{subfigure}%
    \caption{Comparison of the total magnetic body force $\| \fm \|$ (normalized by $\rho_0 R_1 \Omega^2$) for the base case motor
    with steel, copper and aluminum rotors. Notice that the magnetic body force is the Lorentz force $\bm{j} \times \bm{b}$ for the two non-magnetic materials.}
    \label{fig:forces_copalu}
\end{figure}

The results in Figure \ref{fig:forces_copalu} compare the magnetic body force (normalized by $\rho_0 R_1 \Omega^2$) of the base motor for
the different rotor materials. Recall that the magnetic body force is just the Lorentz force for the copper and aluminum rotors, in view of their 
negligible magnetic properties. 
We emphasize again the orders of  magnitude difference  in the magnetic body force between the magnetic (steel) and the non-magnetic (copper, aluminum) materials. The Lorentz forces for the copper and aluminum rotor cases are comparable, given their close electric conductivity (see Table~Table~\ref{tab:values}). Notice however that although the maximum current density is higher in the better conducting copper, the corresponding maximum Lorentz force is higher for the aluminum rotor.

\subsection{Total and elastic stresses}

In order to better assess the influence of the electromagnetic effects on the total $\bm{\sigma}$ and elastic $\sigmae$ stresses, 
we propose to compare them to the purely mechanical (only inertial body forces applied), plane strain elastic stress solution $\sigmai$ for the spinning rotor of the base case motor under angular velocity $\Omega$, a straightforward linear elasticity calculation resulting in the following stress field
\begin{equation}
\displaystyle \sigmasi_{rr} = \frac{\rho_0R_1^2\Omega^2}{8}\left(\frac{3 - 2\nu}{1-\nu} - \frac{3 - 2\nu}{1-\nu}r^2\right)\; , \quad \sigmasi_{r\theta} = 0\; , \quad \sigmasi_{\theta\theta} = \frac{\rho_0R_1^2\Omega^2}{8}\left(\frac{3 - 2\nu}{1-\nu} - \frac{1 + 2\nu}{1-\nu}r^2\right) \; .
\label{eq:elastic-inertia}
\end{equation}
The maximum value for $\sigmasi_{rr}$ and $\sigmasi_{\theta\theta}$ is $[\rho_0 (3 - 2\nu)/8(1-\nu)] (R_1 \Omega)^2$ and occurs at the rotor's center $r=0$. For a more meaningful comparison to the purely mechanical stresses due to inertial effects, all future stress results are normalized by this maximum value, instead of $\rho_0(R_1\Omega)^2$ used thus far.
\begin{figure}[H]
    \centering
    \begin{subfigure}[t]{0.32\linewidth}
        \centering
        \includegraphics[width=\linewidth]{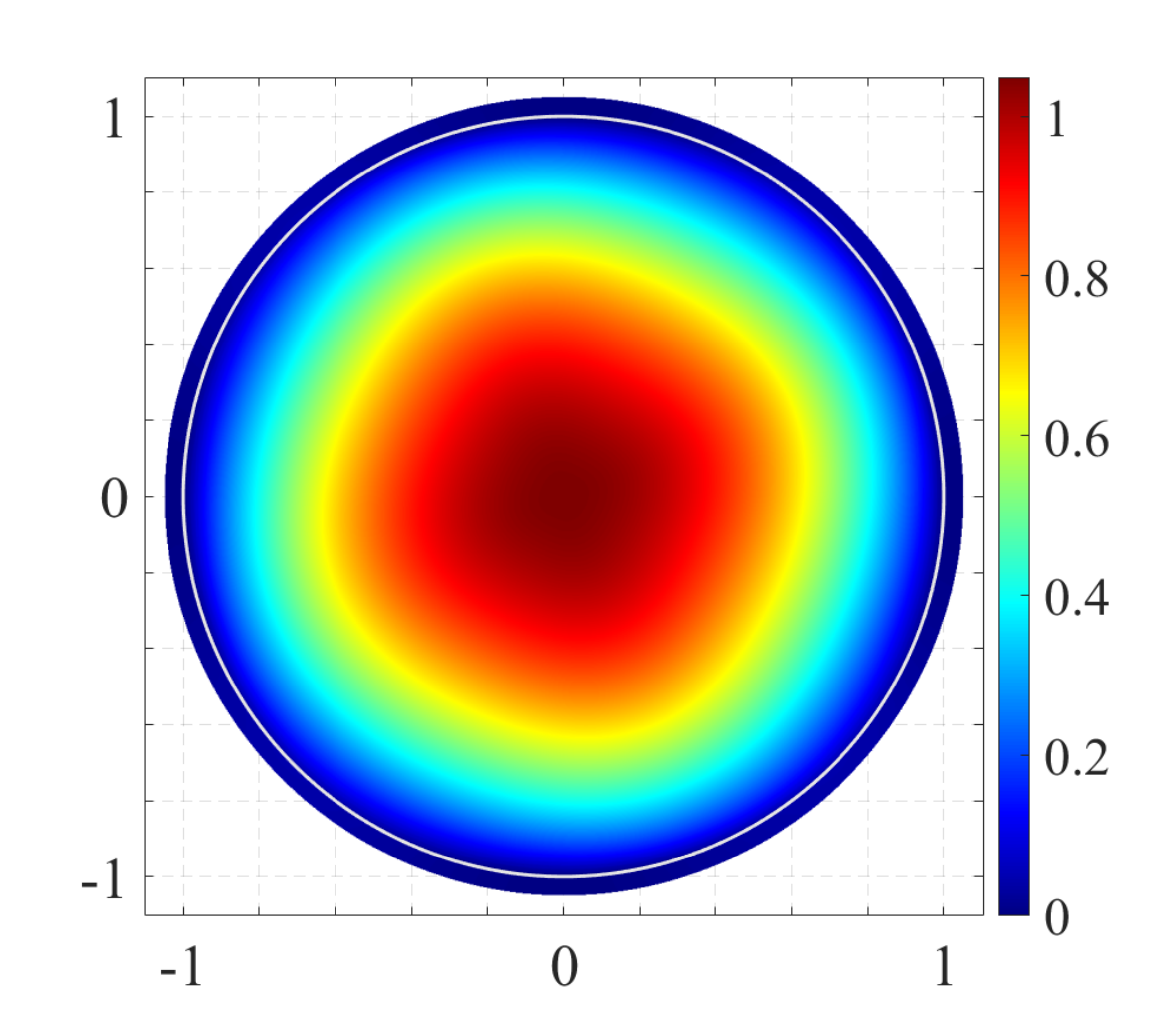}
        \caption{$\sigma_{rr}(r, \Theta)$}
    \end{subfigure}%
    ~ 
    \begin{subfigure}[t]{0.32\linewidth}
        \centering
        \includegraphics[width=\linewidth]{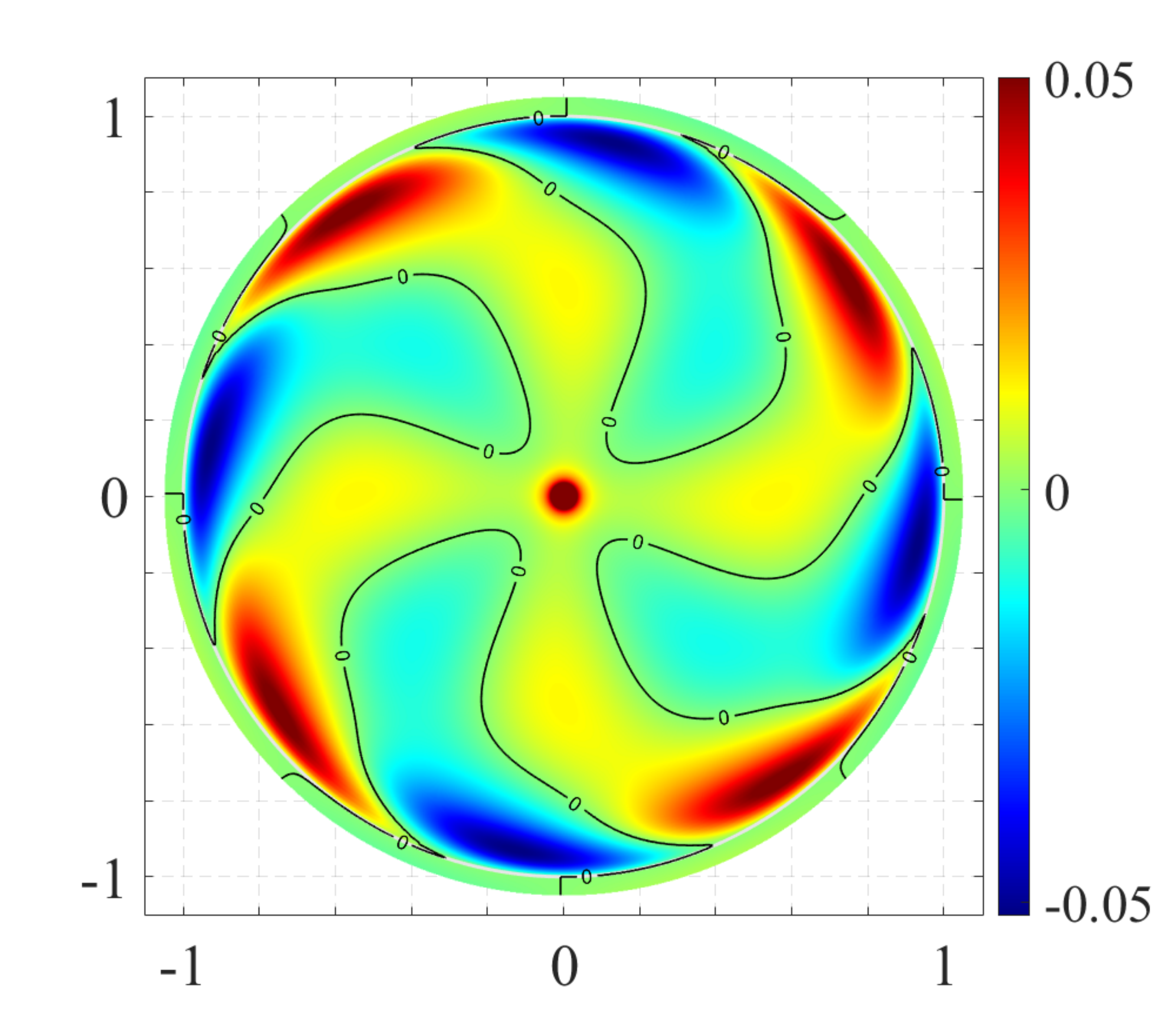}
        \caption{$\sigma_{r\theta}(r, \Theta)$}
    \end{subfigure}
    ~
        \centering
    \begin{subfigure}[t]{0.32\linewidth}
        \centering
        \includegraphics[width=\linewidth]{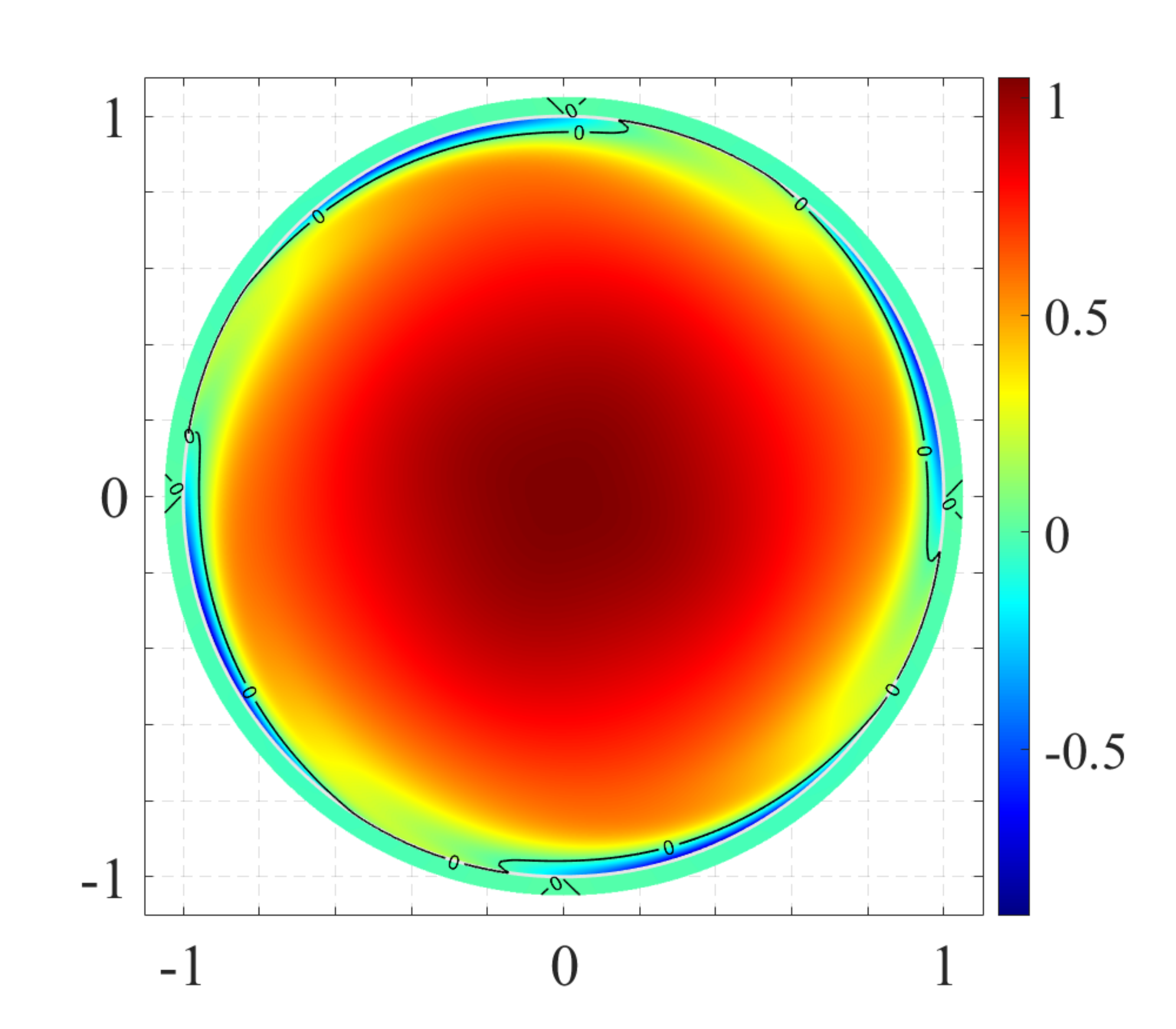}
        \caption{$\sigma_{\theta\theta}(r, \Theta)$}
    \end{subfigure}%
    \caption{Dimensionless total stresses in rotor and airgap (normalized by the maximum inertial stress): (a) normal, (b) shear and (c) hoop, for the base case steel motor.}
    \label{fig:totalstresses}
\end{figure}

The normalized total stress components for the base case motor with the steel rotor are presented in Figure \ref{fig:totalstresses}, -- with the stress fields shown both in the rotor and the airgap -- where one can see the continuity of the normal $\sigma_{rr}$ and shear $\sigma_{r\theta}$ components at the rotor-airgap interface. 

The total normal stress $\sigma_{rr}$ is always positive, never exceeding the maximum, purely inertial value, as seen in Figure~\ref{fig:totalstresses}(a). It monotonically increases away from the rotor's edge and reaches its maximum at the center, region where the electromagnetic effects are negligible,
in contrast to the rotor's edge. The total shear stress $\sigma_{r\theta}$ varies symmetrically between approximately $\pm 5\%$ of the maximum (normal) inertial stress\footnote{The rotor has no shear stresses for the purely inertial loading; plotting the shear stress over the maximum value of the inertial stress (which corresponds to the radial and hoop stresses) allows the comparison of its magnitude with respect to the normal stresses.}, following the angular pattern imposed by the $\cos(2\Theta)$ and $\sin(2\Theta)$ terms. Also notice in Figure~\ref{fig:totalstresses}(b) the singularity in $r=0$ -- truncated in the figure -- due to the external torque applied there. The total hoop stress $\sigma_{\theta\theta}$ is  positive in most of the central domain, where inertial effects dominate, with the same maximum value as for the purely inertial case. The influence of the magnetic field is however evident on the rotor's edge, where a compressive stress of the same absolute value as the maximum inertial stress does appear. 
\begin{figure}[H]
    \centering
    \begin{subfigure}[t]{0.32\linewidth}
        \centering
        \includegraphics[width=\linewidth]{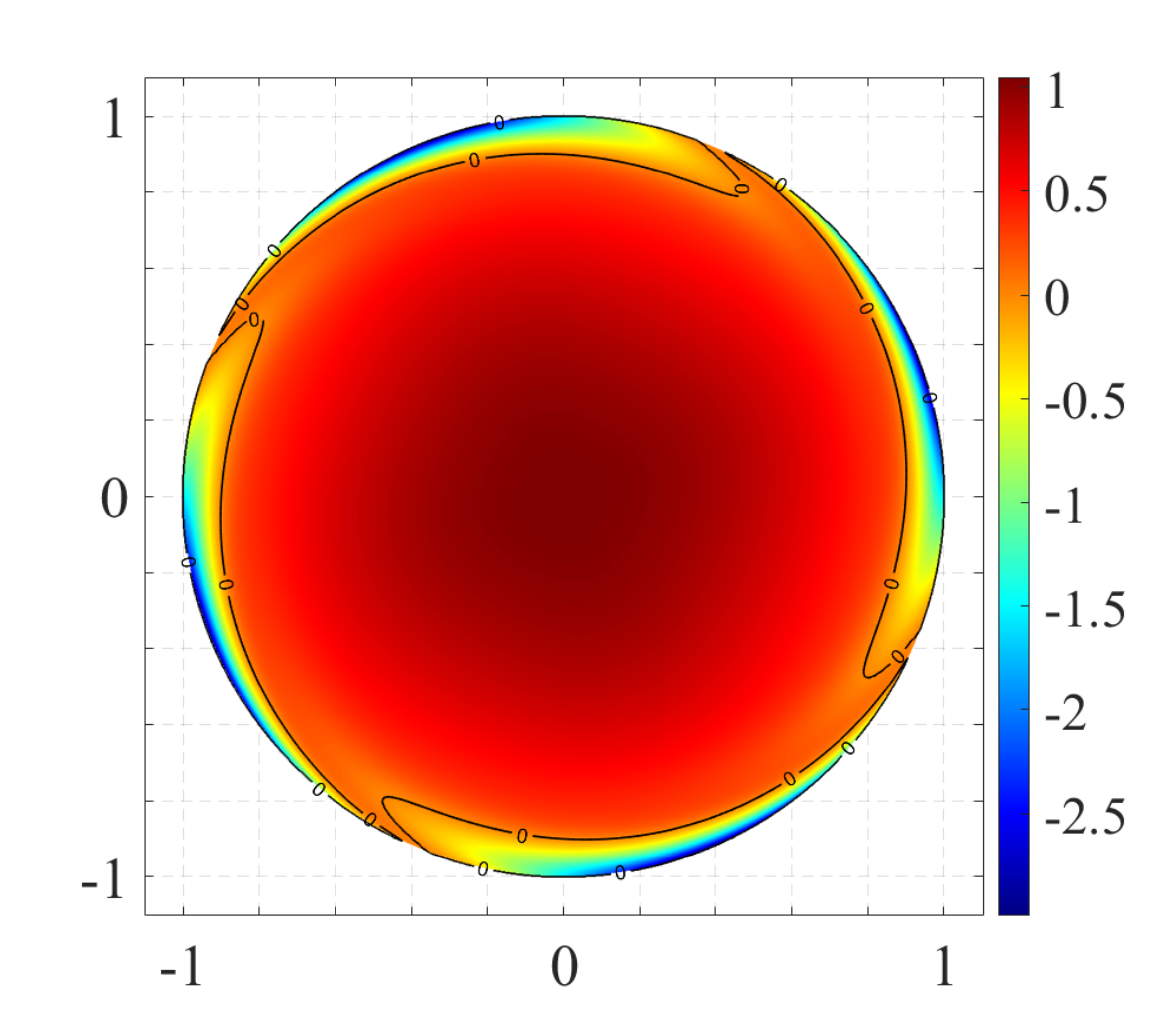}
        \caption{$\sigmase_{rr}(r, \Theta)$}
    \end{subfigure}%
    ~ 
    \begin{subfigure}[t]{0.32\linewidth}
        \centering
        \includegraphics[width=\linewidth]{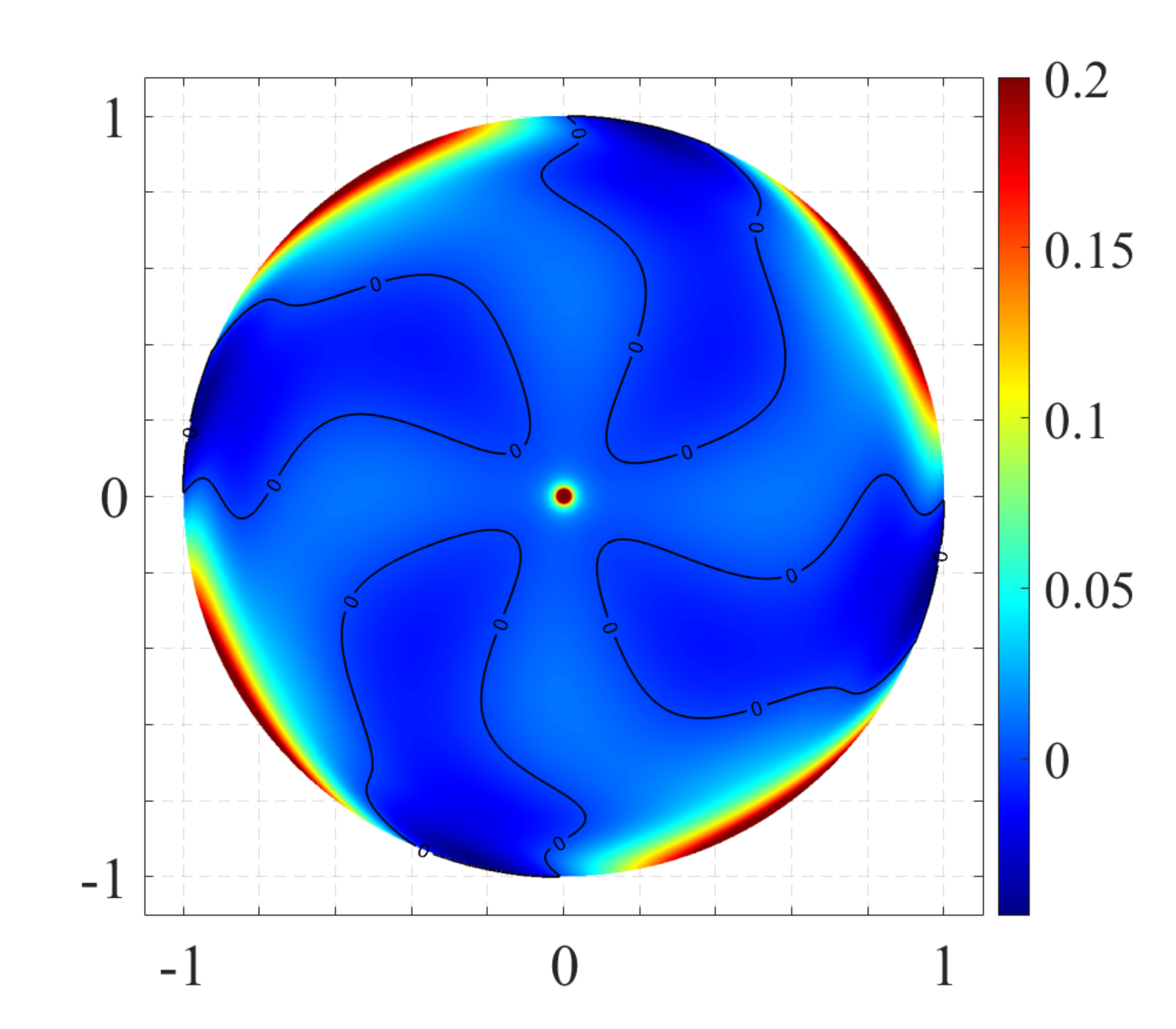}
        \caption{$\sigmase_{r\theta}(r, \Theta)$}
    \end{subfigure}
    ~
        \centering
    \begin{subfigure}[t]{0.32\linewidth}
        \centering
        \includegraphics[width=\linewidth]{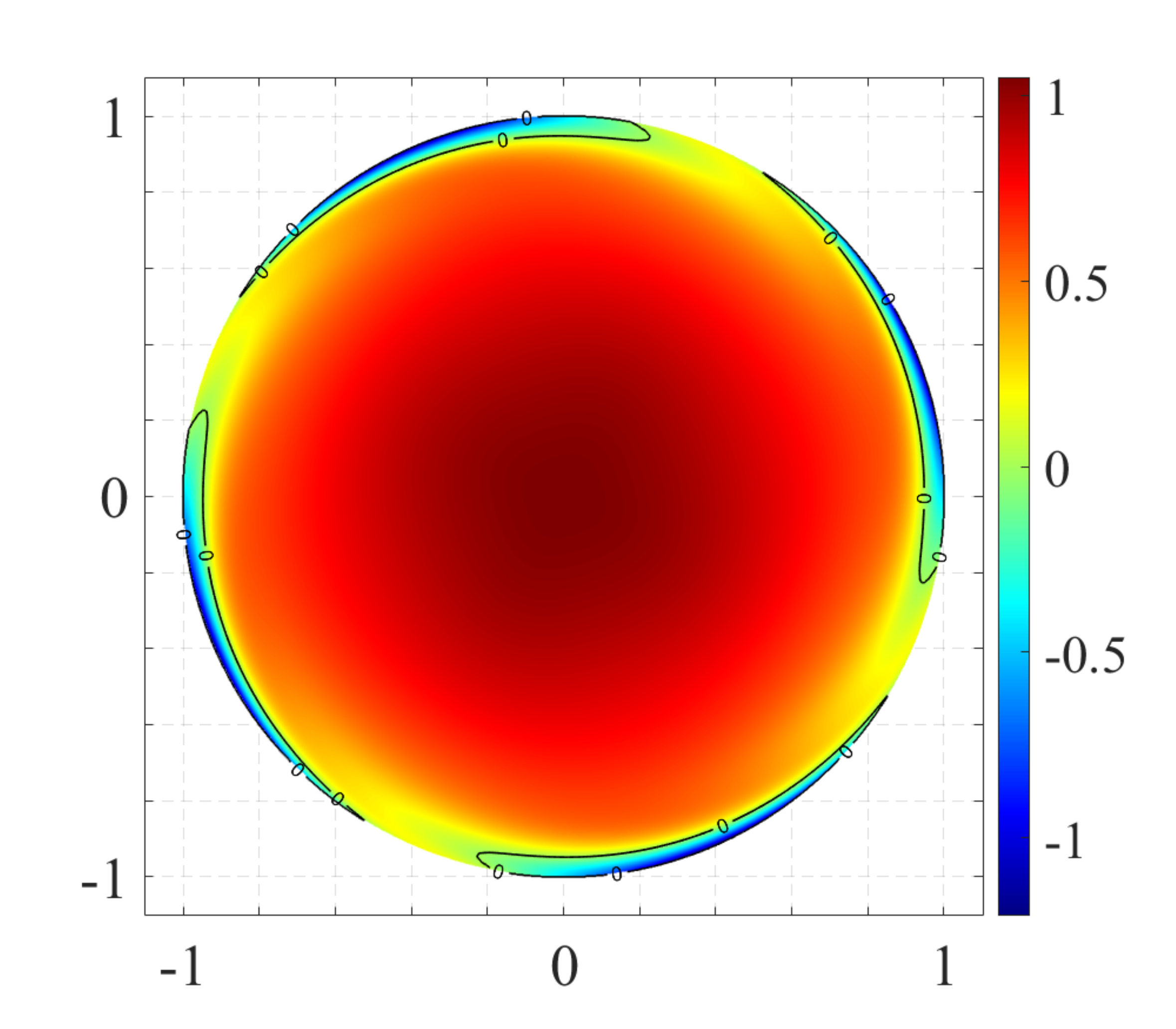}
        \caption{$\sigmase_{\theta\theta}(r, \Theta)$}
    \end{subfigure}%
    \caption{Dimensionless elastic stresses in rotor (normalized by the maximum inertial stress): (a) normal, (b) shear and (c) hoop, for the base case steel motor.}
    \label{fig:elasticstresses}
\end{figure}

The normalized elastic stress $\sigmae$ components in the rotor are given in Figure~\ref{fig:elasticstresses} and differ
significantly from their total stress counterparts $\bm \sigma$, as a simple comparison between Figure~\ref{fig:totalstresses} and Figure~\ref{fig:elasticstresses} shows. The elastic stress components are approximately their inertial counterparts $\sigmai$, given by \eqref{eq:elastic-inertia}, due to the weak magnetic fields at the center of the rotor. However, due to the strong magnetic fields at the rotor boundary, boundary layers develop near its edge resulting
in strong compressive components, up to $250\%$ times for the normal and $125\%$ for the hoop components respectively, higher than the corresponding maximal inertial stress. For the shear stress component, a comparison between Figure~\ref{fig:totalstresses}(b) and Figure~\ref{fig:elasticstresses}(b) shows larger elastic shear stresses, in particular near the rotor's edge, due to the mechanical torque produced.

\subsection{Rotor torque}

\begin{figure}[H]
\centering
    \includegraphics[width=0.45\linewidth]{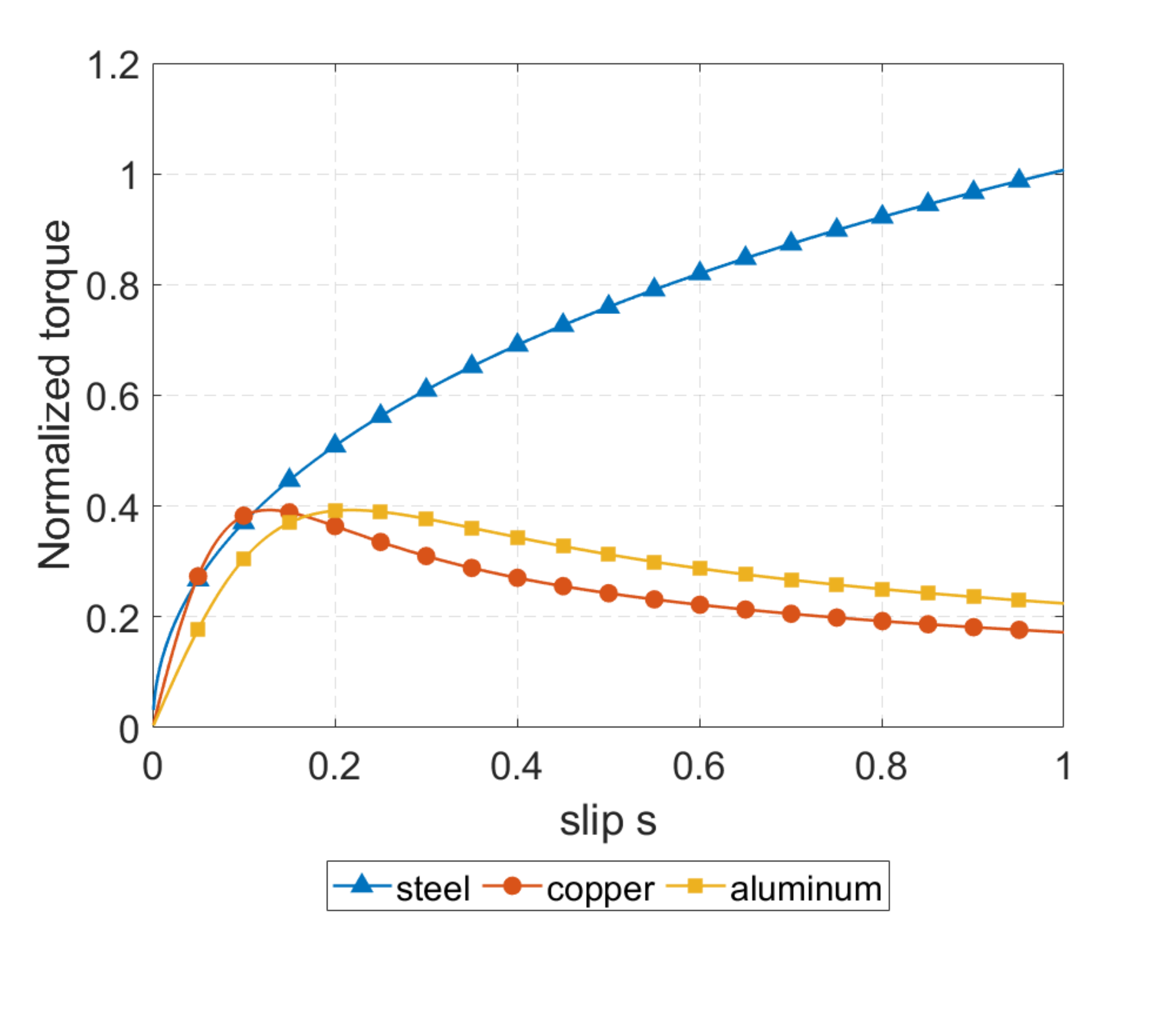}
    \caption{Dimensionless torque $\mathcal T$ (normalized by $\pi \rho_0R_1^4\Omega^2s_0$) vs slip coefficient $s=\omega_r/\omega$ for the base case motor with three different rotor materials.}
    \label{fig:torque}
\end{figure}

The torque $\mathcal T$, normalized\footnote{The normalization quantity is the product of the rotor's area $\pi R_1^2 $ by the electromagnetic stress term $\rho_0R_1^2\Omega^2s_0 = \mu_0\kappa_0^2$.} by $\pi \rho_0R_1^4\Omega^2s_0$, is plotted in Figure~\ref{fig:torque} as a function of the slip coefficient $s$. For low $\kappa_0$ values (where the magnetic field remains below the saturation level for steel for all slip values considered\footnote{As shown in \eqref{sec:results-mag}, the peak value of the magnetic field increases with slip.}) the steel rotor shows higher torque than its copper and aluminum counterparts across almost all the slip range, only slightly dominated by the copper rotor in a region around $5-10\%$ slip. 

For high $\kappa_0$ values, the monotonic increase of the torque as a function of slip for steel  -- due to its linear magnetic response -- is misleading, as saturation may occur, which is not accounted for in the model. In the base case motor, the magnetic field for the steel rotor is already close to saturation for $\kappa_0=1.3\times10^4, s=2\%$ with a value of $1.3$T (see Figure~\ref{fig:magfield_slip}(a)). In this case, it is expected that due to magnetic saturation, the steel torque-slip curve above $s = 2\%$ should be reaching a maximum torque, as is the case for the copper and aluminum rotors.  For $s=5\%$ or higher, the copper motor would produce a larger torque than its steel counterpart.
\section{Conclusion}
\label{sec:conclusion}

Using the direct approach of continuum mechanics, based on \cite{Kovetz2000}, a general framework that couples the electromagnetic, thermal and mechanical effects is derived and subsequently applied to formulate the boundary value problem for electric motors. Particular attention is paid to the derivation of the coupled constitutive equations for isotropic materials under small strain but arbitrary magnetization. As a first application, the theory is employed for the analytical modeling of an idealized asynchronous motor for which we calculate the magnetic, thermal, stress fields and its torque. 
To better assess the influence of magnetization on stresses,  three different rotor materials are examined: electric steel, copper and aluminum and different airgap and slip parameters are considered using realistic geometric and operational regime values (see  \cite{LUBINIM2011}) and material parameters (see \cite{Aydin2017}). Given the linearized magnetic constitutive model adopted  for the sake of an analytical solution, the applied current amplitude is chosen to produce magnetic fields below saturation levels.

Magnetic field results show, as expected, the presence of a boundary layer at the rotor's edge for the steel case and more diffuse patterns for the non-magnetic materials; about two order of magnitude difference is observed in the maximum magnetic field between the magnetic and non-magnetic materials. Comparing the Lorentz, magnetization and the magnetostrictive forces in the steel rotor case we find that the first are negligible (more than three orders of magnitude less for the first compared to the last two). Moreover, magnetostrictive body forces -- resulting from the constitutive coupling between stress and magnetization effects -- although smaller that their magnetic counterparts, are considerably higher than the Lorentz. This is an important finding of our calculations, since the former are usually neglected in the electric motor calculations available in the literature. As expected the magnetic body forces found in the steel rotor are concentrated along a boundary layer and significantly higher than their counterparts 
for the non-magnetic materials that are more diffusely distributed, thus explaining the importance of magnetic rotors for the production of a much higher torque for a given current amplitude, as long as the magnetic field remains below saturation levels.

Due to the realistic thermal parameters used in the calculations, the temperature increase in the rotor is negligible with the temperature maximum appearing in the rotor's center. Significant differences are found in the current density distribution between the magnetic and non-magnetic materials, with a boundary layer appearing in the first and diffuse patterns in the second case.

The analytical solution of the model allows the comparison of the different parts of the
stress tensor (elastic and total) to the purely mechanical stresses due to inertia, revealing the
significant influence of electromagnetic phenomena on the resulting stress state. Although the maximum value of total stress' normal
components never exceed their purely inertial counterparts, the corresponding elastic stress components do so by developing a stress
concentration boundary layer where compressive radial and hoop stresses can be up to three times higher than the maximum inertial value.
Moreover, elastic shear stresses are considerably higher than the total stress and concentrated on a boundary layer about the rotor's edge.

In spite of its limitations, the proposed analytical model shows clearly the importance of correctly accounting for the coupled magneto-mechanical
effects for the accurate calculation of the stress fields. The proposed methodology for solving general boundary value problems is
applicable to more complicated motor geometries and nonlinear constitutive responses that include moderate strains, magnetic saturation
and dissipative effects. For these problems, a numerical approach based on coupled variational principles is necessary (e.g. see \cite{THOMAS2009})
as well as special numerical techniques for the time-dependent aspects of the problem; further research is planned in this direction. \vspace{0.75cm}


{\bf ACKNOWLEDGMENTS}

The work of N.\ H.\ is supported by a Fellowship from the \textit{Andr\'e Citro\"en Chair} of the Ecole Polytechnique.

\section{References}
\bibliographystyle{elsarticle-harv}
\biboptions{authoryear}
\bibliography{bibliography}


\appendix
\section{Isotropic, small strain, arbitrary magnetization constitutive laws}
\label{appendix:smallstrainlinearization}

The derivation of the constitutive laws for an isotropic magnetoelastic material for small strain $\bm \epsilon$, but arbitrary magnetic field $\bm b$,
although straightforward requires lengthy calculations. Although such calculations have been presented in the literature
a long time ago by \cite{Pao1973}, following the early works on magnetoelasticity by \cite{Brown66}, a direct comparison
with our results is not possible due to the different formulations adopted (e.g. different independent variables of the free
energy densities, different definitions of total stress etc.). Moreover, such derivations are not always done consistently in the available literature; a linearized version
of the invariants is often considered, thus violating the objective nature of the free energy since the small strain tensor $\bm \epsilon$ is not objective. 

Derivations are presented here for two different scenarios: the first assumes the most general form of Helmholtz free energy $ \psih(I_k,J_k,T)$ and 
the second is based on  the decoupled form $\psih = \psih_e(I_k) + \psih_m(J_k) + \psih_{th}(T)$ proposed in \eqref{eq:decoupledenergy}. 
In both cases terms in $\bm {\epsilon} \; \bm{ b}$ are kept, providing a more general result than the one presented in \eqref{eq:smallearbitrb}.

\underline{i) General form of free energy $\psi =  \psih(I_1,I_2,I_3,J_1,J_2,J_3,T)$}
Recall that the current configuration expressions for the magnetization and total stress in \eqref{eq:constit_eddycurrent} are
found by differentiating the Helmoltz free energy $\psih(\bm {C}, \bm {B}, T)$. In the case of an isotropic material $\psih(\bm {C}, \bm {B}, T) = \psih(I_1,I_2,I_3,J_1,J_2,J_3,T)$ whose invariants
are expressed in terms of the right Cauchy-Green tensor $\bm C \equiv \bm F^T \contr\bm F $ and $\bm B \equiv \bm {b}\contr \bm {F}$ according to \eqref{eq:decoupledenergy}.

Applying the chain rule of differentiation to the expressions in \eqref{eq:constit_eddycurrent}, one obtains
\begin{equation}
\begin{array}{rl}
	\bm{m} \!\!\! & = \displaystyle - 2{\rho_0\over\sqrt{I_3}}\left( \frac{\partial \psih}{\partial J_1}\bm{I} + \frac{\partial \psih}{\partial J_2}\bm{c} + 
	\frac{\partial  \psih}{\partial J_3}\bm{c}^{2} \right)\contr\bm{b} \; , \vspace{0.2cm} \\
	\bm \sigma  \!\!\! & = \displaystyle 2{\rho_0\over\sqrt{I_3}}\bigg[ \frac{\partial \psih}{\partial I_1}\bm{c} + 
	\frac{\partial \psih}{\partial I_2}(\tr{\bm{c}}\bm{c}-\bm{c}^2) + 
	\frac{\partial \psih}{\partial I_3}\det{\bm{c}}\bm{I} - \frac{\partial \psih}{\partial J_1}\bm{b}\bm{b} + 
	\frac{\partial \psih}{\partial J_3}(\bm{c}\contr\bm{b})(\bm{c}\contr\bm{b}) \bigg]  + \vspace{0.2cm}  \\
	& +  \displaystyle  \frac{1}{\mu_0}\Big(\bm{b} \bm{b} - \frac{1}{2}(\bm{b}\contr\bm{b})\bm{I}\Big) - 
	\Big(\bm{m} \bm{b} + \bm{b} \bm{m} - (\bm{b}\contr\bm{m})\bm{I}\Big) \; ,
\end{array}
\label{eq:newconst}
\end{equation}
where the left Cauchy-Green tensor $\bm c \equiv \bm F \contr\bm F^T$ appears naturally in the constitutive relations \eqref{eq:newconst}. The 
subsequent algebra of small strain linearization is considerably simplified by noting that the invariants involved can be alternatively expressed
in terms of $\bm c$ and $\bm b$ as follows
\begin{equation} 
\begin{array}{l}
	I_1 = \displaystyle \tr{\bm{c}}, \quad I_2 = \half(\tr{\bm{c}}^2 - \tr{\bm{c}\contr\bm{c}}), \quad I_3 = \det{\bm{c}} \; ; 
	\quad \bm c \equiv \bm F\contr\bm F^T \; , \vspace{0.2cm}  \\
	J_1 = \displaystyle \bm{b} \contr \bm{b} = \| \bm{b} \|^2, \quad J_2 = \bm{b}\contr \bm{c} \contr \bm{b}, \quad J_3 = \bm{b}\contr\bm{c}^2\contr\bm{b} \; .
\end{array}
\label{eq:newinvar}
\end{equation}

Expanding the expressions in \eqref{eq:newconst} about $\bm{c} = \bm{I}$ up to the first order in the small strain tensor 
$\bm{\epsilon}~\equiv~(1/2)(\grad\bm{u} + \bm{u}\grad)$, for $\|\bm\epsilon \| \ll 1$, we obtain up to $O(\|\bm\epsilon\|^2)$
\begin{equation}
	\bm{m} \approx \displaystyle \bm{m}( \bm c = \bm{ I}, \bm{b}, T) + \frac{\partial\bm{m}}{\partial \bm{c}}\Big|_{\bm c=\bm I}\dcontr2\bm{\epsilon}, 
	\quad \bm{\sigma} \approx \bm{\sigma}(\bm c = \bm{ I}, \bm{b}, T) + \frac{\partial\bm{\sigma}}{\partial \bm{c}}\Big|_{\bm c=\bm I}\dcontr2\bm{\epsilon} \; ; 
	\quad \bm{c} - \bm {I} \approx 2 \bm\epsilon \; .
	\label{eq:newlinear}
\end{equation}

After lengthy algebraic manipulations of \eqref{eq:newconst} and \eqref{eq:newlinear}, the following expression for the magnetization $\bm{m}$ 
is found involving the scalar quantities $\zeta_i(\| \bm{b}\|)\; , i=1,\cdots, 4$
\footnote{A further simplification can be made for small strains in the expression of $\zeta_2$: since $|-\zeta_1\tr{\bm\epsilon}\bm b| << |-\zeta_1\bm b| $, one
has $\zeta_2 \approx - 4\rho_0\left[ \frac{\partial}{\partial I_1} + 2\frac{\partial}{\partial I_2} + \frac{\partial}{\partial I_3}\right]\left[ \frac{\partial \psih}{\partial J_1} + \frac{\partial \psih}{\partial J_2} + \frac{\partial \psih}{\partial J_3} \right]_{\bm c = \bm I}$.}
\begin{equation}
\begin{array}{rl}
	\bm{m} \!\!\! & = \displaystyle \zeta_1 \bm{b} + \zeta_2 \tr{\bm\epsilon}\bm{b} + 
	\zeta_3 (\bm b\contr\bm\epsilon\contr\bm b)\bm{b} + \zeta_4 \bm{\epsilon}\contr\bm{b} \; ;  \vspace{0.2cm} \\
	\zeta_1(\| \bm{b} \|) \!\!\! & \equiv \displaystyle  - 2\rho_0\left[ \frac{\partial \psih}{\partial J_1} + \frac{\partial \psih}{\partial J_2} + 
	\frac{\partial \psih}{\partial J_3} \right]_{\bm c = \bm I} \; , \vspace{0.2cm} \\
	\zeta_2(\| \bm{b} \|) \!\!\! & \equiv \displaystyle - \zeta_1(\| \bm{b} \|) - 4\rho_0\left[ \frac{\partial}{\partial I_1} + 2\frac{\partial}{\partial I_2} + 
	\frac{\partial}{\partial I_3}\right]\left[ \frac{\partial \psih}{\partial J_1} + \frac{\partial \psih}{\partial J_2} + 
	\frac{\partial \psih}{\partial J_3} \right]_{\bm c = \bm I} \; ,  \vspace{0.2cm} \\
	\zeta_3(\| \bm{b} \|) \!\!\! & \equiv \displaystyle - 4\rho_0\left[ \frac{\partial}{\partial J_2}+ 2\frac{\partial}{\partial J_3}\right]\left[ \frac{\partial \psih}{\partial J_1} + 
	\frac{\partial \psih}{\partial J_2} + \frac{\partial \psih}{\partial J_3} \right]_{\bm c = \bm I} \; ,  \vspace{0.2cm} \\
	\zeta_4(\| \bm{b} \|) \!\!\! & \equiv \displaystyle - 4\rho_0\left[ \frac{\partial \psih}{\partial J_2} + 2\frac{\partial \psih}{\partial J_3}\right]_{\bm c = \bm I} \; .
\end{array}
\label{eq:smallmagnetic}
\end{equation}

The corresponding  small strain linearization expressions yield a total stress $\bm{\sigma}$  as the sum of an elastic $\sigmae$, a magnetic $\sigmam$
and a magnetostrictive $\overset{ms}{\bm{\sigma}}$ (involving terms of the order $\bm {\epsilon} \; \bm{ b}$) component
\begin{equation}
\begin{array}{rl}
	\bm{\sigma} \!\!\! & =\displaystyle \overset{e}{\bm{\sigma}} + \overset{m}{\bm{\sigma}} + \overset{ms}{\bm{\sigma}}\; ;   \vspace{0.2cm}  \\
	\overset{e}{\bm{\sigma}} \!\!\! & \equiv \displaystyle \lambda \tr{\bm\epsilon}\bm{I} + 2G \bm{\epsilon}\; ,   \vspace{0.2cm}  \\
	\overset{m}{\bm{\sigma}} \!\!\! &\equiv \displaystyle \frac{1}{\mu_0}\left[ \bm{b}\bm{b} - \half(\bm{b}\contr\bm{b})\bm{I}\right] - 
	\zeta_1\left[ \bm{b}\bm{b} - (\bm{b}\contr\bm{b})\bm{I}\right] - \frac{\zeta_4}{2}\bm{b}\bm{b} \; ,   \vspace{0.2cm}  \\
	\overset{ms}{\bm{\sigma}} \!\!\! & \equiv \displaystyle  \Sigma_0 \bm{I} + [\Sigma_1\bm{b}\bm{b} + \zeta_2(\bm b\contr\bm b)\bm{I} ] \tr{\bm\epsilon} + 
	[\Sigma_2 \bm{I} + \Sigma_4 \bm{b}\bm{b} + \zeta_3 (\bm{b}\contr\bm{b})\bm{I}](\bm b\contr\bm\epsilon\contr\bm b) + 
	\Sigma_3 [ (\bm b\contr \bm \epsilon)\bm{b} + \bm{b}(\bm \epsilon \contr \bm b) ] \; , \vspace{0.2cm}  \\
	\lambda(\| \bm{b} \|) \!\!\! & \equiv \displaystyle 2\rho_0 \left[ \frac{\partial \psih}{\partial I_3} - \frac{\partial \psih}{\partial I_1}\right]_{\bm c = \bm I} + 
	4\rho_0\left[ \left( \frac{\partial}{\partial I_1} + 2\frac{\partial}{\partial I_2} + \frac{\partial}{\partial I_3}\right)\left(\frac{\partial \psih}{\partial I_1} + 
	2\frac{\partial \psih}{\partial I_2} + \frac{\partial \psih}{\partial I_3}\right)\right]_{\bm c = \bm I}  \; ,  \vspace{0.2cm}   \\
	G(\| \bm{b} \|) \!\!\! &= \displaystyle 2\rho_0\left[ \frac{\partial \psih}{\partial I_1} + \frac{\partial \psih}{\partial I_2} \right]_{\bm c = \bm I}  \; , \vspace{0.2cm}   \\
	\Sigma_0(\| \bm{b} \|) \!\!\! &\equiv \displaystyle 2\rho_0\left[ \frac{\partial \psih}{\partial I_1} + 2\frac{\partial \psih}{\partial I_2} + 
	\frac{\partial \psih}{\partial I_3}\right]_{\bm c = \bm I}  \; ,   \vspace{0.2cm}   \\
	\Sigma_1(\| \bm{b} \|) \!\!\! & \equiv\displaystyle - \zeta_2(\| \bm{b} \|) - \half\zeta_4(\| \bm{b} \|) + \Sigma_2 (\| \bm{b} \|) \; ,   \vspace{0.2cm}  \\
	\Sigma_2(\| \bm{b} \|) \!\!\! & =\displaystyle \zeta_4(\| \bm{b} \|) + 4\rho_0\left[ \left( \frac{\partial}{\partial J_2} + 
	2\frac{\partial}{\partial J_3}\right) \left( \frac{\partial \psih}{\partial I_1} + 2\frac{\partial \psih}{\partial I_2} + 
	\frac{\partial \psih}{\partial I_3} \right) \right]_{\bm c = \bm I} \; ,   \vspace{0.2cm}  \\
	\Sigma_3(\| \bm{b} \|) \!\!\! & \equiv\displaystyle - \zeta_4(\| \bm{b} \|) + 4\rho_0\left[ \frac{\partial \psih}{\partial J_3} \right]_{\bm c = \bm I} \; ,   \vspace{0.2cm}  \\
	\Sigma_4(\| \bm{b} \|) \!\!\! & \equiv\displaystyle - \zeta_3(\| \bm{b} \|) + 4\rho_0\left[ \left( \frac{\partial}{\partial J_2} + 
	2\frac{\partial}{\partial J_3}\right) \left( \frac{\partial \psih}{\partial J_2} + 2\frac{\partial \psih}{\partial J_3} \right) \right]_{\bm c = \bm I} \; ,
\end{array}
\label{eq:smallstress}
\end{equation}
and are expressed in terms of seven magnetic field-dependent coefficients: the two Lam\'e coefficients $\lambda(\| \bm{b}\|)$ and $G(\| \bm{b}\|)$ plus 
five more scalars $\Sigma_i(\| \bm{b}\|)\; , i=0,\cdots, 4$\footnote{A further simplification is possible for small strains: since terms in $\zeta_1\bm\epsilon \bm b\bm b$ (respectively $\zeta_4\bm\epsilon \bm b\bm b$) are negligible in front of terms in $\zeta_1 \bm b\bm b$ (respectively $\zeta_4 \bm b\bm b$), one obtains $\Sigma_1 \approx - \zeta_2 + \Sigma_2$,
$\Sigma_2  \approx 4\rho_0\left[ \left( \frac{\partial}{\partial J_2} + 2\frac{\partial}{\partial J_3}\right) \left( \frac{\partial \psih}{\partial I_1} + 2\frac{\partial \psih}{\partial I_2} + \frac{\partial \psih}{\partial I_3} \right) \right]_{\bm c = \bm I}$ and $\Sigma_3  \approx 4\rho_0\left[ \frac{\partial \psih}{\partial J_3} \right]_{\bm c = \bm I}$}.  This expansion proves that in a first order approximation in $\bm\epsilon$, the coefficients in the expressions for $\bm m$ 
and $\bm \sigma$ depend 
solely on $||\bm b||$. The fact that $\lambda$ and $G$ -- and hence the Young's modulus $E$ -- may depend on $||\bm b||$ is referred to 
as the \textit{$\Delta E$ effect} (see e.g. \cite{Daniel2009}).

\underline{\it ii) Decoupled form of the free energy $\psi = \psih_e(I_1,I_2,I_3) + \psih_m(,J_1,J_2,J_3) + \psih_{th}(T)$}
Under the additional hypothesis of additive decomposition for the specific free energy in \eqref{eq:decoupledenergy},  
one obtains the simplification $\zeta_2 = -\zeta_1$ yielding from \eqref{eq:smallmagnetic} the following expression for the magnetization $\bm{m}$
\begin{equation}
\begin{array}{rl}
	\bm m \!\!\! & = \zeta_1 [1 - \tr{\bm\epsilon}] \bm{b} +  \zeta_3 (\bm b\contr\bm\epsilon\contr\bm b)\bm{b} + 
	\zeta_4 \bm{\epsilon}\contr\bm{b}  \; , \vspace{0.2cm}\\
	\zeta_1(\| \bm{b} \|) \!\!\! & = \displaystyle  - 2\rho_0\left[ \frac{\partial \psih_m}{\partial J_1} + \frac{\partial \psih_m}{\partial J_2} + 
	\frac{\partial \psih_m}{\partial J_3} \right]_{\bm c = \bm I} \; , \vspace{0.2cm} \\
	\zeta_3(\| \bm b \|) \!\!\! & =\displaystyle - 4\rho_0\left[ \frac{\partial}{\partial J_2}+ 2\frac{\partial}{\partial J_3}\right]\left[ \frac{\partial \psih_m}{\partial J_1} + 
	\frac{\partial \psih_m}{\partial J_2} + \frac{\partial \psih_m}{\partial J_3} \right]_{\bm c = \bm I}  \; , \vspace{0.2cm}\\
	\zeta_4(\| \bm b \|) \!\!\! & = \displaystyle - 4\rho_0\left[ \frac{\partial \psih_m}{\partial J_2} + 2\frac{\partial \psih_m}{\partial J_3}\right]_{\bm c = \bm I} \; .
\end{array}
\label{eq:smallmagneticdecoup}
\end{equation}

The corresponding expressions for the elastic $\sigmae$, magnetic $\sigmam$ and magnetostrictive $\overset{ms}{\bm{\sigma}}$ components
of the total stress $\bm \sigma$ simplify from their corresponding counterparts in \eqref{eq:smallstress} into 
\begin{equation}
\begin{array}{rl}
	\overset{e}{\bm{\sigma}}  \!\!\! & = \displaystyle \lambda\tr{\bm\epsilon}\bm{I} + 2G \bm{\epsilon} \; , \vspace{0.2cm}  \\
	\overset{m}{\bm{\sigma}}\!\!\! & = \displaystyle \frac{1}{\mu_0}\left[ \bm{b}\bm{b} - \half(\bm{b}\contr\bm{b})\bm{I}\right] - 
	\zeta_1\left[ \bm{b}\bm{b} - (\bm{b}\contr\bm{b})\bm{I}\right] - \frac{\zeta_4}{2}\bm{b}\bm{b} \; , \vspace{0.2cm}  \\
	\overset{ms}{\bm{\sigma}} \!\!\! & = \displaystyle  [\Sigma_4\bm{b}\bm{b} + \zeta_3(\bm{b}\contr\bm{b})\bm{I}](\bm b\contr\bm\epsilon\contr\bm b) + 
	\Sigma_3\left[ (\bm b\contr\bm\epsilon)\bm{b} + \bm{b}(\bm \epsilon \contr\bm b) \right)] \; ,  
	\end{array}
	\label{eq:smallstressdecoup}
\end{equation}
where the scalars $\zeta_1$, $\zeta_3$, $\zeta_4$ are given in \eqref{eq:smallmagneticdecoup} and $\Sigma_3$ and $\Sigma_4$ 
given in \eqref{eq:smallstress} but with $\psih$ replaced by $\psi_m$.
In deriving \eqref{eq:smallstressdecoup}  from \eqref{eq:smallstress} under the decoupling hypothesis, the pre-stress $\Sigma_0$ and the corresponding Lam\'e coefficients $\lambda, G$ are now constants independent of the magnetic field $\bm b$. It is further assumed that the elastic prestress $\Sigma_0 = 0$. Five functions of $\| \bm b \|$ are thus need to characterize the response of an isotropic, small strain, decoupled-energy, magnetoelastic material: 
$\zeta_1,\zeta_3,\zeta_4,\Sigma_3,\Sigma_4$.

A final remark is in order here to connect the above results to the constitutive equation in \eqref{eq:smallearbitrb} that neglects the magnetostrictive stress component $\overset{ms}{\bm{\sigma}}$. The reason for this simplification
is that for small strains ($\| \bm\epsilon \| \ll 1$) and assuming that the constants appearing in $\sigmam$ and 
$\overset{ms}{\bm{\sigma}}$ are of the same order of magnitude, one deduces that $\| \overset{ms}{\bm{\sigma}} \| \ll 
 \| \sigmam \|$. In the field of dielectric elastomers -- a completely analogous problem where $\bm e \rightarrow \bm b, \ 
\bm p \rightarrow \bm m, \  \varepsilon_0 \rightarrow \mu_0^{-1}$ -- similar results that neglect the coupled terms are justified under the typical hypothesis of small strain and moderate electric field: $\bm \epsilon = O(\zeta)$, $\bm e = O(\sqrt \zeta)$, where $\zeta$ a vanishingly small parameter (e.g. see \cite{Tian2012, Lefevre2017}).
The two coefficients $\zeta_1$ and $\zeta_4$ needed for the determination of $\sigmam$ are related to the magnetic susceptibility $\chi(\| \bm b \|)$ and magnetostrictive coefficient $\Lambda(\| \bm b \|)$ by: 
$\zeta_1(\| \bm b \|) =  \chi(\| \bm b \|) / [\mu_0 (1+\chi(\| \bm b \|))]$ and $\zeta_4(\| \bm b \|) = -2 \Lambda(\| \bm b \|)/[\mu_0 (1+\chi(\| \bm b \|))]$.

\section{Particular and homogeneous solution elastic stress fields}
\label{appendix:stresses}

From \eqref{eq:linksigma-phiV} and \eqref{eq:particular-phi}, the particular solution stress field $\sigmae^{_V}$ components are 
\begin{equation}
\begin{array}{rl} 
	\sigmase^{_V}_{rr} = &\displaystyle\!\!\! V - \half{{1-2\nu}\over{1-\nu}}\left(\frac{2}{r^2}\int_0^r rV_0\mathrm{d}r + (2p-1)r^{^{2p-2}}\int_0^r \frac{V_{cs}}{r^{2p-1}}\mathrm{d}r - \frac{2p+1}{r^{2p+2}}\int_0^r r^{2p+1}V_{cs}\mathrm{d}r \right)  \vspace{0.2cm}  \\
	\sigmase^{_V}_{\theta\theta} = &\displaystyle\!\!\! {\nu V\over{1-\nu}} + \half{{1-2\nu}\over{1-\nu}}\left(\frac{2}{r^2}\int_0^r rV_0\mathrm{d}r - (2p-1)r^{2p-2}\int_0^r \frac{V_{cs}}{r^{2p-1}}\mathrm{d}r + \frac{2p+1}{r^{2p+2}}\int_0^r r^{2p+1}V_{cs}\mathrm{d}r \right)  \vspace{0.2cm}  \\
	\sigmase^{_V}_{r\theta} = &\displaystyle\!\!\! \half{{1-2\nu}\over{1-\nu}}\left((2p-1)r^{2p-2}\int_0^r \frac{V^*_{cs}}{r^{2p-1}}\mathrm{d}r + \frac{2p+1}{r^{2p+2}}\int_0^r r^{2p+1}V^*_{cs}\mathrm{d}r \right)  \vspace{0.2cm}
\end{array}
\end{equation}
where $V^*_{cs} \equiv V_s\cos(2\Theta) - V_c\sin(2\Theta)$ and the $V$ potential components are given by \eqref{eq:potentials}.

From \eqref{eq:linksigma-phih} and \eqref{eq:homogeneous-phi}, the homogeneous solution stress field $\sigmae^{_h}$ components are
\begin{equation}
\begin{array}{rl} 
\sigmase^{_h}_{rr} = & \!\!\! \Phi_{01} + \left((2p-4p^2)\Phi_{c1} r^{2p-2} + (2p + 2 - 4p^2)\Phi_{c2} r^{2p}\right)\cos(2\Theta) \\ 
		&\!\!\! + \left( (2p-4p^2)\Phi_{s1} r^{2p-2} + (2p + 2 - 4p^2)\Phi_{s2} r^{2p} \right)\sin(2\Theta) \vspace{0.2cm} \\ 
\sigmase^{_h}_{\theta\theta} = &\!\!\! \Phi_{01}  + \left( 2p(2p-1)\Phi_{c1} r^{2p-2} + (2p+2)(2p+1)\Phi_{c2} r^{2p} \right)\cos(2\Theta) \\ \vspace{0.2cm}
		&\!\!\! + \left( 2p(2p-1)\Phi_{s1} r^{2p-2} + (2p+2)(2p+1)\Phi_{s2} r^{2p} \right) \sin(2\Theta)  \vspace{0.2cm}   \\ 
\sigmase^{_h}_{r\theta} = &\!\!\! \displaystyle \frac{\Phi_{02}}{r^2} - \left( 2p(2p-1)\Phi_{s1} r^{2p-2} + 2p(2p+1)\Phi_{s2} r^{2p} \right)\cos(2\Theta)  \\
		&\!\!\!+ \left( 2p(2p-1)\Phi_{c1} r^{2p-2} + 2p(2p+1)\Phi_{c2} r^{2p} \right)\sin(2\Theta)  \vspace{0.2cm}
\end{array}
\label{eq:sev}
\end{equation}
Application of the stress boundary condition in \eqref{eq:stress-bc} provides the six $\Phi$ constants of integration in \eqref{eq:sev}.

\section{Experimental determination of the magneto-mechanical coupling coefficient}
\label{appendix:magnetostrictionfit}
\setcounter{figure}{0}

\begin{figure}[H]
\centering
\includegraphics[width=.20\linewidth]{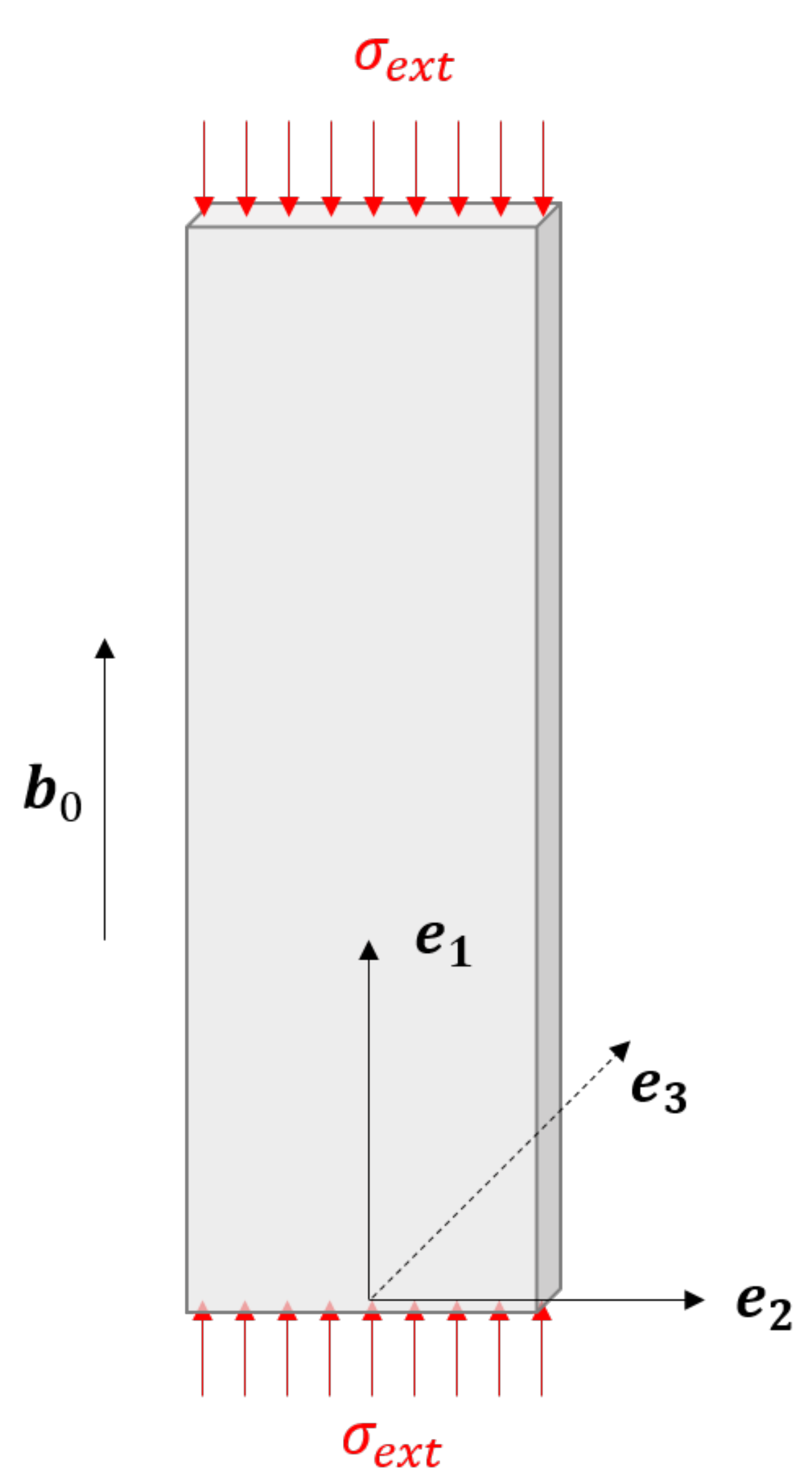}
\caption{Schematics of the magnetostriction setup.}
\label{fig:exp}
\end{figure}

Of all the material constants required for the constitutive model in \eqref{eq:smallearbitrb} only the magneto-mechanical coupling coefficient $\Lambda$
in \eqref{eq:smallearbitrb} is not readily available and needs to be found from experiments. Its determination is based here on results presented
by \cite{Aydin2017} who provide analytical calculations as well as experimental data from \cite{Rekik2014}, 
for the uniaxial magnetostriction vs. the magnetic field for electrical steel samples under different levels of 
mechanical prestress;  a schematic of the setup is depicted in Figure~\ref{fig:exp} based on the description of the typical experimental setup from \cite{Belahcen2006}.

A thin plate of electrical steel is subjected to an external magnetic field $b_0 \base_1$ along its axial direction, resulting in an axial
magnetic field $b_1 = (1 + \chi) b_0$ (assumed uniform) inside the specimen,
where $\chi$ is the material's magnetic susceptibility\footnote{The materials used for holding the plate have no magnetic properties.}.
The plate is also subjected to an externally applied uniaxial stress $\sigma_{ext} \base_{1}\base_{1}$
and hence the total stress $\bm \sigma$ is the sum of the applied stress
and the Maxwell stress in vacuum due to the magnetic field ${\bm b}_0$ 
\begin{equation}
\bm{\sigma}= \sigmae + \sigmam = \sigma_{ext} \base_{1}\base_{1} + \frac{1}{\mu_0}
\left[{\bm{b}}_0 {\bm{b}}_0 - \frac{1}{2}({\bm{b}}_0\contr{\bm{b}}_0)\bm{I}\right]
\label{eq:stress-decompos}
\end{equation}
where the expressions for the elastic and magnetic part of the total stress are given by \eqref{eq:smallearbitrb}.
The corresponding strain and the stress fields in the plate are assumed uniform with edge effects near the corners and edges of the plate neglected. 

Consequently the resulting axial strain $\epsilon_{11}$ is made of an elastic component $\sigma_{ext}/E$ 
plus a component proportional to the square of the magnetic field strength $\zeta_{m} (b_1)^2$, where the
curvature coefficient $\zeta_{m}$ depends on the magnetic constants (susceptibility $\chi$ and magneto-mechanical coupling $\Lambda$). 
A straightforward calculation from  \eqref{eq:stress-decompos} and \eqref{eq:smallearbitrb}, considering
that the specimen's lateral strain is $\epsilon_{22} =  \epsilon_{33}$, gives two
independent equations
\begin{equation}
(\lambda + 2G)\epsilon_{11} + 2\lambda\epsilon_{22} = \sigma_{ext} + {{(b_0)^2} \over {2 \mu_0}} 
\left[1-(1+\chi)^2 - 2 \Lambda(1 + \chi) \right] \; ,\quad
\lambda\epsilon_{11} + 2(\lambda + G)\epsilon_{22} = - {(b_0 \chi)^2 \over 2\mu_0}  \; ,
\label{eq:expsys}
\end{equation}
where the Lam\'e constants are given in terms of Young's modulus $E$ and
Poisson ratio $\nu$ by $G = E/2(1+\nu)$ and $\lambda =\nu E/ (1+\nu)(1-2\nu)$. 
From \eqref{eq:expsys} one obtains the sought relation between the axial strain, the external stress and the magnetic field as well as the
expression for the curvature coefficient $\zeta_m$
\begin{equation}
\epsilon_{11} = {\sigma_{ext}\over E} +\zeta_{m} (b_1)^2\; ; \quad \zeta_{m} = \zeta_{m\chi} + \zeta_{m\Lambda}\; ,\quad 
\zeta_{m\chi}  \equiv \displaystyle -{({1\over 2} -  \nu) \chi^2 + \chi \over  E \mu_0 (1+\chi)^2}\; , 
\  \zeta_{m\Lambda}  \equiv {-\Lambda \over  E \mu_0 (1+\chi)} \; .
\end{equation}
In decomposing the curvature $\zeta_m$ into a magnetic susceptibility $\zeta_{m\chi}$ and a magneto-mechanical 
$\zeta_{m\Lambda}$ component we follow the approach of \cite{Daniel2003}\footnote{In \cite{Daniel2003} and subsequent work by 
this research group by ``\textit{pure  magnetostrictive}'' strains they refer to the strains
due to the magneto-mechanical coupling $\Lambda$.} , where the coefficients $\zeta_{m\chi}$ and $\zeta_{m\Lambda}$ correspond respectively to the magnetic susceptibility $\chi$
and the magneto-mechanical coupling $\Lambda$ parts of the magnetic stress $\sigmam$ defined in \eqref{eq:smallearbitrb}.

For the no external stress case ($\sigma_{ext} = 0$) the data from \cite{Aydin2017}, which are based on the approach
adopted in \cite{Daniel2003}, provide the same magneto-mechanical coupling curvature $\zeta_{m\Lambda}= 2 \; 10^{-6}\; T^{-2}$ for the two materials analyzed. Unfortunately, the values for $\nu$ associated to these materials are not reported there. We assume typical values for steel: $\nu = 0.34$, $E=183$GPa and  a magnetic susceptibility $\chi = 4\; 10^3$, resulting in $\Lambda \approx - 1.8\times 10^3$ which is used in our calculations, as seen in Table~\ref{tab:values}.

\end{document}